%% file: manuscript.tex
\documentclass[final,3p, times,11pt]{elsarticle}
\input{math_commands.tex}

\usepackage{float}
\usepackage[pagewise]{lineno}
\usepackage[dvipsnames]{xcolor}
\usepackage{booktabs}       
\usepackage{framed}
\usepackage{multicol}
\usepackage{amssymb}
\usepackage{todonotes}     
\usepackage[utf8]{inputenc}
\usepackage[ruled, vlined]{algorithm2e}
\usepackage{graphicx}
\usepackage{subcaption}
\usepackage{tabularx}
\usepackage{amsmath}
\usepackage{multirow}
\usepackage{setspace}
\usepackage{microtype}
\usepackage{tabu}
\usepackage{tcolorbox}
\usepackage[export]{adjustbox}
\usepackage[percent]{overpic}
\usepackage{threeparttable}  
\usepackage{tikz}
\usetikzlibrary{arrows.meta}

\DeclareMathAlphabet{\mathcal}{OMS}{cmsy}{m}{n}
\DeclareMathAlphabet\mathbfcal{OMS}{cmsy}{b}{n}

\usepackage{jabbrv}

\usepackage{siunitx} 
\usepackage[colorlinks=true]{hyperref}
\usepackage[noabbrev]{cleveref}

\makeatletter
\renewcommand\paragraph{\@startsection{paragraph}{4}{\z@}%
  {3.25ex \@plus1ex \@minus.2ex}%
  {-1em}%
  {\normalfont\normalsize\bfseries\itshape}}
\makeatother

\usepackage[normalem]{ulem}
\usepackage{color}

\newcommand{\FEsq}{FE$^{\,2}\,$}
\newcommand{\tr}{\mathrm{tr}\,}

\def\appendixname{\!\!}

\makeatletter
\AddToHook{cmd/appendix/before}{\def\cref@section@alias{appendix}\def\cref@subsection@alias{appendix}}
\makeatother

\graphicspath{{figs/}}
\journal{}
\begin{document}
\hypersetup{
    urlcolor=magenta,
    citecolor=blue,
    linkcolor=red
    }
\begin{frontmatter}
\title{Physics-Informed Reduced-Order Operator Learning\\ for Hyperelasticity in Continuum Micromechanics}

\author{Hamidreza Eivazi\corref{cor}}
\author{Henning Wessels}

\address{Division Data-Driven Modeling of Mechanical Systems, Institute of Applied Mechanics, Technische Universität Braunschweig, Pockelsstr. 3, 38106 Braunschweig, Germany}

\cortext[cor]{Corresponding author}
\nonumnote{{Email addresses:} \texttt{\{h.eivazi,h.wessels\}@tu-braunschweig.de} (HE, HW)}

\begin{abstract}
Physics-informed operator learning is an attractive candidate for surrogate modeling of microstructures, especially in multiscale finite-element simulations. Its practical use, however, is often limited by the high cost of loss evaluation. We address this bottleneck by combining the Equilibrium Neural Operator (EquiNO) with the QR-based discrete empirical interpolation method (Q-DEIM). EquiNO learns only the modal coefficients of reduced displacement-fluctuation and first Piola-Kirchhoff stress representations built from periodic and divergence-free bases, thereby enforcing periodicity and mechanical equilibrium by construction. Q-DEIM then identifies a small set of spatial points through a column-pivoted QR factorization of the stress basis and restricts constitutive evaluations during training to these points alone. This makes full-batch second-order optimization practical for three-dimensional representative volume elements (RVEs). Homogenized first Piola--Kirchhoff stresses are recovered directly from the offline-averaged reduced stress modes, without the need to reconstruct the full stress field at inference time.
We validate the framework on two three-dimensional finite-strain hyperelastic RVEs. Q-DEIM reduces the per-step training cost by roughly three orders of magnitude relative to full-field loss evaluation, while reduced homogenization achieves speed-up factors of order $10^3$ to $10^4$ over direct full-field computations. Despite relying on only a small number of offline snapshot loading paths for basis construction, the method accurately interpolates and extrapolates both microscopic stress fields and homogenized stresses, with prediction quality improving systematically as more snapshots are added. These results demonstrate that the integration of constrained reduced representations, hyper-reduced loss evaluation, and reduced homogenization makes physics-informed operator learning substantially more practical for three-dimensional computational micromechanics.
\end{abstract}


\end{frontmatter}

\section{Introduction}
\label{sec:introduction}

Heterogeneous materials derive their effective behavior from complex local deformation and stress fields evolving on the microscale. Numerical homogenization and related multiscale frameworks therefore require the repeated solution of boundary-value problems on representative microstructures, which quickly becomes computationally prohibitive in many-query settings and for nonlinear material behavior \citep{HASSANI1998,GEERS2010,MATOUS2017,MieheKoch2002}. This motivates surrogate models that can approximate not only homogenized quantities but also the underlying local fields in a physically meaningful way.
Operator learning provides a natural framework for such surrogates, where the goal is to learn mappings between function spaces rather than pointwise input-output relations \citep{kovachki2023neural,lu2021learning,li2020fourier}. In practice, however, many successful operator-learning methods are closely connected to reduced-order modeling (ROM): they represent families of solutions in low-dimensional latent spaces, reduced bases, or learned manifolds. Linear constructions such as POD-DeepONet and PCA-based operator learners make this connection explicit \citep{lu2022poddeeponet,bhattacharya2020model}, while more recent works combine operator learning with nonlinear model reduction and manifold decoders to better capture solution sets that are poorly approximated by linear subspaces \citep{Eivazi2024rom,Chen2024,Seidman2022}. 
A low-dimensional state representation alone does not guarantee computational efficiency. For nonlinear problems, the cost of evaluating constitutive responses, residuals, and quadrature terms can still scale with the full discretization. This is the motivation behind hyper-reduction techniques such as DEIM and Q-DEIM \citep{deim,qDEIM}, empirical cubature and related optimized integration rules \citep{Hernandez2017,vanTuijl2019}, and reduced multiscale formulations tailored to computational homogenization \citep{Hernandez2014,Hernandez2020,Raschi2021,Lange2025}. Across these developments, a common theme is that efficient nonlinear surrogates require both reduced representations and reduced evaluation of the underlying physics.

In parallel, neural networks have become an important tool for constitutive modeling. Physics-augmented constitutive surrogates use invariants, energy-based outputs, and architectural constraints to encode objectivity, symmetry, thermodynamic consistency, or convexity by design \citep{linkaetal2021,KLEIN2022,Klein2023,LINDEN2023,rosenkranz2024}. These ideas are especially important for anisotropic hyperelasticity, where structural tensors provide a principled description of material symmetry. Recent works formulate anisotropic hyperelastic surrogates through generalized structural tensors and isotropic tensor functions \citep{Kalina2025} and systematically construct irreducible integrity bases for anisotropic hyperelasticity from the structural tensor concept \citep{Riemer2025}. Moreover, recent constitutive learners such as constitutive Kolmogorov--Arnold networks (CKANs) show that these models can combine expressive nonlinear approximations with symbolic interpretability \citep{Abdolazizi2025}. Reduced representations have also been used directly within learned material models, for example through POD-based plasticity surrogates \citep{Huang2020}. These developments highlight the importance of physics-structured learning, but they primarily target constitutive response modeling rather than the full microscale field operator.

A complementary route is to learn the parametric microscale field response directly. Physics-informed operator learning combines operator networks with governing equations, weak forms, or discretized residuals so that local deformation and stress fields can be predicted while respecting the underlying mechanics \citep{RAISSI2019,Goswami_VDeepONet,Eshaghi_vino,Rezaei2024fol,HARANDI2025}. This perspective is especially attractive for microstructures because it retains access to local fields, interfaces naturally with homogenization, and connects directly to reduced representations of the solution manifold. Yet the corresponding ideas are often developed separately: operator learning emphasizes parametric field mappings, ROM emphasizes low-dimensional structure, hyper-reduction emphasizes efficient nonlinear evaluation, and neural constitutive modeling emphasizes material symmetry and physical admissibility.

\paragraph{Our Contribution} In this paper, we combine these threads in an extension of the Equilibrium Neural Operator (EquiNO) \citep{equino} to finite-strain hyperelastic microstructures. The proposed framework learns only the modal coefficients of periodic displacement-fluctuation and stress representations, enforces periodicity and equilibrium by construction through the reduced basis, and uses Q-DEIM to hyper-reduce the physics-informed loss. In this way, operator learning, reduced-order modeling, and hyper-reduction are integrated into a single physics-informed surrogate for efficient prediction of local fields and homogenized responses in three-dimensional microstructures. The main innovations of the present work are summarized as follows:
\begin{itemize}
    \item The EquiNO framework is extended to three-dimensional finite-strain hyperelastic RVEs while retaining periodicity and equilibrium by construction through the reduced basis representation.
    \item The combination of EquiNO and Q-DEIM reduces the training-side loss evaluation by roughly three orders of magnitude relative to full-field evaluation, which also makes full-batch training of a physics-informed operator practical in the present three-dimensional setting.
    \item With the proposed Q-DEIM reduction, physics-informed training on $233$ unsupervised loading paths takes approximately half the time required for one finite-element simulation of a single loading path with the periodic homogenization solver.
    \item Homogenized stresses are evaluated directly from the reduced stress representation by averaging the modes offline, which enables very fast computation of macroscopic quantities without reconstructing the full-field stress.
\end{itemize}

\paragraph{Structure of the paper} After the introduction, the methodology is presented in \cref{sec:methodology}. The numerical results and discussion are provided in \cref{sec:results-and-discussion}, and \cref{sec:conclusions} closes the paper with a summary and outlook. The source code, data, trained models, and supplementary materials associated with this study will be made available on GitHub.

\section{Physics-informed learning for continuum micromechanics}
\label{sec:methodology}

In this section, we present the methodology used to learn the finite-strain hyperelastic microscale response of heterogeneous RVEs with EquiNO. The central idea is to learn only the modal coefficients of reduced fluctuation and stress representations, while periodicity and equilibrium are enforced by construction through the chosen bases. We first introduce the governing equations of periodic homogenization, then formulate the physics-informed operator-learning problem, discuss the Q-DEIM-based hyper-reduction of the loss, describe the training procedure, and finally summarize the reduced homogenization step used to provide macroscopic quantities for \FEsq\ simulations.

\subsection{Governing equations of hyperelasticity in micromechanics}
\label{subsec:governing-equations}

The microscale problem is formulated as a finite-strain periodic homogenization problem for concurrent \FEsq\ computations. Instead of evaluating a constitutive law directly at the macroscale integration point, an initial-boundary-value problem is solved on a representative volume element (RVE) to obtain the local fields and, through homogenization, the effective stress and tangent response. In this work, we consider quasi-static hyperelastic constituents and restrict attention to periodic boundary conditions at the microscale. For a discussion of alternative boundary conditions in computational homogenization, we refer to \citet{MieheKoch2002}.

Let $\Omega \subset \mathbb{R}^3$ denote the RVE in the reference configuration,
\begin{equation}
    \Omega=\left\{\vx=\left\{x_1,x_2,x_3\right\}^T \in \mathbb{R}^3 \left\lvert\, -\frac{L_i}{2} \leq x_i \leq \frac{L_i}{2}, \, i=1,2,3 \right.\right\},
\end{equation}
where $L_i$ are the edge lengths of the unit cell. The heterogeneous microstructure is composed of matrix and fiber phases such that
\begin{equation}
    \Omega=\Omega^{m}\cup\Omega^{f}, \qquad \Omega^{m}\cap\Omega^{f}=\varnothing.
\end{equation}
Here, $\Omega^{m},\Omega^{f}\subset\mathbb{R}^3$ denote the matrix and fiber domains, respectively.

For a prescribed macroscopic deformation gradient $\bar{\mF}\in\mathbb{R}^{3\times 3}$, the motion of the RVE is decomposed into an affine part and a periodic fluctuation field,
\begin{equation}
    \vvarphi(\vx)=\bar{\mF}\,\vx + \tilde{\vu}(\vx),
\end{equation}
which implies the local deformation gradient
\begin{equation}
    \label{eq:kinematic_finite}
    \mF(\vx)=\nabla \vvarphi(\vx)=\bar{\mF}+\nabla \tilde{\vu}(\vx), \qquad \vx \in \Omega.
\end{equation}
Here, $\vvarphi:\Omega\to\mathbb{R}^3$ denotes the deformation mapping, $\tilde{\vu}:\Omega\to\mathbb{R}^3$ is the displacement fluctuation field, and $\mF:\Omega\to\mathbb{R}^{3\times 3}$ is the deformation gradient. Periodicity is enforced on the fluctuation field. Hence, for two corresponding points $\vx^+\in\partial\Omega^+$ and $\vx^-\in\partial\Omega^-$ on opposite boundaries, we require
\begin{subequations}
\begin{equation}
    \label{eq:bc_fluctuation}
    \tilde{\vu}(\vx^+) = \tilde{\vu}(\vx^-),
\end{equation}
\begin{equation}
    \label{eq:bc_motion}
    \vvarphi(\vx^+) - \vvarphi(\vx^-) = \bar{\mF}\left(\vx^+ - \vx^-\right),
\end{equation}
\begin{equation}
    \label{eq:bc_zero_mean}
    \int_\Omega \tilde{\vu}\, \mathrm{d}\Omega = \vzero,
\end{equation}
\end{subequations}
where $\vx^\pm\in\mathbb{R}^3$, $\partial\Omega^\pm\subset\mathbb{R}^3$, and $\vzero\in\mathbb{R}^3$; \cref{eq:bc_zero_mean} removes rigid-body modes.

The periodic homogenization problem consists of finding $\tilde{\vu}$ such that the balance of linear momentum is satisfied in the reference configuration,
\begin{equation}
    \label{eq:balanceMom}
    \nabla \cdot \mP(\vx) = \vzero, \qquad \vx \in \Omega,
\end{equation}
neglecting body forces and inertia. The stress measure $\mP$ is the first Piola--Kirchhoff stress tensor and follows from a hyperelastic constitutive law
\begin{equation}
    \label{eq:constitutive_hyper}
    \mP(\vx)=\dfrac{\partial \Psi}{\partial \mF}(\vx,\mF),
\end{equation}
where $\mP:\Omega\to\mathbb{R}^{3\times 3}$ and the stored-energy density $\Psi:\Omega\times\mathbb{R}^{3\times 3}\to\mathbb{R}$ is defined phase-wise as
\begin{equation}
    \Psi(\vx,\mF)=
    \begin{cases}
      \Psi^{m}(\mF), & \vx \in \Omega^{m}, \\
      \Psi^{f}(\mF), & \vx \in \Omega^{f}.
    \end{cases}
\end{equation}
In this study, both phases are modeled as isotropic compressible hyperelastic materials with
\begin{equation}
    \Psi^\alpha(\mF)=
    \frac{\mu_\alpha}{2}\left(I_1-\ln I_3-3\right)
    +\frac{\lambda_\alpha}{4}\left(I_3-\ln I_3-1\right), \qquad \alpha\in\{m,f\},
\end{equation}
where $\lambda_\alpha,\mu_\alpha\in\mathbb{R}_{>0}$ are the Lam\'e parameters and
\begin{equation}
    \mC=\mF^T\mF, \qquad I_1=\tr(\mC), \qquad I_3=\det(\mC).
\end{equation}
Thus, $\mC\in\mathbb{R}^{3\times 3}$ and $I_1,I_3\in\mathbb{R}$.
Therefore, for each prescribed macroscopic deformation gradient $\bar{\mF}$, the microscale boundary-value problem determines the fluctuation field $\tilde{\vu}$ as well as the local deformation and stress fields, which are subsequently used to construct the surrogate operator and the homogenized response.

\subsection{Physics-informed operator learning via EquiNO}

The goal of operator learning is to approximate the solution operator of the microscale boundary-value problem. In the present setting, this operator maps a prescribed macroscopic deformation gradient $\bar{\mF}$ to the corresponding local fields on the RVE. In particular, we seek an approximation of the form
\begin{equation}
    \gG_{\vtheta} : (\bar{\mF}, \vx) \mapsto \left(\tilde{\vu}(\vx), \mP(\vx)\right),
\end{equation}
where $\gG_{\vtheta}:\mathbb{R}^{3\times 3}\times\Omega\to\mathbb{R}^3\times\mathbb{R}^{3\times 3}$ is a parameterized model and $\vtheta\in\mathbb{R}^{n_\theta}$ collects its trainable parameters. Unlike conventional supervised operator-learning approaches, which are typically trained from paired input-output data, our objective is to identify this operator from a limited set of snapshots and the governing equations introduced in \cref{subsec:governing-equations}.
EquiNO \cite{equino} follows the POD-DeepONet \cite{lu2022poddeeponet} viewpoint that the field dependence on the spatial coordinate is represented through a reduced basis, whereas the dependence on the loading parameter is represented through learned modal coefficients. Let $p_{\tilde{\vu}}\in\mathbb{N}$ and $p\in\mathbb{N}$ denote the numbers of retained fluctuation and stress modes, respectively. Using the reduced bases introduced above, we define two branch networks,
\begin{subequations}
\begin{equation}
    \vb_{\tilde{\vu}}(\bar{\mF}) \approx \gN_{\tilde{\vu}}(\bar{\mF}; \vtheta_{\tilde{\vu}}),
\end{equation}
\begin{equation}
    \vb_{\mP}(\bar{\mF}) \approx \gN_{\mP}(\bar{\mF}; \vtheta_{\mP}),
\end{equation}
\end{subequations}
which predict the coefficients of the fluctuation and stress bases, respectively. The notation $\gN_{\tilde{\vu}}(\bar{\mF};\vtheta_{\tilde{\vu}})$ and
$\gN_{\mP}(\bar{\mF};\vtheta_{\mP})$ denotes the complete branch maps from the
macroscopic deformation gradient to the modal coefficients. In the implementation,
$\bar{\mF}$ is first transformed to the macroscopic Green--Lagrange strain tensor
$\bar{\mE} = \frac{1}{2}(\bar{\mF}^{T}\bar{\mF}-\mI)$, and the six independent
components of $\bar{\mE}$ are then used as the input features of the neural networks. Here, $\vb_{\tilde{\vu}}(\bar{\mF}),~\gN_{\tilde{\vu}}(\bar{\mF}; \vtheta_{\tilde{\vu}})\in\mathbb{R}^{p_{\tilde{\vu}}}$ and $\vb_{\mP}(\bar{\mF}),~\gN_{\mP}(\bar{\mF}; \vtheta_{\mP})\in\mathbb{R}^{p}$. This yields the reduced representations
\begin{subequations}
\begin{equation}
    \tilde{\vu}(\vx, \bar{\mF}) = \mPhi_{\tilde{\vu}}^T(\vx)\, \gN_{\tilde{\vu}}(\bar{\mF}; \vtheta_{\tilde{\vu}}),
\end{equation}
\begin{equation}
    \tilde{\mF}(\vx, \bar{\mF}) = \bar{\mF} + \nabla \mPhi_{\tilde{\vu}}^T(\vx)\, \gN_{\tilde{\vu}}(\bar{\mF}; \vtheta_{\tilde{\vu}}),
\end{equation}
\begin{equation}
    \tilde{\mP}(\vx, \bar{\mF}; \vtheta_{\mP}) = \mPhi_{\mP}^T(\vx)\, \gN_{\mP}(\bar{\mF}; \vtheta_{\mP}).
\end{equation}
\end{subequations}
The corresponding bases satisfy $\mPhi_{\tilde{\vu}}(\vx)\in\mathbb{R}^{p_{\tilde{\vu}}\times 3}$ and $\mPhi_{\mP}(\vx)\in\mathbb{R}^{p\times r_P}$ with $r_P=9$, so that $\tilde{\vu}(\vx,\bar{\mF})\in\mathbb{R}^3$ and $\tilde{\mF}(\vx,\bar{\mF}),\tilde{\mP}(\vx,\bar{\mF};\cdot)\in\mathbb{R}^{3\times 3}$.
The defining idea of EquiNO is that the fluctuation field is expanded in periodic modes and the first Piola--Kirchhoff stress in divergence-free stress modes. As a result, the predicted fluctuation field satisfies the periodic boundary conditions by construction, while the projected stress field
\begin{equation}
    \tilde{\mP}(\vx,\bar{\mF};\vtheta_{\mP}) = \mPhi_{\mP}^T(\vx)\, \gN_{\mP}(\bar{\mF}; \vtheta_{\mP})
\end{equation}
satisfies the balance of linear momentum by construction, up to discretization errors inherited from the snapshot basis.

To incorporate the constitutive relation, we compute a second stress field from the kinematically admissible deformation gradient,
\begin{equation}
    \tilde{\mP}_{\tilde{\vu}}(\vx, \bar{\mF}; \vtheta_{\tilde{\vu}})
    =
    \frac{\partial \Psi}{\partial \mF}
    \left(
    \bar{\mF} + \nabla \mPhi_{\tilde{\vu}}^T(\vx)\, \gN_{\tilde{\vu}}(\bar{\mF}; \vtheta_{\tilde{\vu}})
    \right).
\end{equation}
This constitutive stress field satisfies the kinematic relation and the constitutive law, while the projected stress field satisfies equilibrium. Training EquiNO therefore amounts to matching these two stress representations over a set of unsupervised macroscopic deformation gradients $\{\bar{\mF}_i^{\,f}\}_{i=1}^{N_f}$,
\begin{equation}
    \label{eq:loss}
    \gL(\vtheta_{\tilde{\vu}},\vtheta_{\mP})
    =
    \dfrac{1}{N_f\,m\,r_P}
    \sum_{i=1}^{N_f}
    \left\|
    \tilde{\mP}_{\tilde{\vu}}(\vzeta,\bar{\mF}_i^{\,f};\vtheta_{\tilde{\vu}})
    -
    \tilde{\mP}(\vzeta,\bar{\mF}_i^{\,f};\vtheta_{\mP})
    \right\|_2^2,
\end{equation}
where $N_f,m,r_P\in\mathbb{N}$, with $m$ denoting the number of spatial evaluation points and $r_P$ the number of stress components. Moreover, $\gL\in\mathbb{R}_{\ge 0}$ is a scalar stress-consistency loss. In contrast to soft-constrained physics-informed operator networks, this formulation does not require separate penalty terms for equilibrium and boundary conditions. Instead, these constraints are hard-enforced through the reduced representation itself, and the optimization is driven by a single stress-consistency loss.

\subsection{Hyper-reduction of the physics-informed loss}
\label{sec:hyper-reduction}
The reduced bases introduced in the previous subsection are obtained by POD from a limited set of microscale snapshots; the detailed basis construction follows our previous EquiNO formulation in \citet{equino}. In the present work, the key additional ingredient is a hyper-reduction of the stress-consistency loss. Although the POD representation already reduces the number of unknowns, evaluating the loss over all spatial points of a three-dimensional microstructure remains expensive. To address this issue, we apply the QR-based discrete empirical interpolation method (Q-DEIM) \citep{qDEIM} to identify a small set of representative spatial points, referred to as magic points, at which the loss is enforced.

Let $\mPhi_{\mP}\in\mathbb{R}^{(m\times r_P)\times p}$ denote the stacked POD basis of the first Piola--Kirchhoff stress field sampled at the $m$ spatial points $\{\vx_j\}_{j=1}^m$, where $r_P=9$ for three-dimensional problems and $p$ is the number of retained stress modes. We reshape this basis into a matrix in which each row corresponds to one spatial point and all stress components and modes are grouped column-wise, resulting in $\widehat{\mPhi}_{\mP}\in\mathbb{R}^{m\times (r_P p)}$. Applying Q-DEIM to the reshaped basis amounts to computing a column-pivoted QR factorization of its transpose,
\begin{equation}
\label{eq:qdeim_qr}
    \widehat{\mPhi}_{\mP}^{\,T}\,\Pi = \mQ\,\mR,
\end{equation}
where $\Pi\in\mathbb{R}^{m\times m}$ is a permutation matrix, $\mQ\in\mathbb{R}^{(r_P p)\times (r_P p)}$, and $\mR\in\mathbb{R}^{(r_P p)\times m}$. The Q-DEIM sampling indices are then given by the first $q$ pivots,
\begin{equation}
\label{eq:qdeim_indices}
    \mathcal{I}_q = \{\,\pi_1,\ldots,\pi_q\,\}, 
    \qquad \{\vx_{\pi_\ell}\}_{\ell=1}^q \subset \{\vx_j\}_{j=1}^m,
\end{equation}
where $q$ denotes the number of selected magic points. In the present work, we choose $q=p$ unless stated otherwise, although the number of Q-DEIM points is not required to coincide with the number of POD modes in general. These indices define a sampling operator $\mS_{\mathcal{I}_q}\in\{0,1\}^{q\times m}$ that extracts the selected spatial locations.

Using this operator, the stress consistency term is evaluated only at the Q-DEIM points instead of over the full spatial discretization. In this representation, the predicted stress field is viewed pointwise as an $m\times r_P$ array, and the operator $\mS_{\mathcal{I}_q}$ selects spatial rows while retaining all $r_P$ stress components at each selected point. For a set of unsupervised macroscopic deformation gradients $\{\bar{\mF}_i^{\,f}\}_{i=1}^{N_f}$, the reduced loss reads
\begin{equation}
\label{eq:loss_qdeim}
    \gL_q(\vtheta_{\tilde{\vu}},\vtheta_{\mP})
    =
    \dfrac{1}{N_f\,q\,r_P}
    \sum_{i=1}^{N_f}
    \left\|
    \mS_{\mathcal{I}_q}\big(\tilde{\mP}(\vzeta,\bar{\mF}_i^{\,f};\vtheta_{\tilde{\vu}})
    -
    \tilde{\mP}(\vzeta,\bar{\mF}_i^{\,f};\vtheta_{\mP})\big)
    \right\|_2^2 .
\end{equation}
Here, $q\in\mathbb{N}$ and $\gL_q\in\mathbb{R}_{\ge 0}$. This hyper-reduced formulation substantially decreases the cost of training while preserving the dominant spatial structure encoded in the stress basis. In practice, Q-DEIM makes the finite-strain three-dimensional setting computationally tractable without altering the reduced-order representation itself.

\subsection{Data generation and training}
\label{subsec:data-generation-training}

EquiNO \citep{equino} is identified in a self-supervised manner from the governing equations and a set of snapshot loading states. Following the sampling strategy described in \citet{Kalina2025}, we sample in the six-dimensional space
\begin{equation}
    \mathcal{S}:=(\bar{\lambda}_1,\bar{\lambda}_2,\bar{J},\theta_1,\theta_2,\theta_3)
    \in \mathbb{R}_{>0}\times \mathbb{R}_{>0}\times \mathbb{R}_{>0}\times \mathbb{R}\times \mathbb{R}\times \mathbb{R}
    ,
\end{equation}
by Latin Hypercube Sampling (LHS) \citep{LHS}. For each sampled state, a loading path from $\bar{\mF}=\mI$ to the final deformation is generated by linearly interpolating in $\bar{\lambda}_1$, $\bar{\lambda}_2$, and $\bar{J}$ within $n_{i}$ increments while keeping $\theta_1$, $\theta_2$, and $\theta_3$ fixed. The resulting deformation states are then used to evaluate the stress-consistency objective in \cref{eq:loss} or, in the hyper-reduced setting, \cref{eq:loss_qdeim}.

The optimization is performed with the Adam optimizer \citep{adam} and the limited-memory Broyden--Fletcher--Goldfarb--Shanno (L-BFGS) algorithm \citep{liu1989} in full-batch mode. This is feasible because Q-DEIM reduces the number of spatial evaluation points entering the loss and thereby lowers the cost of each optimization step sufficiently for three-dimensional finite-strain RVEs. Accordingly, the coefficient networks are trained directly with deterministic full-batch optimization until the loss converges.

\subsection{Computational homogenization}
\label{subsec:homogenization}

For computational homogenization in the finite-strain setting, the quantity of interest is the homogenized first Piola--Kirchhoff stress tensor. For a prescribed macroscopic deformation gradient $\bar{\mF}$, it is defined as the volume average of the microscopic stress field,
\begin{equation}
    \bar{\mP}(\bar{\mF}) = \frac{1}{|\Omega|}\int_{\Omega} \mP(\vx,\bar{\mF})\, \mathrm{d}\Omega .
\end{equation}
Thus, $\bar{\mP}(\bar{\mF})\in\mathbb{R}^{3\times 3}$.
Using the finite element discretization of the RVE, this average is evaluated by numerical quadrature as
\begin{equation}
    \bar{\mP}(\bar{\mF}) \approx \frac{1}{|\Omega|}\sum_{i=1}^{m} \omega_i\, \mP(\vzeta_i,\bar{\mF}),
    \label{eq:homogenization}
\end{equation}
where $\omega_i\in\mathbb{R}_{>0}$ and $\vzeta_i\in\Omega$ denote the quadrature weights and integration points, respectively.

In EquiNO, homogenization is performed directly on the equilibrium-preserving reduced stress representation. It is a post-processing step, only considered after training. Averaging the stress basis offline yields
\begin{equation}
    \bar{\mPhi}_{\mP}
    =
    \frac{1}{|\Omega|}\int_{\Omega} \mPhi_{\mP}(\vx)\, \mathrm{d}\Omega
    \approx
    \frac{1}{|\Omega|}\sum_{i=1}^{m} \omega_i\, \mPhi_{\mP}(\vzeta_i),
\end{equation}
with $\bar{\mPhi}_{\mP}\in\mathbb{R}^{p\times r_P}$, and therefore
\begin{equation}
    \bar{\mP}(\bar{\mF})
    =
    \bar{\mPhi}_{\mP}^T\, \vb_{\mP}(\bar{\mF})
    \approx
    \bar{\mPhi}_{\mP}^T\, \gN_{\mP}(\bar{\mF};\vtheta_{\mP}).
    \label{eq:Ghomogenization}
\end{equation}
Hence, the homogenized response can be computed directly from the predicted modal coefficients without reconstructing the full local stress field over the RVE. This yields an efficient reduced-order evaluation of the macroscopic constitutive response. The material tangent is then obtained as
\begin{equation}
    \bar{\sA} = \frac{\partial \bar{\mP}}{\partial \bar{\mF}},
\end{equation}
where $\bar{\sA}\in\mathbb{R}^{3\times 3\times 3\times 3}$ is the fourth-order material tangent. This quantity can be computed by automatic differentiation of \cref{eq:Ghomogenization}. Thus, EquiNO provides the homogenized first Piola--Kirchhoff stress and tangent quantities required by the macroscale \FEsq\ solver in reduced-order form.

\section{Results and discussion}
\label{sec:results-and-discussion}

In this section, we assess the performance of EquiNO in a sequence of numerical studies. We first introduce the considered microstructures, material parameters, loading conditions and evaluation metrics. We then discuss the computational efficiency of the reduced loss evaluation and reduced homogenization, before assessing in-range and out-of-range generalization for representative RVEs in terms of local and homogenized responses.

\subsection{Problem setup}
\label{subsec:problem-setup}
\begin{figure}[htb]
    \centering
    \hfill
    \begin{subfigure}{0.45\textwidth}
        \centering
        \includegraphics[height=0.55\textwidth]{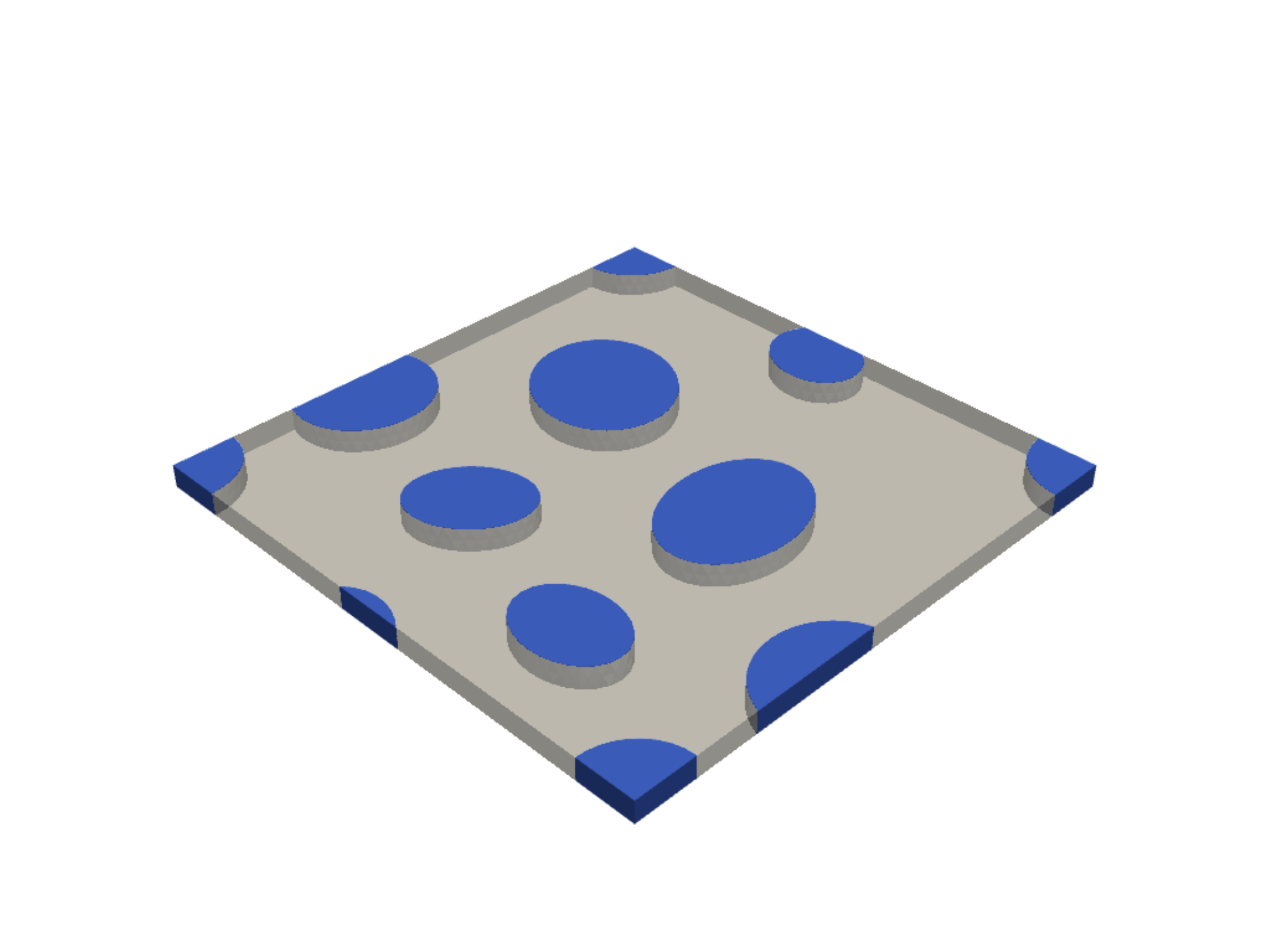}
        \caption{Stochastic fibers.}
        \label{fig:subplot1}
    \end{subfigure}
    \hfill
    \begin{subfigure}{0.45\textwidth}
        \centering
        \includegraphics[height=0.55\textwidth]{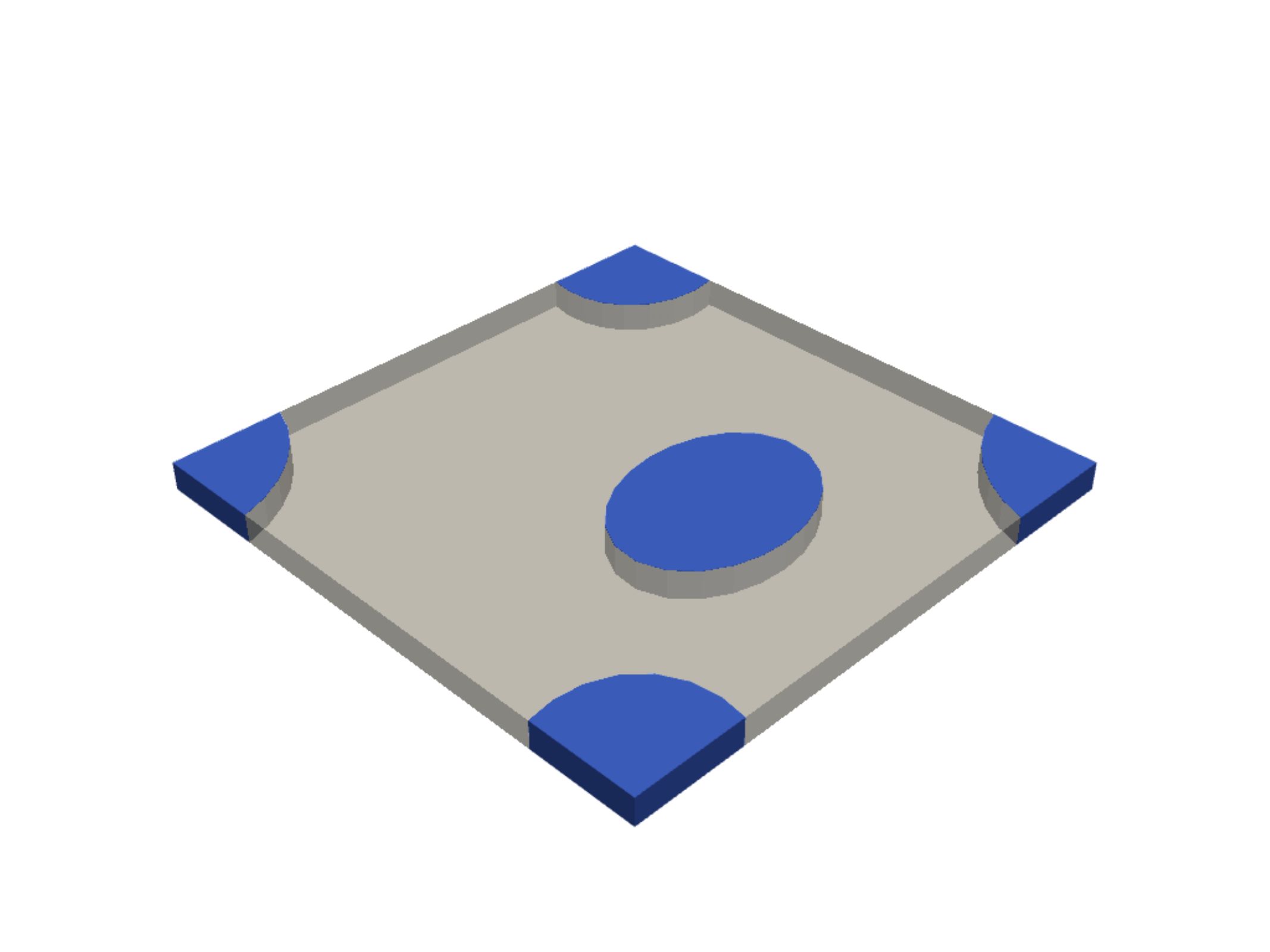}
        \caption{Hexagonal fibers.}
        \label{fig:subplot2}
    \end{subfigure}
    \hfill~
    
    \caption{Microstructures considered in the study. Blue and grey colors indicate the fiber and matrix phases, respectively. The fiber volume fractions are $\phi=26\%$ and $\phi=21\%$ from left to right.}
    \label{fig:rve_tests}
\end{figure}

Following the setting of \citet{Kalina2025}, we consider two two-phase soft composites shown in \cref{fig:rve_tests}: an RVE with stochastic fiber distribution and an RVE with hexagonal fiber arrangement. The corresponding fiber volume fractions are $\phi=26\%$ and $\phi=21\%$, respectively. The geometries and periodic meshes are generated with the Python interface of Gmsh \citep{GeuzaineRemacle2009}, and the corresponding reference solutions are obtained with FEniCS/DOLFIN \citep{Alnaes2015,LoggWells2010} by solving the finite-strain periodic homogenization problem introduced in \cref{sec:methodology}.
The stochastic-fiber mesh contains 50,282 nodes and 27,031 quadratic tetrahedral elements, whereas the hexagonal-fiber mesh contains 44,802 nodes and 23,443 quadratic tetrahedral elements. We use four quadrature points per element and take these quadrature points as the spatial collocation points for the operator-learning and field evaluation. This results in 108,124 collocation points for the stochastic-fiber RVE and 93,772 collocation points for the hexagonal-fiber RVE before Q-DEIM reduction. These discretizations provide high-resolution, high-fidelity full-field data for operator-learning and evaluation.
All phases are modeled as compressible isotropic hyperelastic materials. As in \citet{Kalina2025}, we use the two-parameter neo-Hookean energy proposed by \citet{Ciarlet1988}. Unless stated otherwise, the matrix phase is characterized by $E_m=1.0$ and $\nu_m=0.40$, whereas the fiber phase, that is, fibers or particles, uses $E_i=10.0$ and $\nu_i=0.44$.

\subsection{Deformation sampling}

\begin{figure}[p]
    \centering
    \begin{subfigure}{0.47\textwidth}
        \centering
        \includegraphics[width=\textwidth]{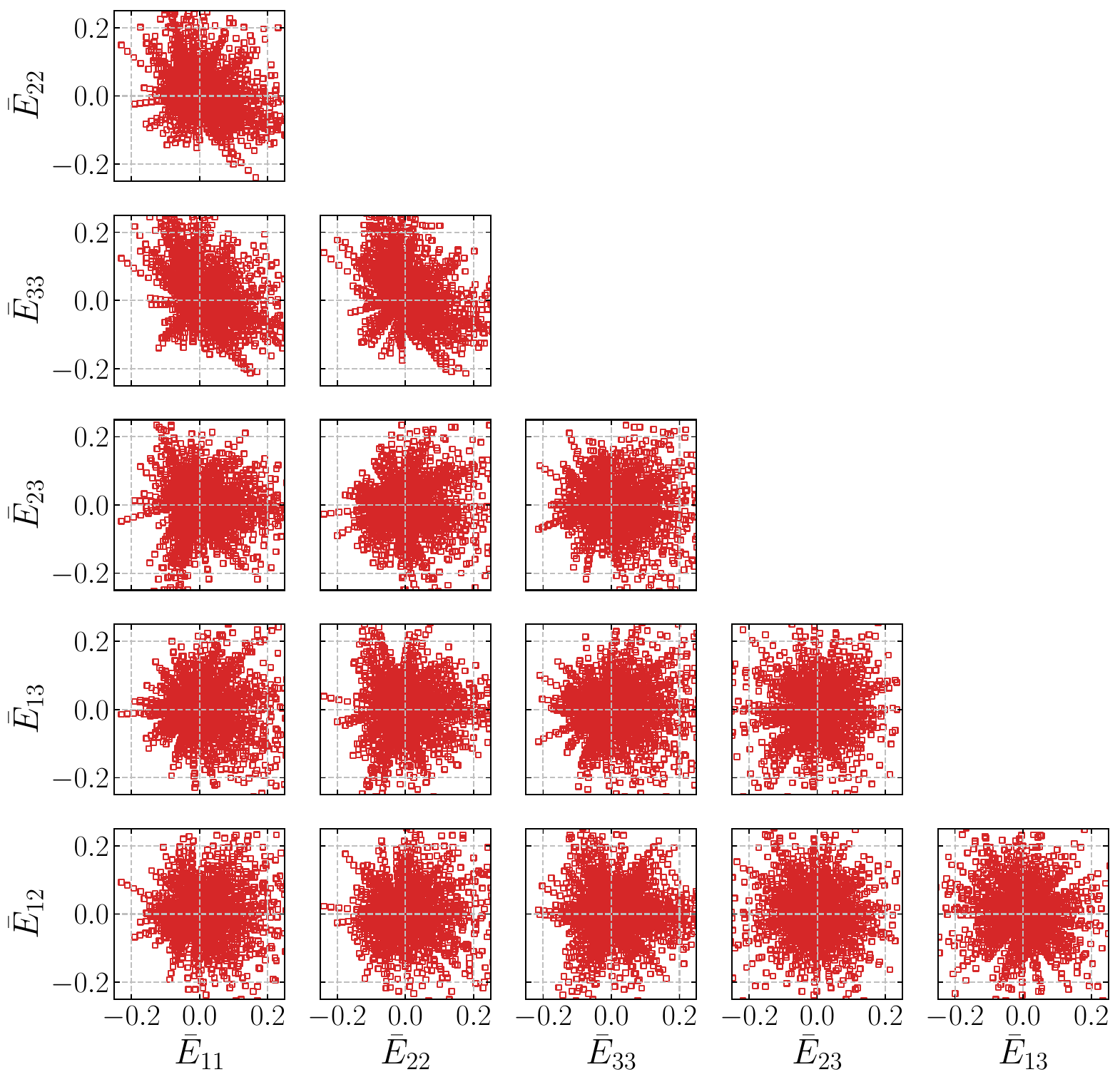}
        \caption{Unsupervised samples.}
        \label{fig:fbar_samples}
    \end{subfigure}
    \vspace{0.5cm}
    
    \begin{subfigure}{0.47\textwidth}
        \centering
        \includegraphics[width=\textwidth]{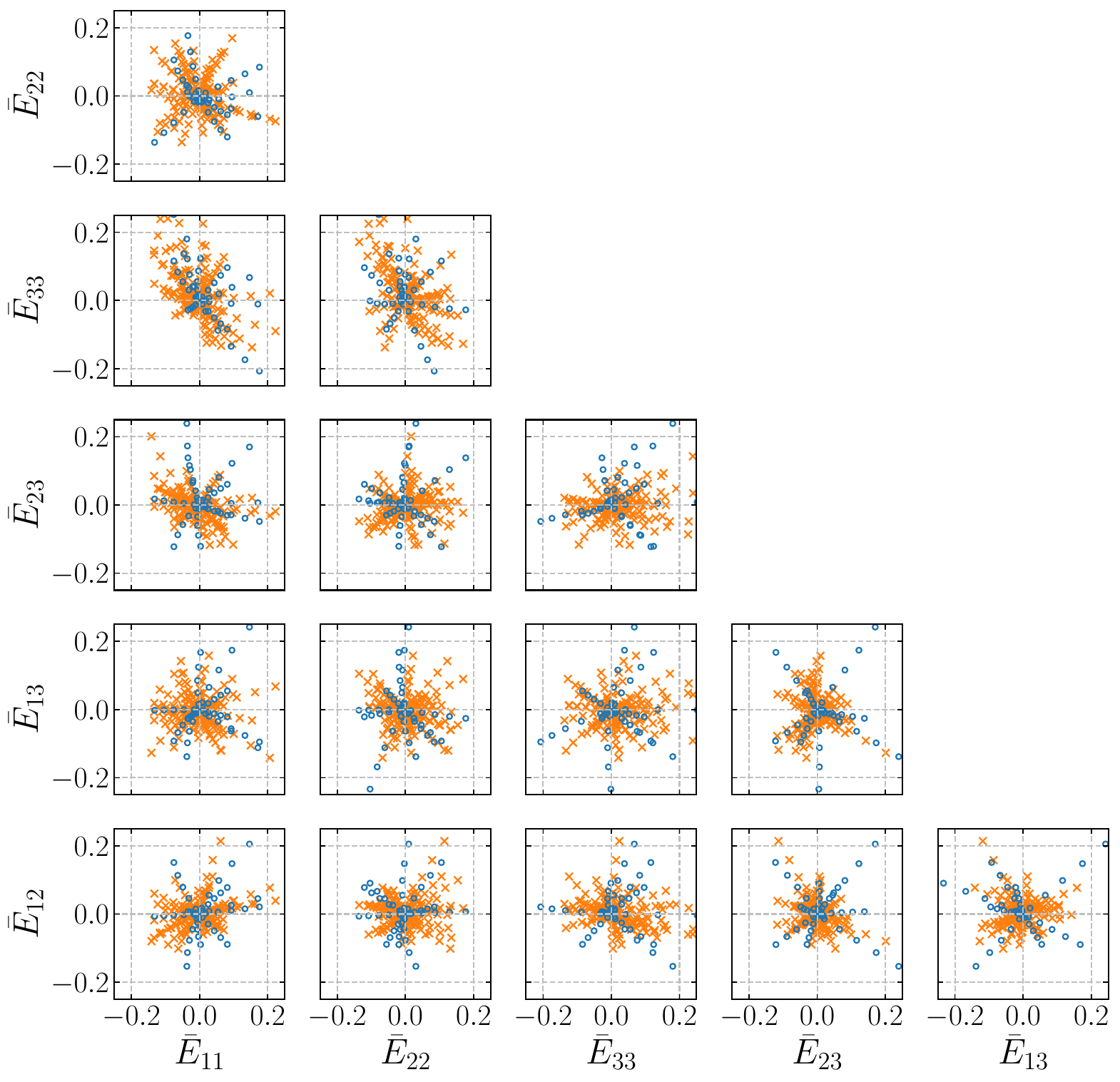}
        \caption{In-range setting.}
        \label{fig:fbar_sto}
    \end{subfigure}
    \hfil
    \begin{subfigure}{0.47\textwidth}
        \centering
        \includegraphics[width=\textwidth]{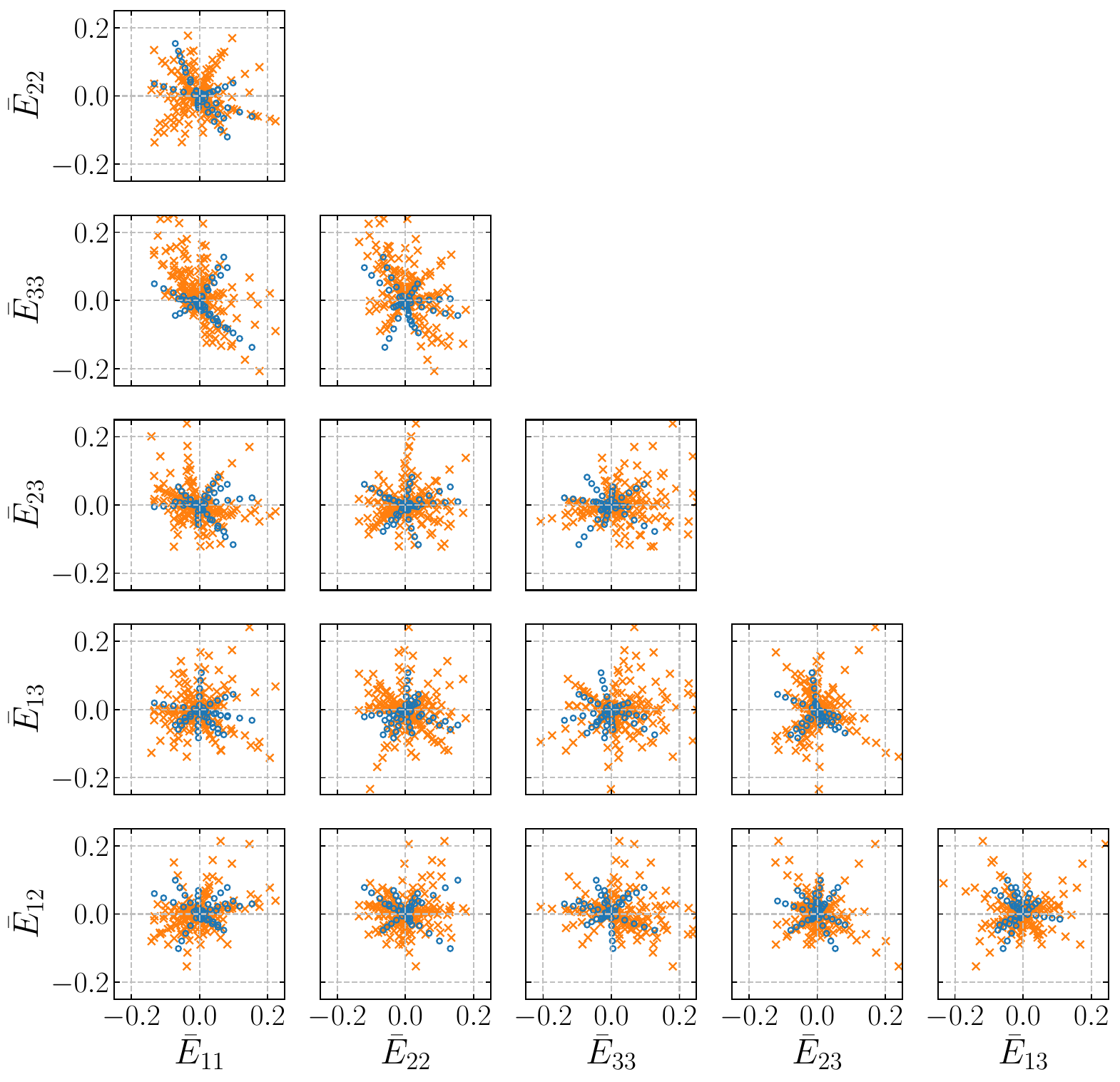}
        \caption{Out-of-range setting.}
        \label{fig:fbar_sto_ext}
    \end{subfigure}
    \caption{Unsupervised loading paths (red), snapshot loading paths (blue) and unseen test loading paths (orange) in the macroscopic Green--Lagrange strain space. The left panel corresponds to the unsupervised samples, while the middle and right panels corresponds to the in-range and out-of-range settings with $\beta=0.4$, respectively. Both in-range and out-of-range settings use $n_p=10$ snapshot loading paths.}
    \label{fig:fbar_sto_cases}
\end{figure}

We generate two disjoint sets of loading paths with the strategy discussed in \cref{subsec:data-generation-training}. The first contains $n_p^f$ unsupervised paths used for physics-informed training, whereas the second contains $n_p^s$ supervised paths for POD basis construction and testing. Each path is discretized into $n_i$ increments, resulting in $N^f=n_p^f n_i^f$ sampled deformation states for training and $N^s=n_p^s n_i^s$ states for the supervised snapshot and test pool. After applying the filtering procedure proposed in \citet{Kalina2025} to remove duplicate states and reduce the sample set, we retain $n_p^f=233$ unsupervised paths and $n_p^s=50$ supervised paths. Of the supervised paths, $n_p$ are used to construct the POD bases and the remaining $n_p^t=n_p^s-n_p$ are reserved for testing. In all cases, we use $n_i^f=10$ and $n_i^s=5$ increments per path.
Therefore, for each RVE, a data set $\gD$ containing $N=n_pn_i^s$ full-field snapshots,
\begin{equation}
    \mathcal{D} = \left\{\tilde{\vu}^{\,i}(\vxi,\bar{\mF}^{\,i}), \mP^{i}(\vzeta,\bar{\mF}^{\,i})\right\}_{i=1}^{N},
\end{equation}
are considered for constructing the POD bases, where the symbols $\vxi,\vzeta\in\Omega$ denote the spatial sampling locations of the displacement and stress snapshots, respectively.
\Cref{fig:fbar_sto_cases} compares the loading paths in the macroscopic Green--Lagrange strain space for the in-range and out-of-range settings with $n_p=10$ snapshot loading paths and $n_p^t=40$ test loading paths. The snapshot, test, and unsupervised training paths are shown in blue, orange, and red, respectively, with the unsupervised set containing $n_p^f=233$ loading paths used for physics-informed training.

In the in-range setting shown in the middle panel, the snapshot loading paths are selected from the full sampled loading space, that is, no additional restriction on the strain norm $\|\bar{\mE}\|$ is imposed when constructing the POD basis. In the out-of-range setting shown in the right panel, the snapshot loading paths are restricted to states satisfying $\|\bar{\mE}\|\le \beta\,\|\bar{\mE}\|_{\max}$, where $\|\bar{\mE}\|_{\max}$ denotes the maximum strain norm over the sampled loading space. Test paths containing states beyond this threshold are then used to quantify out-of-range generalization. 

\subsection{Evaluation metrics}
\label{app:error-metrics}

The microscopic errors reported in the main text are mean relative $L_2$ errors. For an evaluation set with $N_{\mathrm{eval}}$ samples, a predicted field $\tilde{\mX}^i$, and the corresponding reference field $\mX^i$, we define
\begin{equation}
    \varepsilon_{\mX}
    =
    \frac{1}{N_{\mathrm{eval}}}
    \sum_{i=1}^{N_{\mathrm{eval}}}
    \frac{\left\|{\mX}^{i}-\tilde{\mX}^{i}\right\|_2}
    {\left\|\mX^{i}\right\|_2} \times 100,
    \qquad \mX \in \{\mF,\mP\}.
\end{equation}
For the microscopic fields $\mF$ and $\mP$, each sample is vectorized before evaluating the Euclidean norms, $\varepsilon_{\mP}, \varepsilon_{\mF}$, respectively, so that all element, quadrature-point, and tensor components contribute to the same relative error measure. The metric $\varepsilon_{\mP}$ therefore quantifies the full-field stress-prediction error in the same spatially resolved sense as the loss introduced in \eqref{eq:loss}. Analogously, we report $\varepsilon_{\mP_q}$, which is the relative error only evaluated at Q-DEIM points, c.f. \eqref{eq:loss_qdeim}.  Learning curves report the evolution of $\varepsilon_{\mP_q}$ over the optimization, whereas $\varepsilon_{\mP_q}^{\mathrm{train}}$ denotes its value at the final training step. We report $\varepsilon_{\mP_q}^{\mathrm{train}}$ to make the final training residual comparable in scale with the stress errors $\varepsilon_{\mP}$ reported for unseen test states.

For a homogenized quantity $\mY\in\{\bar{\mP},\bar{\sA}\}$, we collect all components over the evaluation set in the vectors $\vy_{\mY},\tilde{\vy}_{\mY}\in\mathbb{R}^{d_{\mY}}$, where $d_{\mY}$ is the total number of flattened entries. The coefficient of determination and normalized mean absolute error are then defined compactly as
\begin{equation}
    R^2_{\mY}
    =
    1
    -
    \frac{\left\|\vy_{\mY}-\tilde{\vy}_{\mY}\right\|_2^2}
    {\left\|\vy_{\mY}-\mu_{\mY}\vone\right\|_2^2},
    \qquad
    \mu_{\mY}=\frac{1}{d_{\mY}}\sum_{k=1}^{d_{\mY}} y_{\mY,k},
\end{equation}
\begin{equation}
    \mathrm{NMAE}_{\mY}
    =
    \frac{
    \left\|\vy_{\mY}-\tilde{\vy}_{\mY}\right\|_1/d_{\mY}
    }
    {
    \max_{1\le k\le d_{\mY}} y_{\mY,k}
    -
    \min_{1\le k\le d_{\mY}} y_{\mY,k}
    } \times 100.
\end{equation}
Thus, $R^2_{\mY}\in(-\infty,1]$ and $\mathrm{NMAE}_{\mY}\in\mathbb{R}_{\ge 0}$. The NMAE normalizes the mean absolute error by the range of the reference values and is therefore reported in percent.

To quantify the computational benefit of the reduced evaluations, we report two forward-pass speed-up factors, both measured for one full-batch evaluation. The first factor,
\begin{equation}
    s_{\mathrm{QDEIM}}
    =
    \frac{t_{\mathrm{full}}^{\mathrm{fb}}}{t_{\mathrm{QDEIM}}^{\mathrm{fb}}},
\end{equation}
measures the reduction in the cost of the physics-informed loss evaluation when the constitutive response is computed at all spatial points of the RVE versus only at the selected Q-DEIM points. Since this evaluation is carried out at every training step, $s_{\mathrm{QDEIM}}$ quantifies the training-side forward-pass speed-up induced by Q-DEIM. The second factor,
\begin{equation}
    s_{\mathrm{hom}}
    =
    \frac{t_{\mathrm{full}\rightarrow\mathrm{hom}}^{\mathrm{fb}}}{t_{\mathrm{hom}}^{\mathrm{fb}}},
\end{equation}
measures the speed-up obtained when the homogenized first Piola--Kirchhoff stress is computed directly from the branch-stress coefficients and the pre-averaged stress modes, instead of first reconstructing the full-field stress and then homogenizing it. This quantity is particularly relevant for deployment in multiscale \FEsq\ simulations, where the microscale model is queried repeatedly at macroscale quadrature points and Newton iterations. In that setting, avoiding the reconstruction of local full-field stresses can substantially reduce the constitutive-query cost seen by the macroscale solver. In summary, the first factor $s_{\mathrm{QDEIM}}$ measures the speed-up relevant to the training, and the second factor $s_{\mathrm{hom}}$ indicates the speed-up in the prediction of homogenized first Piola--Kirchhoff stresses. 

\subsection{Computational efficiency}
\label{subsec:efficiency}

As shown later in \cref{tab:errors_compact,tab:errors_out_range}, the proposed reduced evaluations consistently yield speed-up factors of order $10^3$ for $s_{\mathrm{QDEIM}}$ and of order $10^3$ to $10^4$ for $s_{\mathrm{hom}}$ relative to the corresponding full-field computations. These gains are central for operator learning of RVEs in three-dimensional finite-strain settings, because otherwise the repeated full-field constitutive evaluations required during physics-informed training dominate the computational cost. 
The reported values are empirical timing ratios rather than asymptotic complexity estimates, and we present them directly as rounded dimensionless factors in the tables. 

\begin{figure}[ht]
    \centering
    \includegraphics[width=0.6\textwidth]{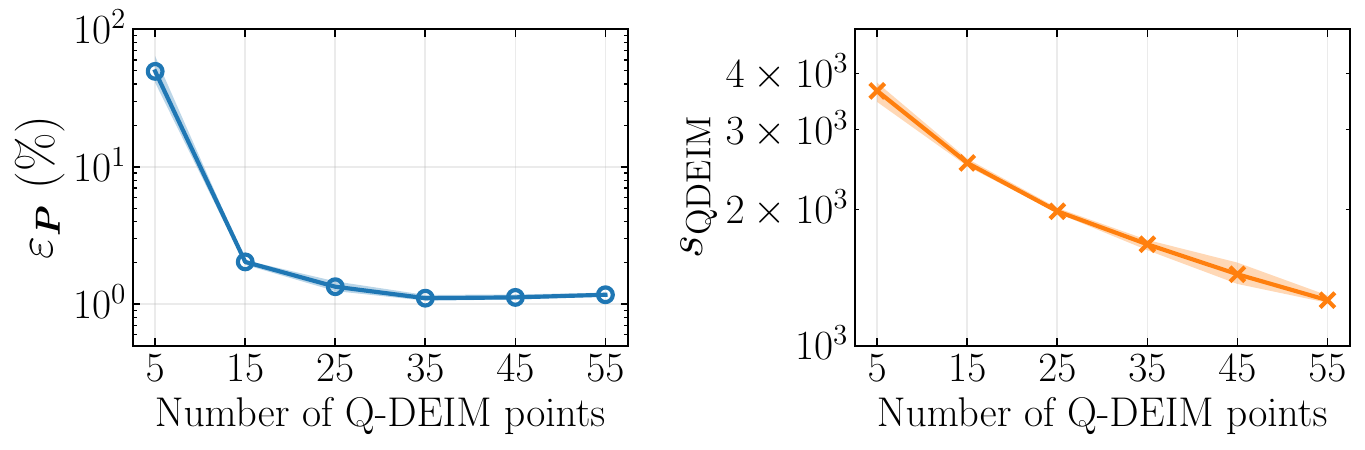}
    \caption{Mean relative $L_2$ errors (left) and speed-up factors (right) for the stochastic RVE versus the number of Q-DEIM points in the in-range setting with $n_p=20$ and a $2\times64$ network discussed in \cref{subsec:in-range-generalization}.}
    \label{fig:magic_points_sweep_sto}
\end{figure}

\Cref{fig:magic_points_sweep_sto} shows the effect of the number of Q-DEIM points on the accuracy, measured in terms of $\varepsilon_{\mP}$, and the speed-up factor $s_{\mathrm{QDEIM}}$ for the stochastic-fiber RVE. Similar trends are observed for the hexagonal-fiber RVE and therefore not reported here. The number of snapshot load paths is $n_p=20$, and the number of POD modes has been determined to $p=38$ such that the ratio between the cumulative sum and the total sum of singular values exceeds $0.9999$. The figure illustrates that for  $q\geq 35$, the error $\varepsilon_{\mP}$ is on the order of 1\% and does not decrease further. Using fewer Q-DEIM points leads to higher errors, while using more points does not significantly improve accuracy but reduces the speed-up factor.

The models reported in this study are trained on a MacBook Pro equipped with an Apple M2 chip, using CPU execution only. With the proposed Q-DEIM reduction, physics-informed training of EquiNO on $233$ unsupervised loading paths takes on the order of ten minutes for the considered three-dimensional RVE examples. For reference, solving one loading path with the present finite-element periodic homogenization solver takes approximately eighteen minutes. Thus, after the limited supervised snapshot set has been generated for basis construction, the physics-informed training can incorporate a large number of additional loading states at a cost below that of a single loading-path finite-element simulation.

\begin{figure}[ht]
    \centering
    \includegraphics[width=0.3\textwidth]{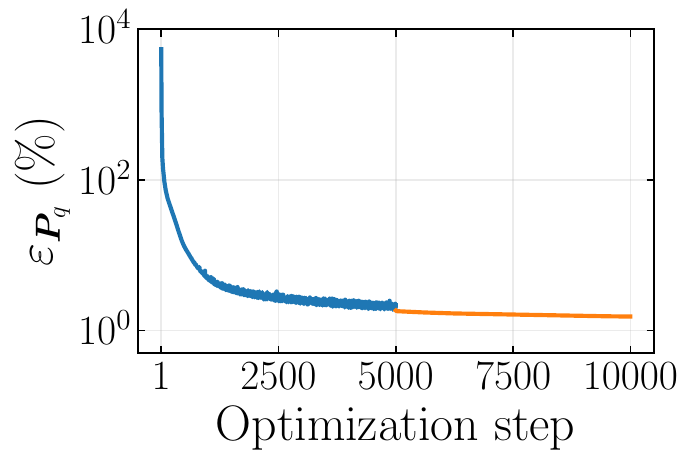}
    ~~~~
    \includegraphics[width=0.3\textwidth]{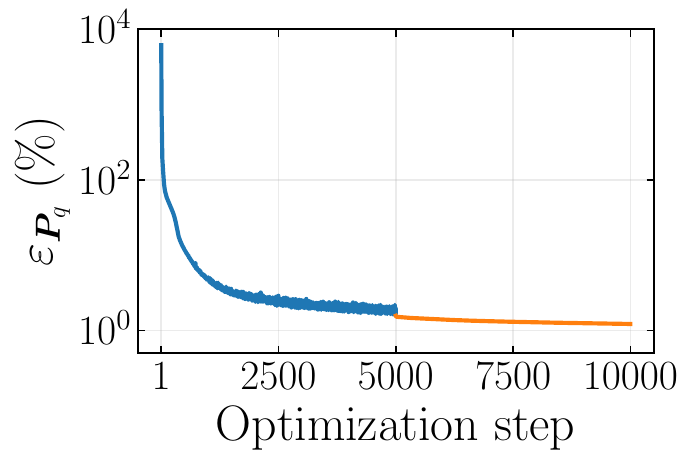}
    \caption{Representative learning curve for the stochastic-fiber (left) and hexagonal-fiber (right) RVEs in the out-of-range setting with $n_p=20$, $\beta=0.4$, and a $2\times64$ network discussed in \cref{subsec:out-of-range-generalization}. Blue shows the optimization steps with Adam, while orange shows the optimization steps with L-BFGS.}
    \label{fig:learning_curve_hex}
\end{figure}

\Cref{fig:learning_curve_hex} illustrates a representative optimization history for EquiNO with the Q-DEIM-reduced loss for both RVEs. The error is shown in terms of the mean relative $L_2$ error for the predicted first Piola--Kirchhoff stress field at the Q-DEIM points, denoted by $\varepsilon_{\mP_q}$. 
It can be observed that the loss decreases smoothly and steadily over the optimization, with a significant drop in the early iterations. The final error level is around $1\%$ for the predicted first Piola--Kirchhoff stress field at the Q-DEIM points for the unsupervised input loading states used for training, which is consistent with the results reported for the test loading states in \cref{tab:errors_compact,tab:errors_out_range}. This indicates that the physics-informed training effectively guides the two stress representations toward the unique solution of the governing equations, first rapidly during the initial optimization phase and then toward an accurate converged state. The corresponding component-wise learning curves are reported in \cref{app:supplementary-results}; see \cref{fig:learning_curve_components}.

The reduced loss evaluation makes it feasible to optimize the model in full-batch mode with L-BFGS over a large set of sampled loading states. This is important in the present three-dimensional finite-strain setting, because a full-field loss evaluation would require storing and processing constitutive quantities at all spatial points for all sampled boundary conditions within each optimization step. For sufficiently large batches, this full-field formulation becomes memory-intensive and, in practice, is often prohibitive, especially on GPUs. By restricting the loss evaluation to the selected Q-DEIM points, the memory footprint and per-step cost are reduced enough to make deterministic full-batch optimization practical. 

The following subsections show that the computational advantages of Q-DEIM hyperreduction are achieved without compromising accuracy. In particular, we evaluate in-range and out-of-range generalization. In-range generalization refers to test loading paths that remain within the prescribed loading range used to generate the snapshot paths, whereas out-of-range generalization refers to test paths that extend beyond the restricted snapshot range used for basis construction.

\subsection{In-range generalization}
\label{subsec:in-range-generalization}

\paragraph{Stochastic-fiber RVE:} 
\Cref{tab:errors_compact} summarizes the in-range results for the {stochastic-fiber RVE} for different numbers of snapshot loading paths $n_p$, the resulting POD dimensions $p$, and different network sizes.
\begin{table}[ht]
    \begin{center}
        \resizebox{\textwidth}{!}{
        \begin{footnotesize}
            \begin{threeparttable}
                \caption{In-range generalization errors for the stochastic-fiber RVE for different numbers of snapshot loading paths $n_p$, the corresponding POD dimensions $p$, and different network sizes. The last two columns report the forward-pass speed-up factors defined in \cref{subsec:efficiency}.}
                \label{tab:errors_compact}
                \begin{tabular}{lllccccccccc}
                    \toprule
                    \multicolumn{3}{c}{Test setup} & \multicolumn{3}{c}{Full-field errors} & \multicolumn{4}{c}{Homogenized errors} & \multicolumn{2}{c}{Speed-up factors} \\
                    \cmidrule(lr){1-3}\cmidrule(lr){4-6}\cmidrule(lr){7-10}\cmidrule(lr){11-12}
                    $n_p$ & $p$ & NN size & $\varepsilon_{\mF}$ (\%) & $\varepsilon_{\mP}$ (\%) & $\varepsilon_{\mP_q}^{\mathrm{train}}$ (\%) & $R^2_{\bar{\mP}}$ & $\mathrm{NMAE}_{\bar{\mP}}$ (\%) & $R^2_{\bar{\sA}}$ & $\mathrm{NMAE}_{\bar{\sA}}$ (\%) & $s_{\mathrm{QDEIM}}$ & $s_{\mathrm{hom}}$ \\
                    \midrule
        10 & 23 & $1 \times 16$ & 0.30 & 5.16 & 6.35 & 0.9906 & 0.32 & 0.9829 & 0.78 & $3\times 10^{3}$ & $4\times 10^{4}$ \\
        10 & 23 & $1 \times 32$ & 0.21 & 2.77 & 3.75 & 0.9968 & 0.18 & 0.9920 & 0.56 & $3\times 10^{3}$ & $3\times 10^{4}$ \\
        10 & 23 & $1 \times 64$ & 0.18 & 2.07 & 2.70 & 0.9984 & 0.13 & 0.9942 & 0.46 & $2\times 10^{3}$ & $2\times 10^{4}$ \\
        10 & 23 & $2 \times 16$ & 0.23 & 3.08 & 4.15 & 0.9964 & 0.20 & 0.9925 & 0.56 & $3\times 10^{3}$ & $3\times 10^{4}$ \\
        10 & 23 & $2 \times 32$ & 0.18 & 2.14 & 2.72 & 0.9986 & 0.13 & 0.9952 & 0.44 & $2\times 10^{3}$ & $2\times 10^{4}$ \\
        10 & 23 & $2 \times 64$ & 0.17 & 1.86 & 2.33 & 0.9986 & 0.12 & 0.9959 & 0.39 & $2\times 10^{3}$ & $8\times 10^{3}$ \\
        \midrule
        20 & 38 & $1 \times 16$ & 0.31 & 5.06 & 6.20 & 0.9899 & 0.30 & 0.9845 & 0.77 & $2\times 10^{3}$ & $6\times 10^{4}$ \\
        20 & 38 & $1 \times 32$ & 0.19 & 2.57 & 3.08 & 0.9979 & 0.15 & 0.9942 & 0.49 & $2\times 10^{3}$ & $4\times 10^{4}$ \\
        20 & 38 & $1 \times 64$ & 0.13 & 1.96 & 2.28 & 0.9988 & 0.10 & 0.9953 & 0.42 & $2\times 10^{3}$ & $2\times 10^{4}$ \\
        20 & 38 & $2 \times 16$ & 0.25 & 4.47 & 5.33 & 0.9936 & 0.24 & 0.9862 & 0.71 & $2\times 10^{3}$ & $4\times 10^{4}$ \\
        20 & 38 & $2 \times 32$ & 0.13 & 1.79 & 2.12 & 0.9990 & 0.10 & 0.9960 & 0.40 & $2\times 10^{3}$ & $3\times 10^{4}$ \\
        20 & 38 & $2 \times 64$ & 0.09 & 1.09 & 1.10 & 0.9997 & 0.05 & 0.9976 & 0.30 & $1\times 10^{3}$ & $1\times 10^{4}$ \\
                    \bottomrule
                \end{tabular}
            \end{threeparttable}
        \end{footnotesize}
        }
    \end{center}
\end{table}
For each snapshot set, the number of retained modes is chosen such that the ratio between the cumulative sum and the total sum of singular values exceeds $0.9999$, which leads here to $p=23$ for $n_p=10$ and $p=38$ for $n_p=20$. The microscopic accuracy is assessed by $\varepsilon_{\mF}$ and $\varepsilon_{\mP}$, whereas the macroscopic response is evaluated by $R^2_{\bar{\mP}}$ and $\mathrm{NMAE}_{\bar{\mP}}$; all metrics are defined in \cref{app:error-metrics}. The table also reports $\varepsilon_{\mP_q}^{\mathrm{train}}$, the mean relative $L_2$ error at the Q-DEIM points on the training states. The last two columns report the forward-pass speed-up factors $s_{\mathrm{QDEIM}}$ and $s_{\mathrm{hom}}$ introduced in \cref{subsec:efficiency}.
\begin{figure}[htb]
    \centering
    \begin{subfigure}{0.2\textwidth}
        \centering
        \includegraphics[height=0.9\textwidth]{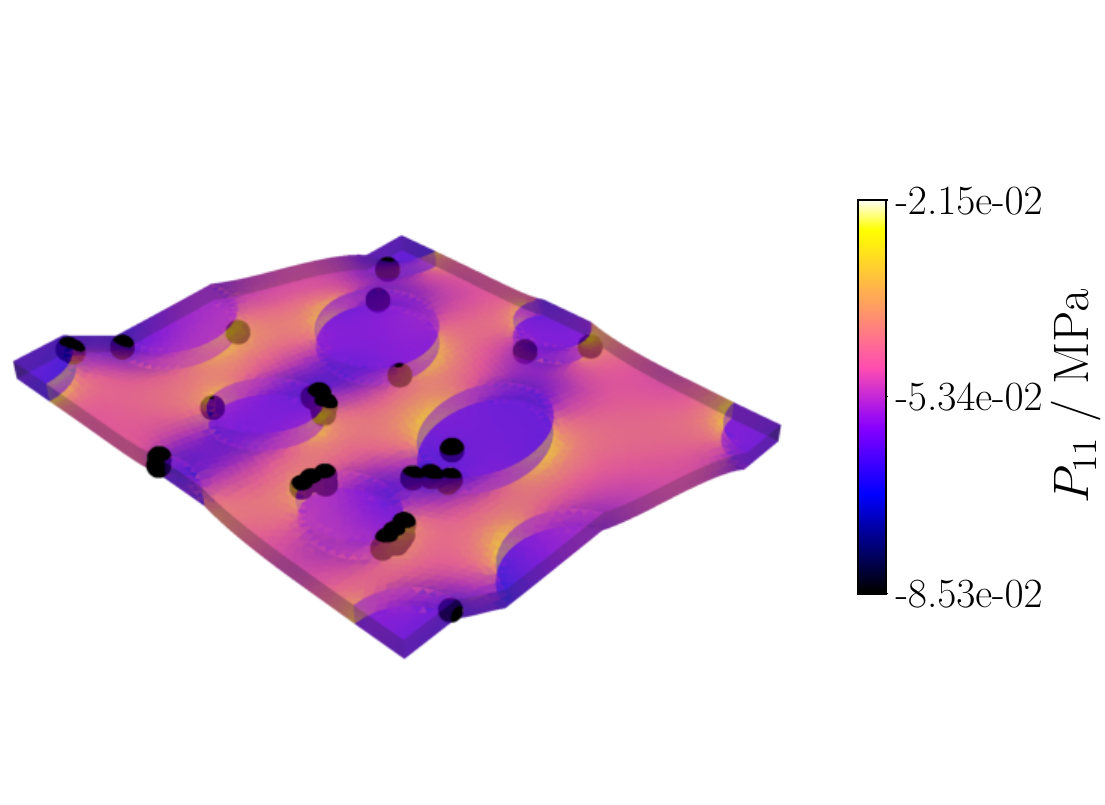}
    \end{subfigure}
    \hfil
    \begin{subfigure}{0.2\textwidth}
        \centering
        \includegraphics[height=0.9\textwidth]{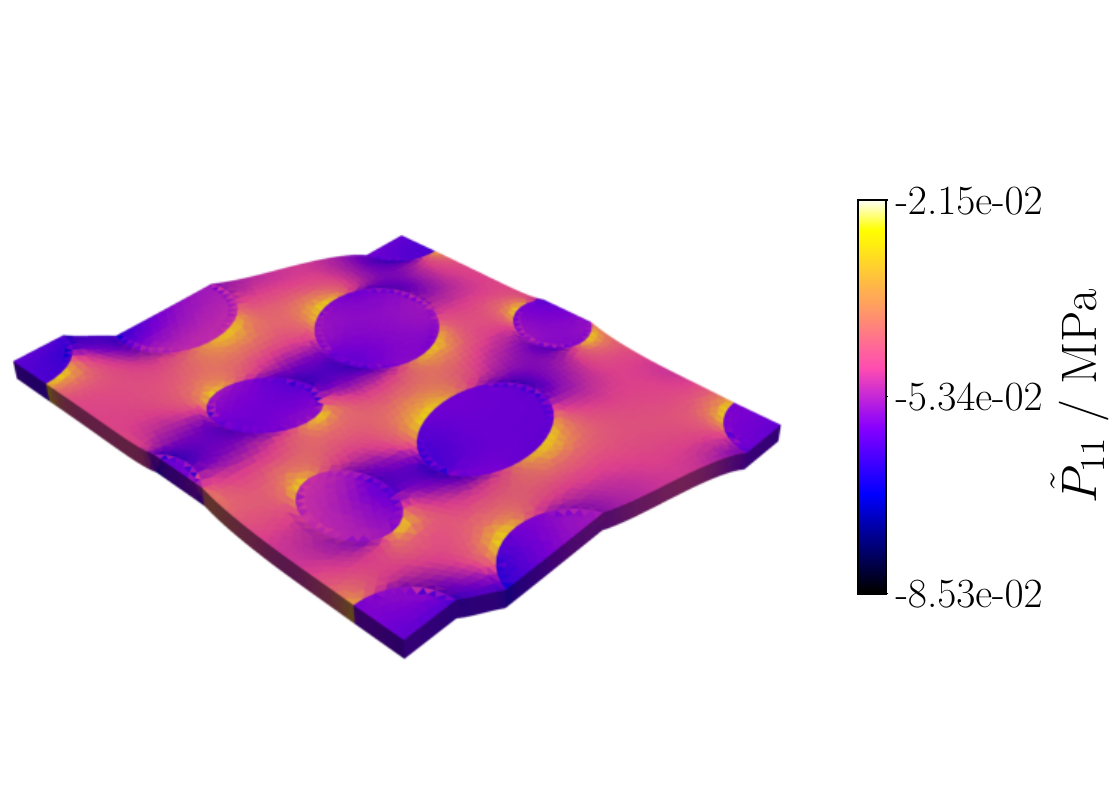}
    \end{subfigure}
    \hfil
    \begin{subfigure}{0.2\textwidth}
        \centering
        \includegraphics[height=0.9\textwidth]{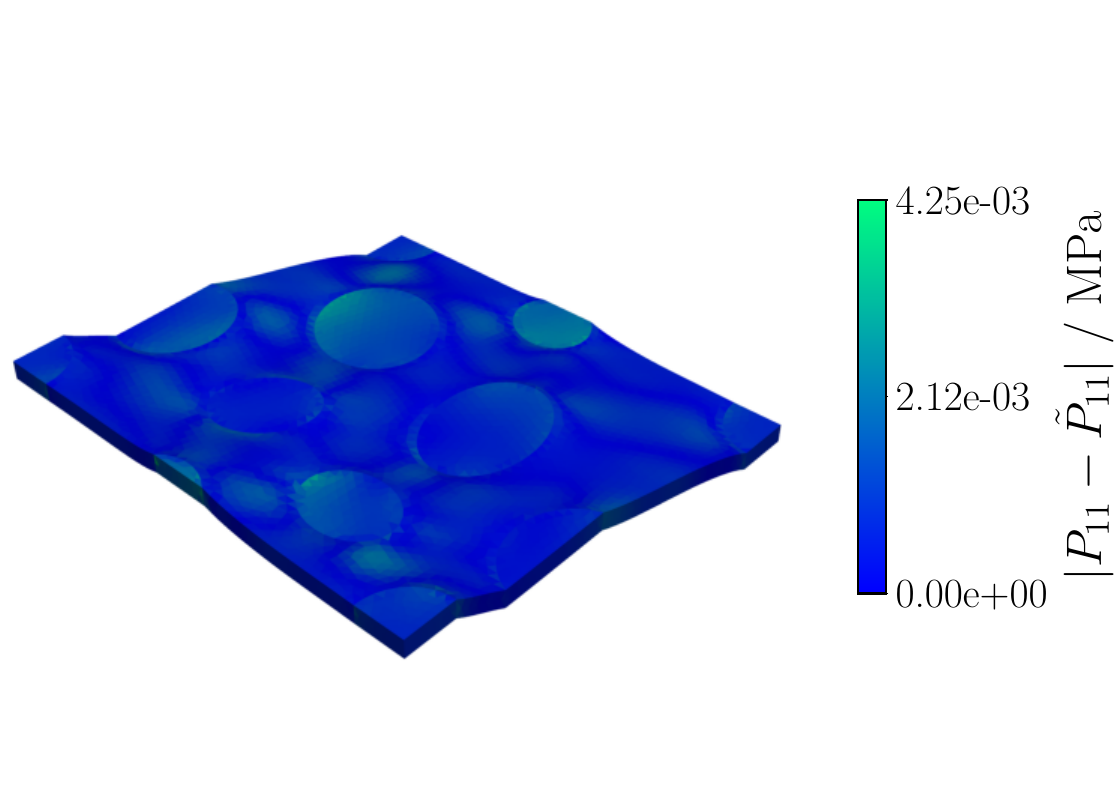}
    \end{subfigure}
    \hfil

    \begin{subfigure}{0.2\textwidth}
        \centering
        \includegraphics[height=0.9\textwidth]{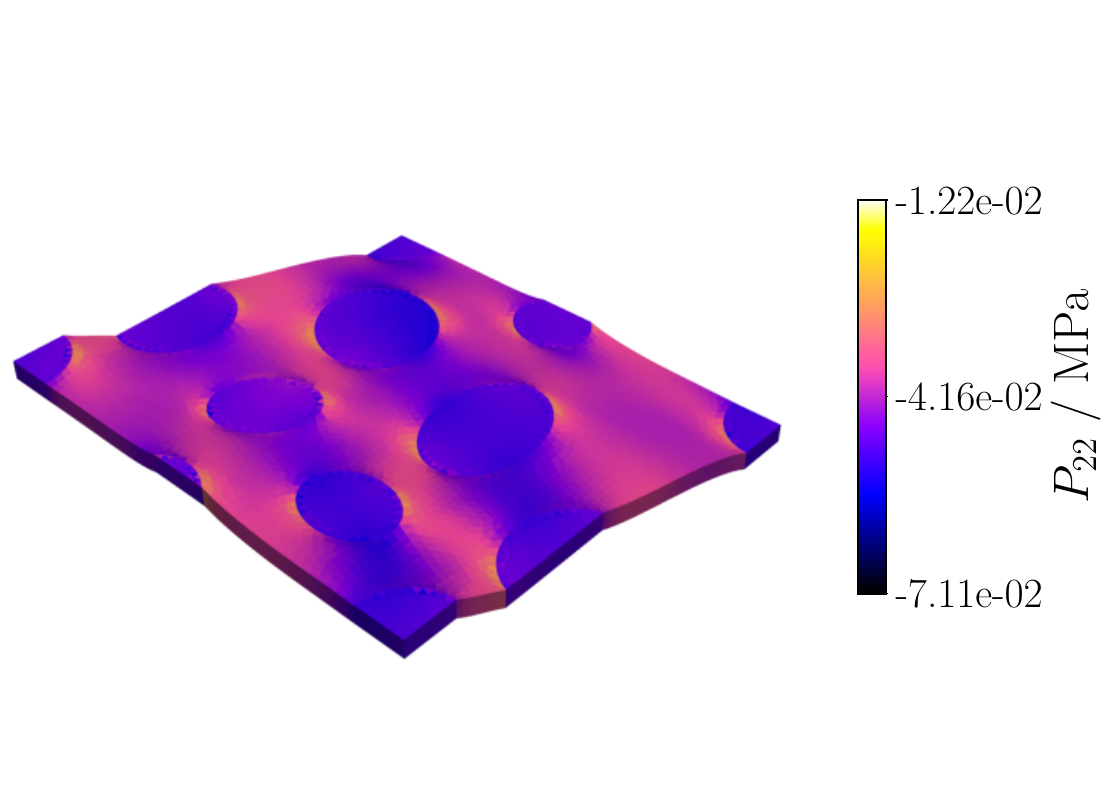}
    \end{subfigure}
    \hfil
    \begin{subfigure}{0.2\textwidth}
        \centering
        \includegraphics[height=0.9\textwidth]{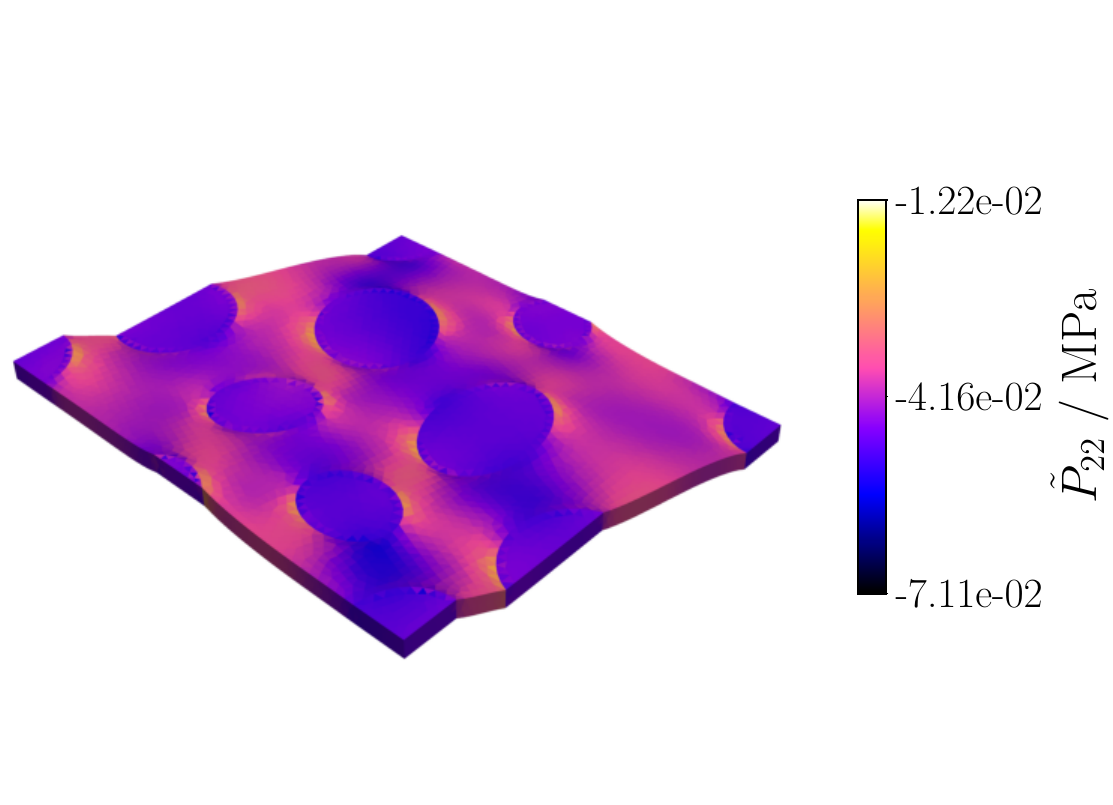}
    \end{subfigure}
    \hfil
    \begin{subfigure}{0.2\textwidth}
        \centering
        \includegraphics[height=0.9\textwidth]{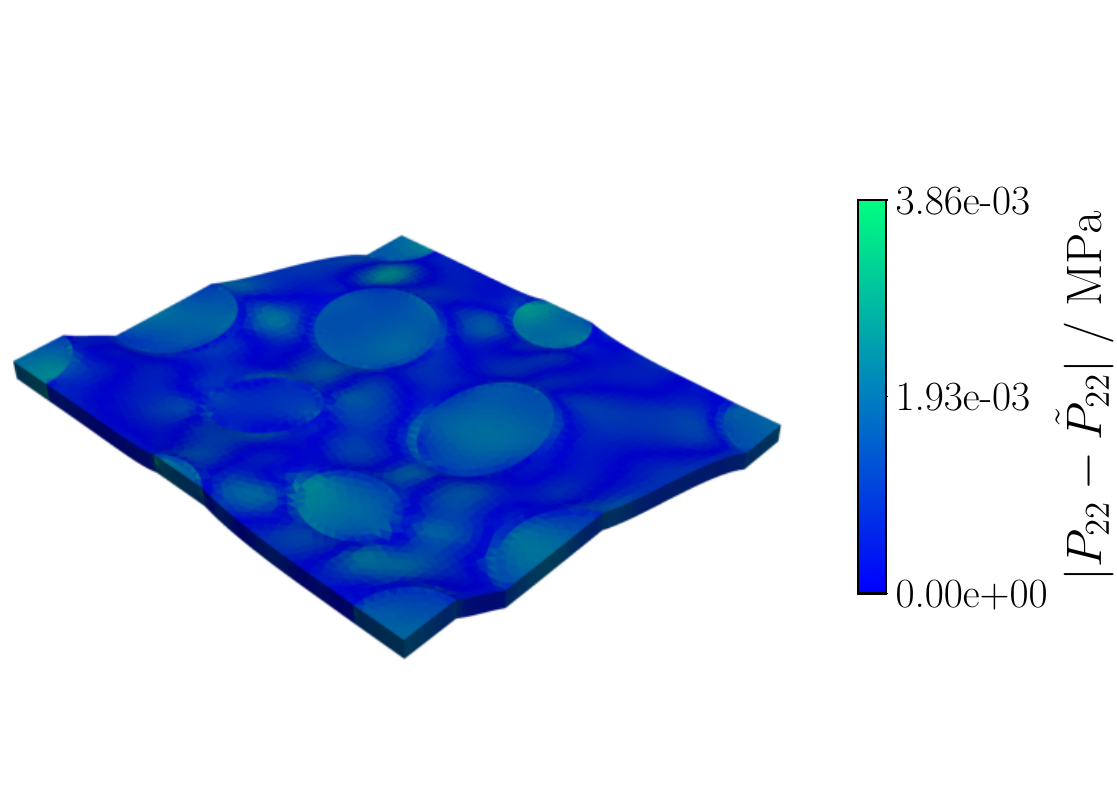}
    \end{subfigure}
    \hfil

    \begin{subfigure}{0.2\textwidth}
        \centering
        \includegraphics[height=0.9\textwidth]{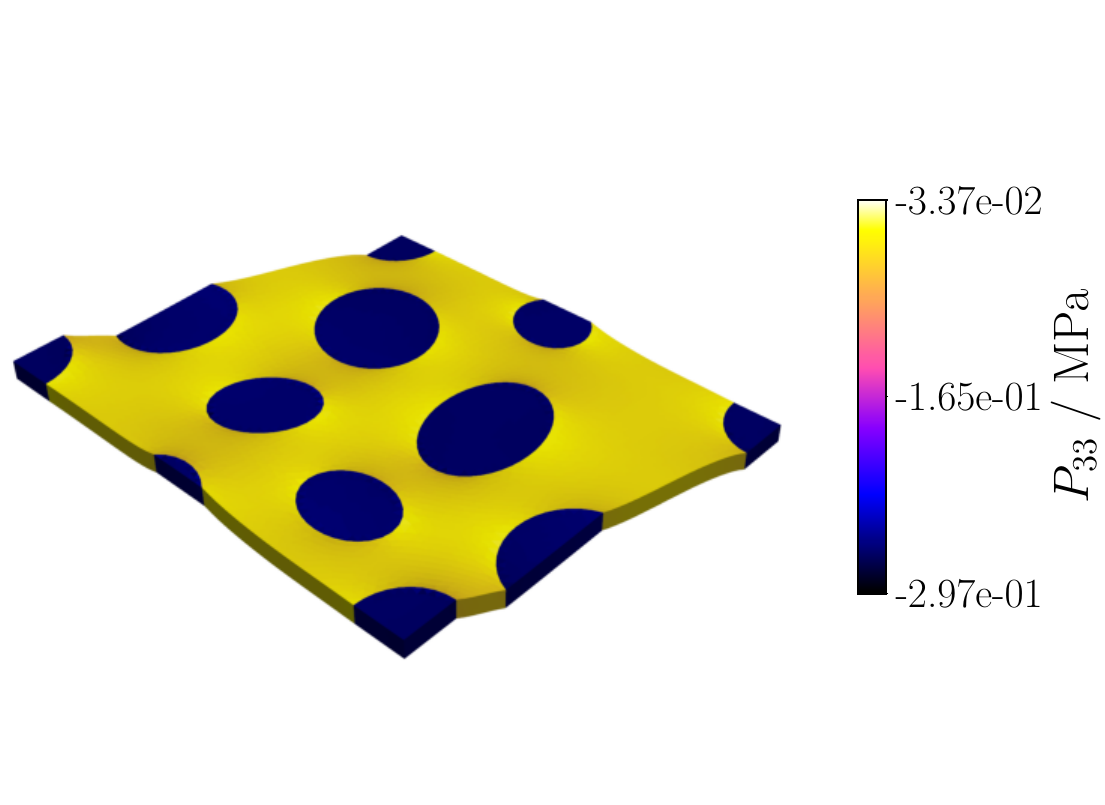}
    \end{subfigure}
    \hfil
    \begin{subfigure}{0.2\textwidth}
        \centering
        \includegraphics[height=0.9\textwidth]{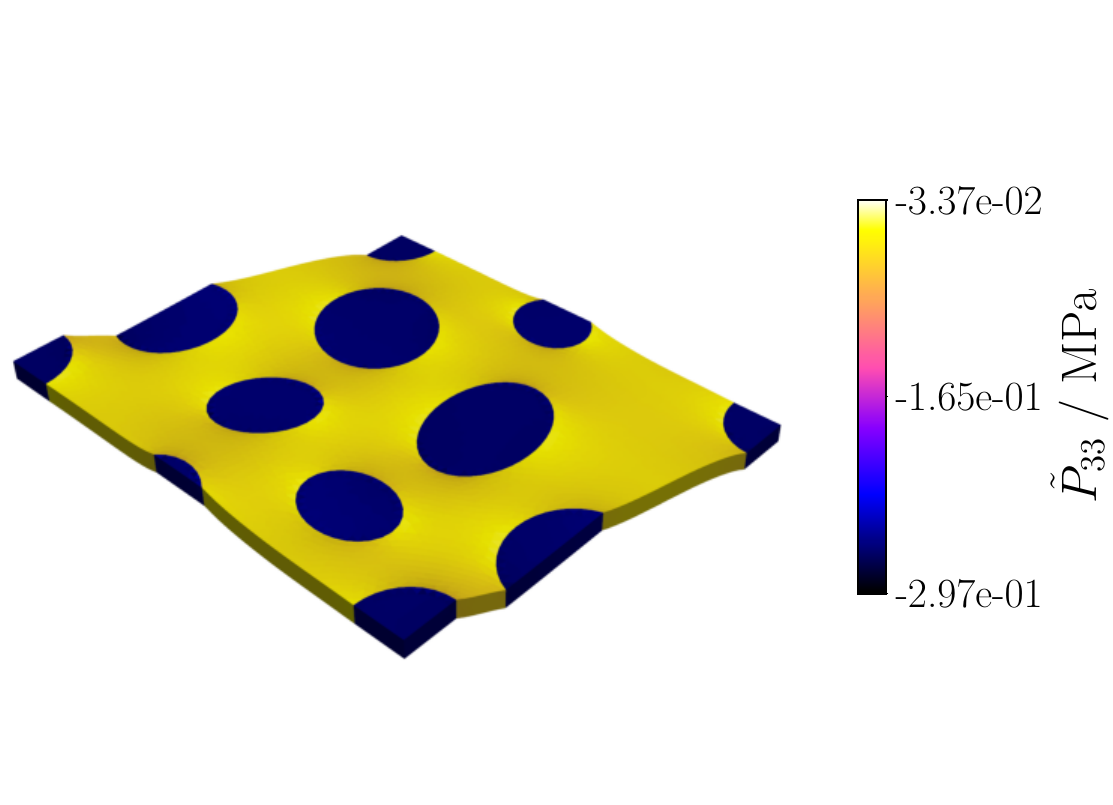}
    \end{subfigure}
    \hfil
    \begin{subfigure}{0.2\textwidth}
        \centering
        \includegraphics[height=0.9\textwidth]{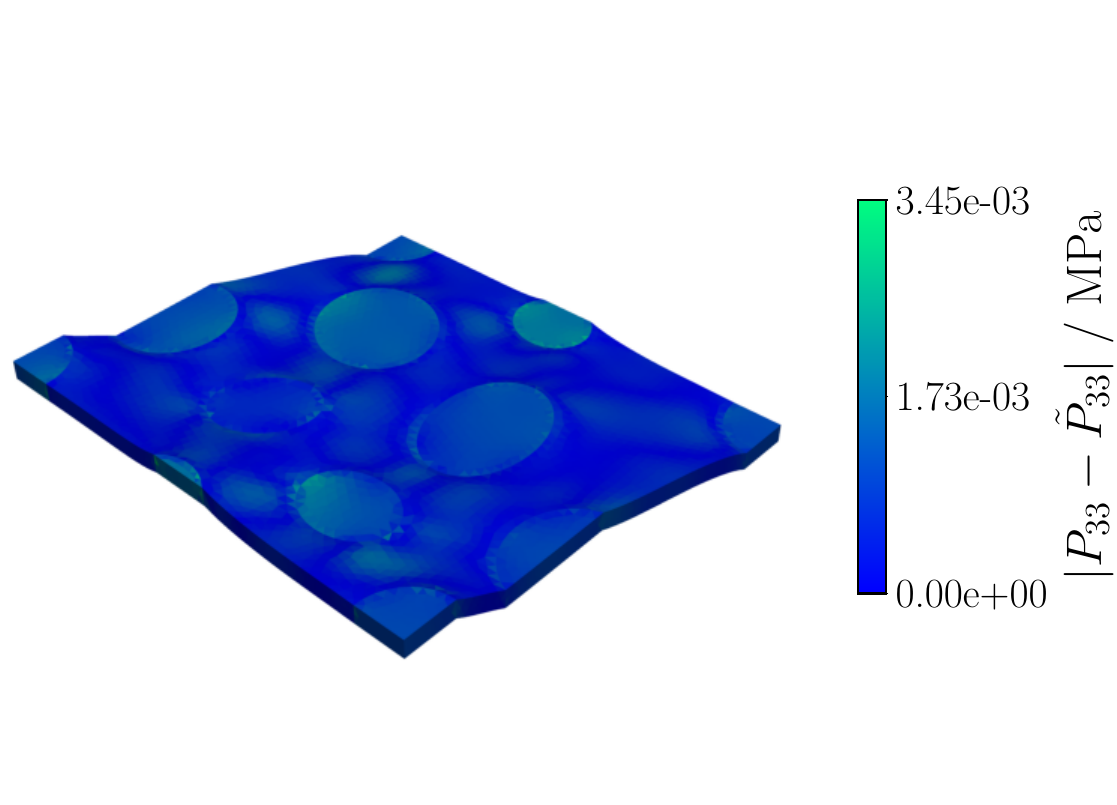}
    \end{subfigure}
    \hfil
    \caption{Comparison of reference, predicted, and absolute-error fields for the microscopic first Piola--Kirchhoff stress components of the stochastic-fiber RVE for the test sample with the median relative $L_2$ error among all in-range test samples. Columns from left to right show the reference solution, the EquiNO prediction, and the absolute error. Rows from top to bottom correspond to $P_{11}$, $P_{22}$, and $P_{33}$. Results are shown for the best in-range model with $n_p=20$, $p=38$, and a $2\times64$ network. The Q-DEIM points are illustrated in the top left panel.}
    \label{fig:sto_vis}
\end{figure}
The results for $n_p=10$ already show that accurate surrogates can be obtained from a very limited offline snapshot set. In particular, the $1\times32$ network trained with only ten snapshot loading paths reaches $\varepsilon_{\mF}=0.21\%$, $\varepsilon_{\mP}=2.77\%$, $R^2_{\bar{\mP}}=0.9968$, and $\mathrm{NMAE}_{\bar{\mP}}=0.18\%$, while still providing speed-up factors of order $10^3$ for $s_{\mathrm{QDEIM}}$ and order $10^4$ for $s_{\mathrm{hom}}$. At fixed $n_p$, increasing the network width improves the prediction quality more clearly than adding an additional hidden layer. The best overall performance in \cref{tab:errors_compact} is obtained for $n_p=20$ with the $2\times64$ network, which yields $R^2_{\bar{\mP}}=0.9997$ together with the smallest microscopic and macroscopic errors, a Q-DEIM speed-up of order $10^3$, and a reduced-homogenization speed-up close to $10^4$.
Results corresponding to this model are shown in \cref{fig:sto_vis}. The sample displayed in \cref{fig:sto_vis} is the test case with the median relative $L_2$ error among all test samples. Its predicted local stress fields $P_{11}$, $P_{22}$, and $P_{33}$ closely reproduce the reference solutions, while the corresponding absolute errors remain small throughout the RVE. The Q-DEIM points are illustrated in the top left panel of \cref{fig:sto_vis}. Samples with minimum and maximum relative $L_2$ errors are shown in \cref{app:supplementary-results}.

The homogenized material tangent is also recovered accurately for the same best stochastic-fiber model. Here, $\bar{\sA}$ is obtained by differentiating the homogenized first Piola--Kirchhoff stress predicted by the reduced stress model, that is, from the branch-stress coefficients and the homogenized POD modes, with respect to the macroscopic deformation gradient. The tangent therefore provides a stricter test of the reduced representation. For the $n_p=20$, $p=38$, $2\times64$ model, the tangent prediction reaches $R^2_{\bar{\sA}}=0.9976$ and $\mathrm{NMAE}_{\bar{\sA}}=0.30\%$.

\paragraph{Hexagonal-fiber RVE:} 
\Cref{tab:errors_compact_hex} summarizes the corresponding in-range results for the {hexagonal-fiber RVE}. The POD truncation criterion gives $p=22$ for $n_p=10$ and $p=37$ for $n_p=20$, so the larger snapshot set again leads to a richer stress basis. 
\begin{table}[ht]
    \begin{center}
        \resizebox{\textwidth}{!}{
        \begin{footnotesize}
            \begin{threeparttable}
                \caption{In-range generalization errors for the hexagonal-fiber RVE for different numbers of snapshot loading paths $n_p$, the corresponding POD dimensions $p$, and different network sizes. The last two columns report the forward-pass speed-up factors defined in \cref{subsec:efficiency}.}
                \label{tab:errors_compact_hex}
                \begin{tabular}{lllccccccccc}
                    \toprule
                    \multicolumn{3}{c}{Test setup} & \multicolumn{3}{c}{Full-field errors} & \multicolumn{4}{c}{Homogenized errors} & \multicolumn{2}{c}{Speed-up factors} \\
                    \cmidrule(lr){1-3}\cmidrule(lr){4-6}\cmidrule(lr){7-10}\cmidrule(lr){11-12}
                    $n_p$ & $p$ & NN size & $\varepsilon_{\mF}$ (\%) & $\varepsilon_{\mP}$ (\%) & $\varepsilon_{\mP_q}^{\mathrm{train}}$ (\%) & $R^2_{\bar{\mP}}$ & $\mathrm{NMAE}_{\bar{\mP}}$ (\%) & $R^2_{\bar{\sA}}$ & $\mathrm{NMAE}_{\bar{\sA}}$ (\%) & $s_{\mathrm{QDEIM}}$ & $s_{\mathrm{hom}}$ \\
                    \midrule
        10 & 22 & $1 \times 16$ & 0.27 & 4.48 & 6.04 & 0.9918 & 0.32 & 0.9815 & 0.79 & $2\times 10^{3}$ & $4\times 10^{4}$ \\
        10 & 22 & $1 \times 32$ & 0.19 & 2.97 & 4.05 & 0.9945 & 0.23 & 0.9913 & 0.56 & $2\times 10^{3}$ & $3\times 10^{4}$ \\
        10 & 22 & $1 \times 64$ & 0.16 & 2.60 & 3.69 & 0.9967 & 0.19 & 0.9925 & 0.54 & $2\times 10^{3}$ & $2\times 10^{4}$ \\
        10 & 22 & $2 \times 16$ & 0.21 & 3.51 & 4.94 & 0.9937 & 0.27 & 0.9879 & 0.71 & $2\times 10^{3}$ & $3\times 10^{4}$ \\
        10 & 22 & $2 \times 32$ & 0.15 & 2.38 & 3.25 & 0.9975 & 0.16 & 0.9946 & 0.49 & $2\times 10^{3}$ & $2\times 10^{4}$ \\
        10 & 22 & $2 \times 64$ & 0.14 & 2.24 & 3.06 & 0.9975 & 0.16 & 0.9946 & 0.47 & $1\times 10^{3}$ & $6\times 10^{3}$ \\
        \midrule
        20 & 37 & $1 \times 16$ & 0.32 & 7.49 & 10.68 & 0.9719 & 0.50 & 0.9578 & 1.21 & $2\times 10^{3}$ & $5\times 10^{4}$ \\
        20 & 37 & $1 \times 32$ & 0.21 & 4.65 & 6.35 & 0.9902 & 0.30 & 0.9835 & 0.78 & $2\times 10^{3}$ & $3\times 10^{4}$ \\
        20 & 37 & $1 \times 64$ & 0.15 & 2.95 & 4.14 & 0.9962 & 0.19 & 0.9912 & 0.56 & $2\times 10^{3}$ & $2\times 10^{4}$ \\
        20 & 37 & $2 \times 16$ & 0.30 & 6.19 & 9.15 & 0.9813 & 0.42 & 0.9722 & 1.00 & $2\times 10^{3}$ & $4\times 10^{4}$ \\
        20 & 37 & $2 \times 32$ & 0.14 & 2.71 & 4.05 & 0.9965 & 0.19 & 0.9903 & 0.60 & $2\times 10^{3}$ & $2\times 10^{4}$ \\
        20 & 37 & $2 \times 64$ & 0.09 & 1.37 & 2.08 & 0.9990 & 0.10 & 0.9951 & 0.41 & $1\times 10^{3}$ & $8\times 10^{3}$ \\
                    \bottomrule
                \end{tabular}
            \end{threeparttable}
        \end{footnotesize}
        }
    \end{center}
\end{table}
As in the stochastic-fiber case, the table reports full-field errors, homogenized-stress and homogenized-tangent errors, the training error $\varepsilon_{\mP_q}^{\mathrm{train}}$ at the Q-DEIM points, and the forward-pass speed-up factors $s_{\mathrm{QDEIM}}$ and $s_{\mathrm{hom}}$.
\begin{figure}[htb]
    \centering
    \begin{subfigure}{0.2\textwidth}
        \centering
        \includegraphics[height=0.9\textwidth]{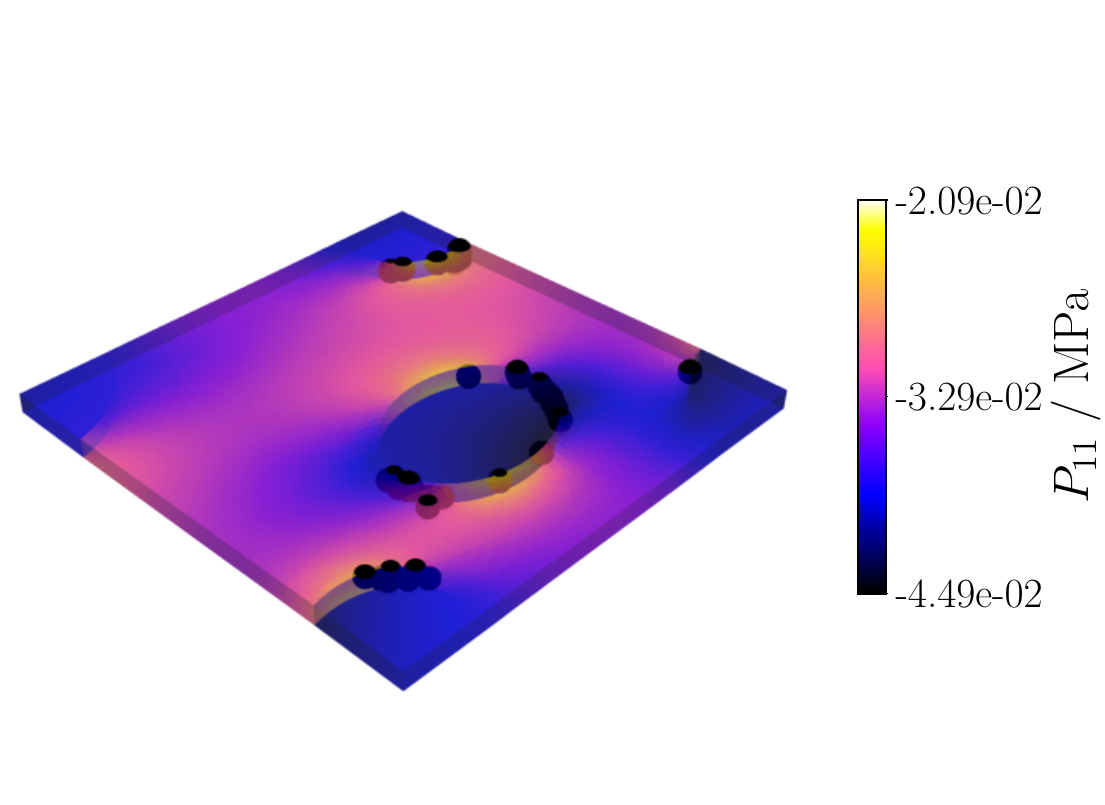}
    \end{subfigure}
    \hfil
    \begin{subfigure}{0.2\textwidth}
        \centering
        \includegraphics[height=0.9\textwidth]{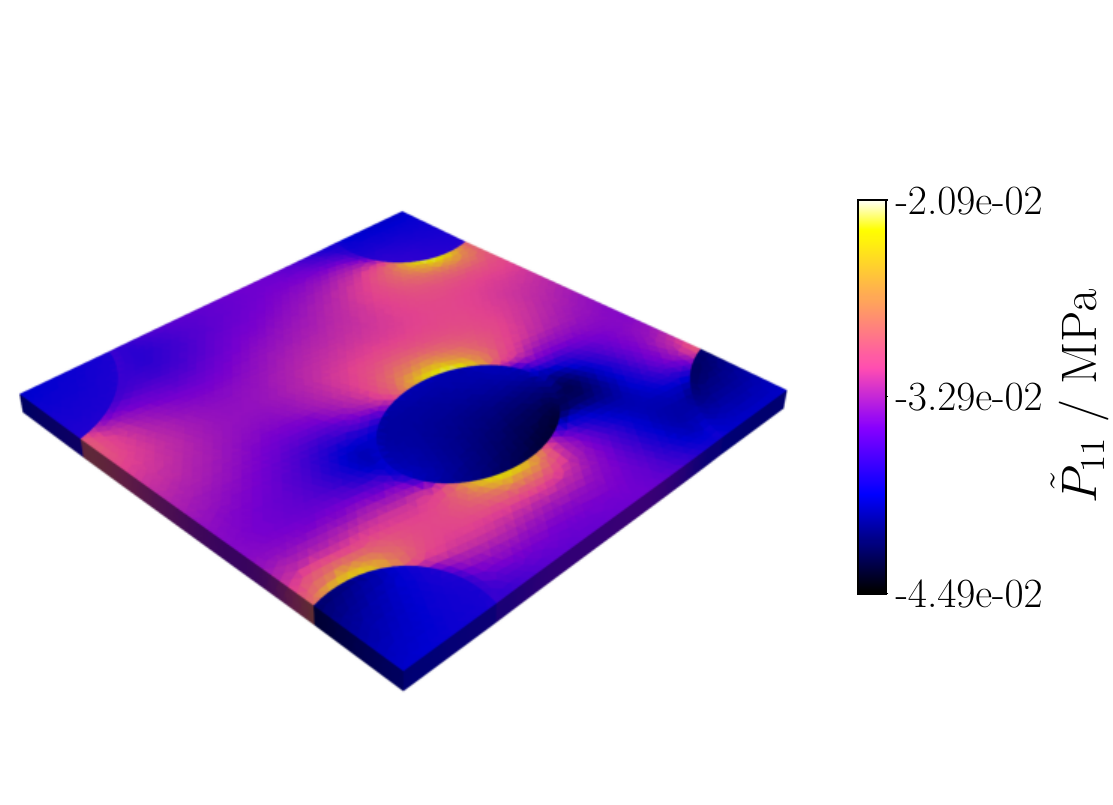}
    \end{subfigure}
    \hfil
    \begin{subfigure}{0.2\textwidth}
        \centering
        \includegraphics[height=0.9\textwidth]{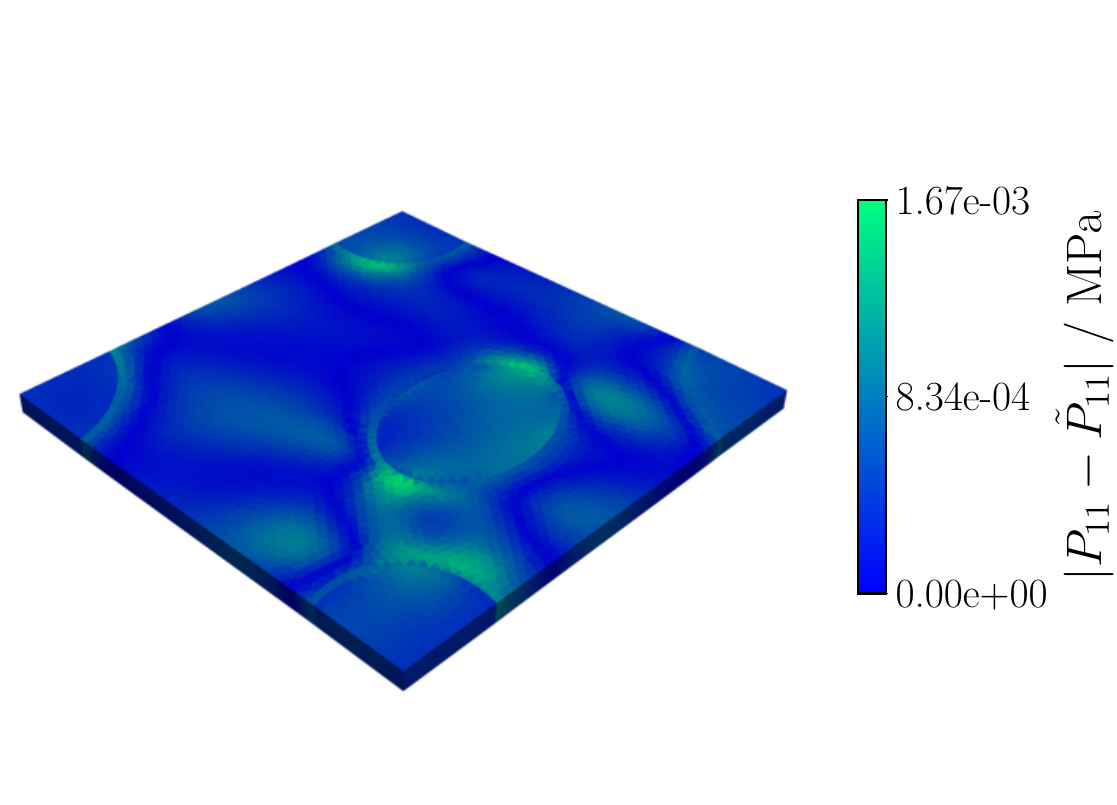}
    \end{subfigure}
    \hfil

    \begin{subfigure}{0.2\textwidth}
        \centering
        \includegraphics[height=0.9\textwidth]{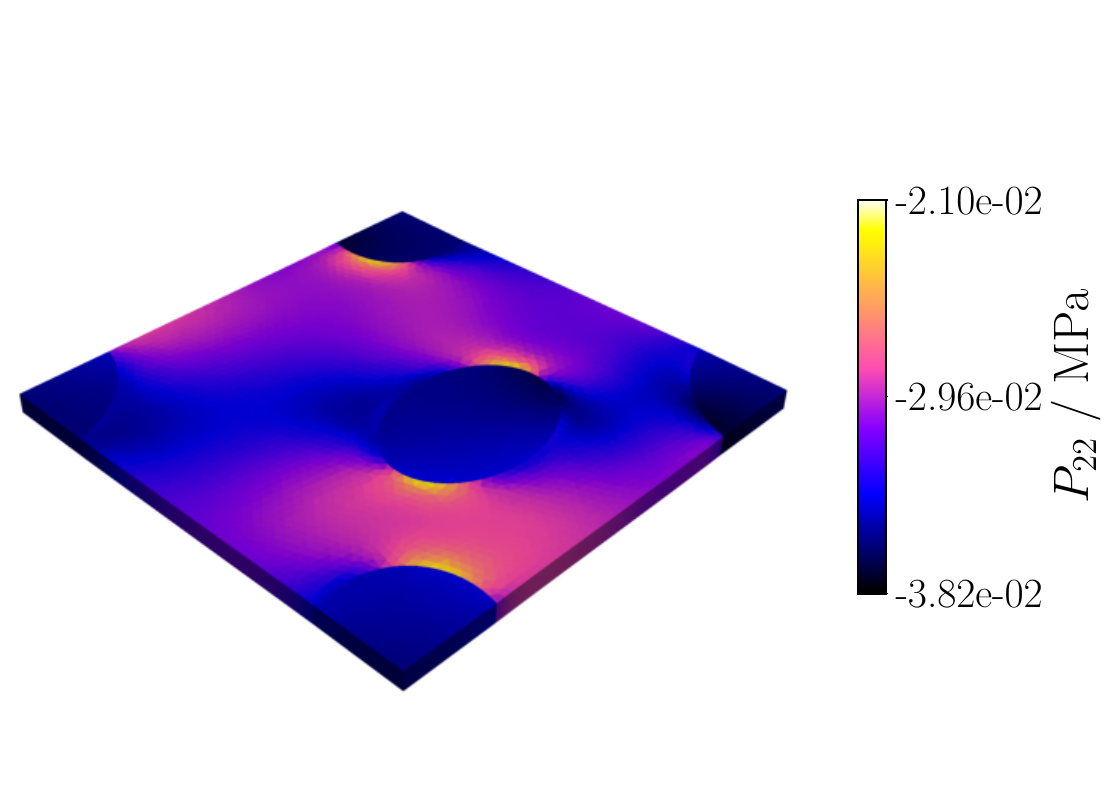}
    \end{subfigure}
    \hfil
    \begin{subfigure}{0.2\textwidth}
        \centering
        \includegraphics[height=0.9\textwidth]{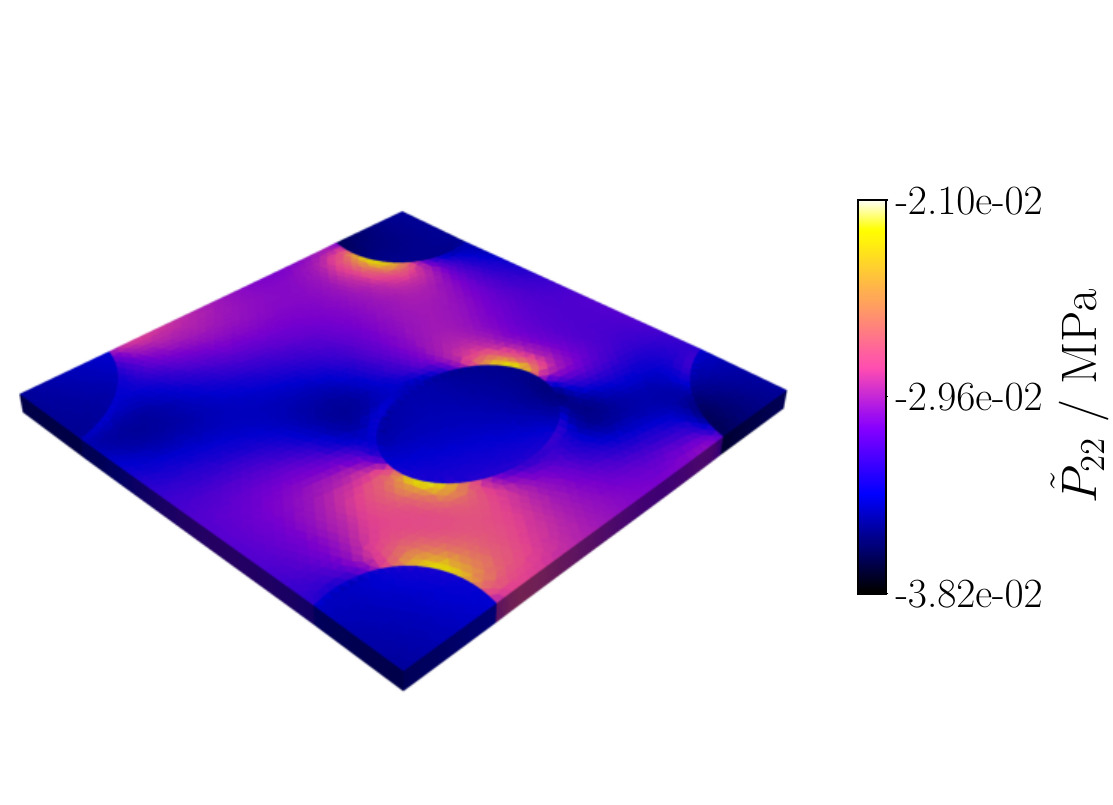}
    \end{subfigure}
    \hfil
    \begin{subfigure}{0.2\textwidth}
        \centering
        \includegraphics[height=0.9\textwidth]{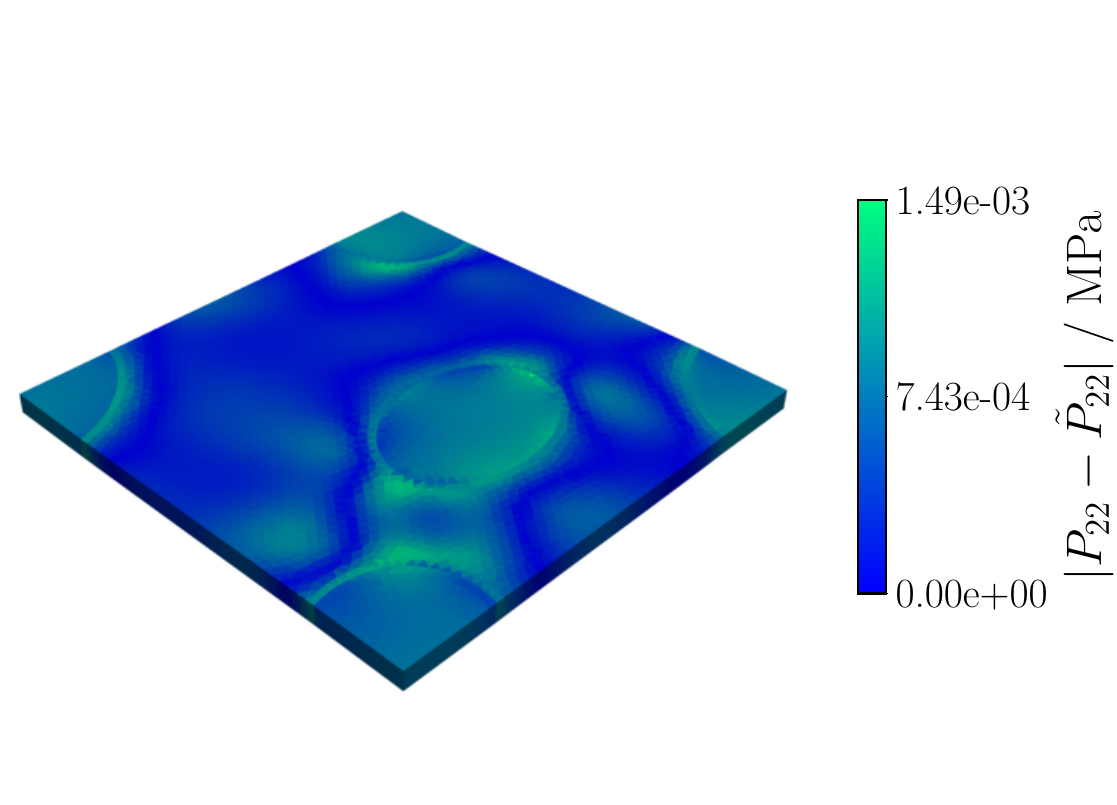}
    \end{subfigure}
    \hfil

    \begin{subfigure}{0.2\textwidth}
        \centering
        \includegraphics[height=0.9\textwidth]{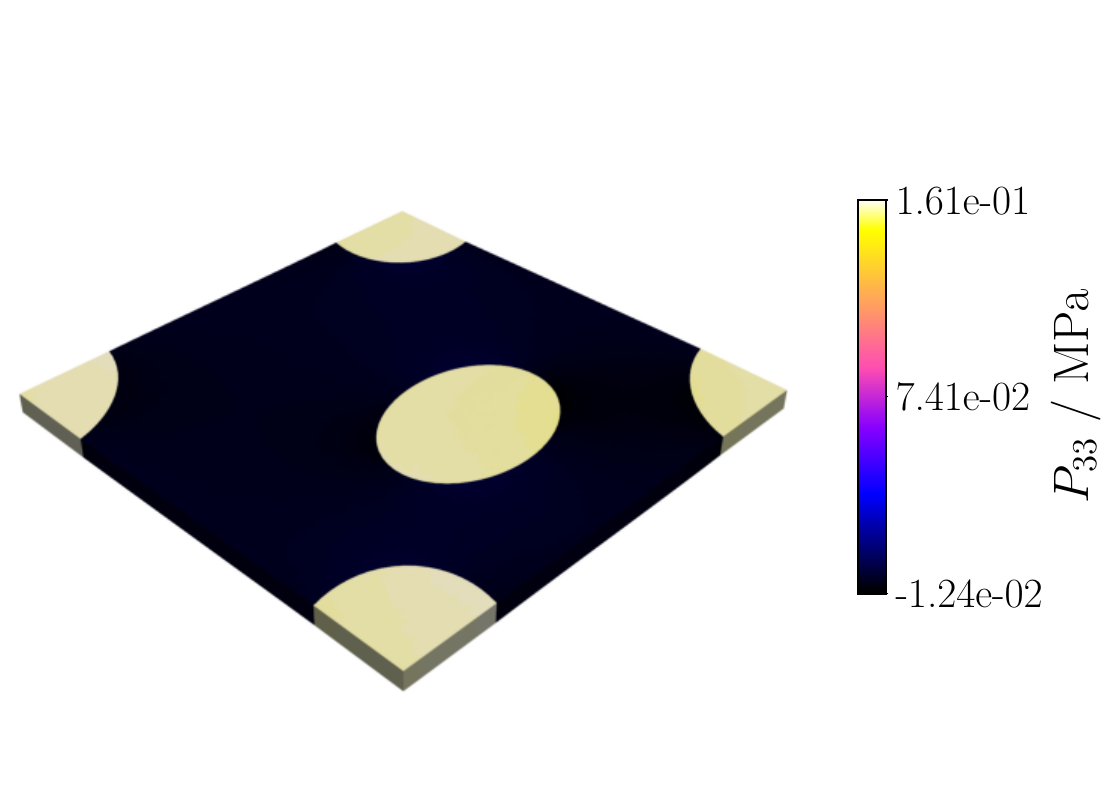}
    \end{subfigure}
    \hfil
    \begin{subfigure}{0.2\textwidth}
        \centering
        \includegraphics[height=0.9\textwidth]{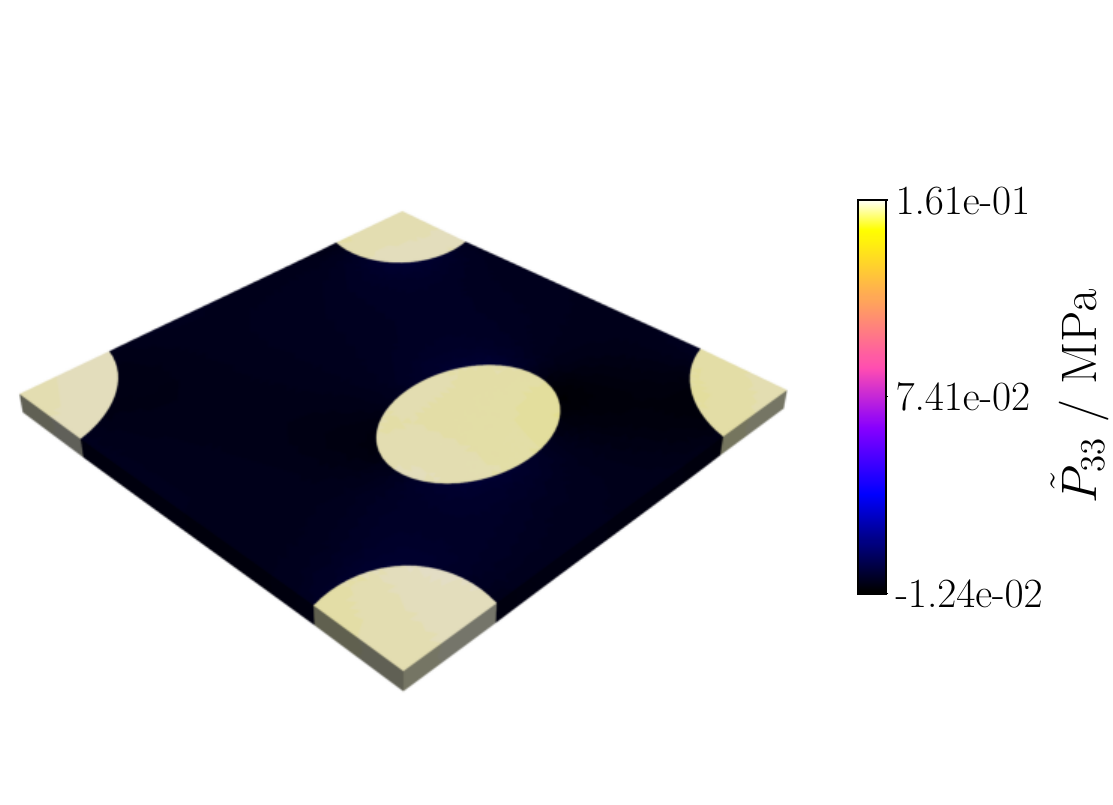}
    \end{subfigure}
    \hfil
    \begin{subfigure}{0.2\textwidth}
        \centering
        \includegraphics[height=0.9\textwidth]{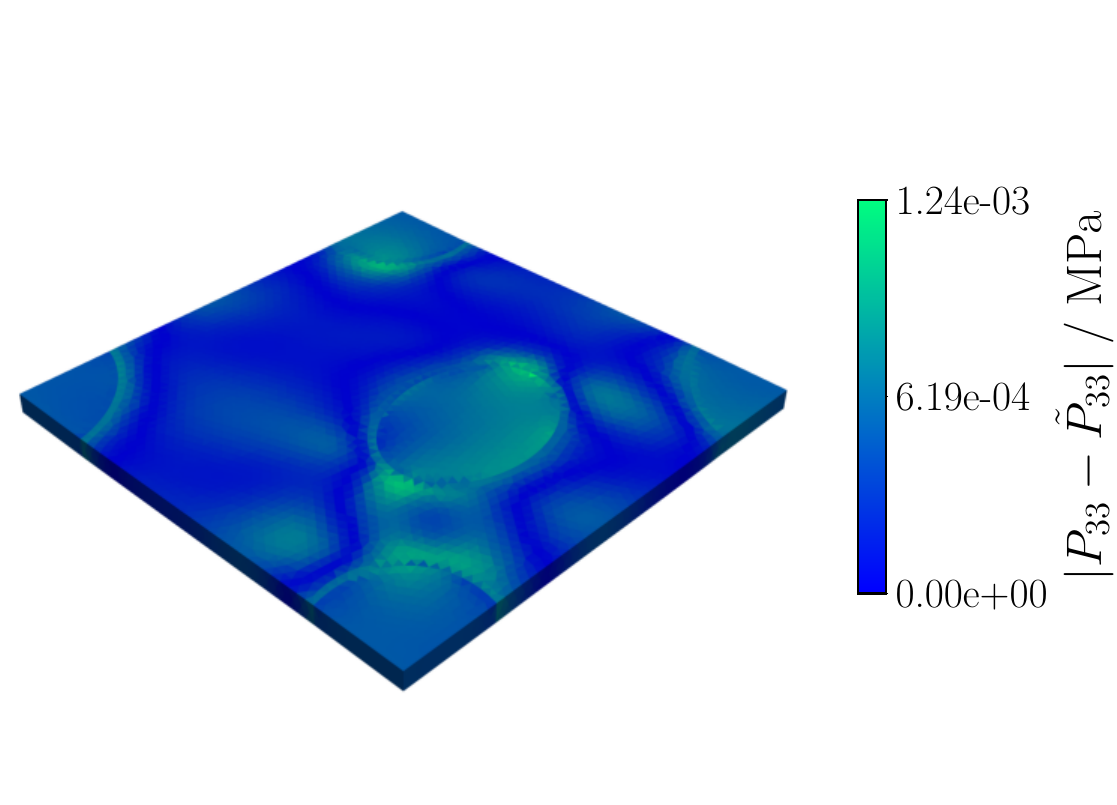}
    \end{subfigure}
    \hfil
    \caption{Comparison of reference, predicted, and absolute-error fields for the microscopic first Piola--Kirchhoff stress components of the hexagonal-fiber RVE for the test sample with the median relative $L_2$ error among all in-range test samples. Columns from left to right show the reference solution, the EquiNO prediction, and the absolute error. Rows from top to bottom correspond to $P_{11}$, $P_{22}$, and $P_{33}$. Results are shown for the best in-range model with $n_p=20$, $p=37$, and a $2\times64$ network. The Q-DEIM points are illustrated in the top left panel.}
    \label{fig:hex_vis}
\end{figure}
The hexagonal-fiber results show that accurate predictions are also obtained for the more structured RVE. With only $n_p=10$ snapshot loading paths, the $1\times32$ network reaches $\varepsilon_{\mF}=0.19\%$, $\varepsilon_{\mP}=2.97\%$, $R^2_{\bar{\mP}}=0.9945$, and $\mathrm{NMAE}_{\bar{\mP}}=0.23\%$, while retaining speed-up factors of order $10^3$ for $s_{\mathrm{QDEIM}}$ and order $10^4$ for $s_{\mathrm{hom}}$. For $n_p=20$, the smaller networks show larger errors because the retained basis dimension increases to $p=37$, but the wider two-layer network resolves the additional modes effectively. The best performance in \cref{tab:errors_compact_hex} is obtained by the $n_p=20$, $p=37$, $2\times64$ model, with $\varepsilon_{\mF}=0.09\%$, $\varepsilon_{\mP}=1.37\%$, $R^2_{\bar{\mP}}=0.9990$, and $\mathrm{NMAE}_{\bar{\mP}}=0.10\%$.
Results corresponding to this best hexagonal-fiber model are shown in \cref{fig:hex_vis}. The displayed sample is the test case with the median relative $L_2$ error among all in-range test samples. The predicted stress components $P_{11}$, $P_{22}$, and $P_{33}$ reproduce the reference fields across the fiber arrangement, and the absolute-error fields remain small relative to the stress variations in the RVE.
The homogenized material tangent is also predicted accurately for the same model. For the $n_p=20$, $p=37$, $2\times64$ model, the tangent prediction reaches $R^2_{\bar{\sA}}=0.9951$ and $\mathrm{NMAE}_{\bar{\sA}}=0.41\%$.

\subsection{Out-of-range generalization}
\label{subsec:out-of-range-generalization}

For the out-of-range study, we restrict the snapshot loading paths to the region $\|\bar{\mE}\|\le 0.4\,\|\bar{\mE}\|_{\max}$ and evaluate the trained model on test paths that extend beyond this threshold. The corresponding loading-path distribution is shown in the right panel of \cref{fig:fbar_sto_cases} for the case of $n_p=10$, where the blue paths remain confined to the inner region of the strain space, while the orange paths probe larger deformation states outside the snapshot domain.
\begin{table}[bh]
    \begin{center}
        \resizebox{\textwidth}{!}{
        \begin{footnotesize}
            \begin{threeparttable}
                \caption{Out-of-range generalization errors for the stochastic-fiber and hexagonal-fiber RVEs for $\beta=0.4$ using the $2\times64$ network and different numbers of snapshot loading paths $n_p$. The table reports full-field errors, homogenized first Piola--Kirchhoff stress and material-tangent errors, and the forward-pass speed-up factors defined in \cref{subsec:efficiency}.}
                \label{tab:errors_out_range}
                \begin{tabular}{llllccccccccc}
                    \toprule
                    \multicolumn{4}{c}{Test setup} & \multicolumn{3}{c}{Full-field errors} & \multicolumn{4}{c}{Homogenized errors} & \multicolumn{2}{c}{Speed-up factors} \\
                    \cmidrule(lr){1-4}\cmidrule(lr){5-7}\cmidrule(lr){8-11}\cmidrule(lr){12-13}
                    RVE & $n_p$ & $p$ & NN size & $\varepsilon_{\mF}$ (\%) & $\varepsilon_{\mP}$ (\%) & $\varepsilon_{\mP_q}^{\mathrm{train}}$ (\%) & $R^2_{\bar{\mP}}$ & $\mathrm{NMAE}_{\bar{\mP}}$ (\%) & $R^2_{\bar{\sA}}$ & $\mathrm{NMAE}_{\bar{\sA}}$ (\%) & $s_{\mathrm{QDEIM}}$ & $s_{\mathrm{hom}}$ \\
                    \midrule
                    Sto. & 10 & 22 & $2 \times 64$ & 0.19 & 1.92 & 2.46 & 0.9987 & 0.09 & 0.9958 & 0.33 & $2\times 10^{3}$ & $8\times 10^{3}$ \\
                    Sto. & 20 & 37 & $2 \times 64$ & 0.11 & 1.15 & 1.15 & 0.9994 & 0.05 & 0.9968 & 0.27 & $2\times 10^{3}$ & $1\times 10^{4}$ \\
                    \midrule
                    Hex. & 10 & 22 & $2 \times 64$ & 0.24 & 3.51 & 5.03 & 0.9821 & 0.24 & 0.9768 & 0.68 & $2\times 10^{3}$ & $7\times 10^{3}$ \\
                    Hex. & 20 & 35 & $2 \times 64$ & 0.10 & 1.07 & 1.16 & 0.9995 & 0.05 & 0.9969 & 0.29 & $1\times 10^{3}$ & $8\times 10^{3}$ \\
                    \bottomrule
                \end{tabular}
            \end{threeparttable}
        \end{footnotesize}
        }
    \end{center}
\end{table}
The quantitative out-of-range results are summarized in \cref{tab:errors_out_range} for both stochastic-fiber and hexagonal-fiber RVEs. The table reports the same full-field errors $\varepsilon_{\mF}$ and $\varepsilon_{\mP}$, the training error $\varepsilon_{\mP_q}^{\mathrm{train}}$ at the Q-DEIM points, and the homogenized errors for the first Piola--Kirchhoff stress and material tangent, measured by $R^2_{\bar{\mP}}$, $\mathrm{NMAE}_{\bar{\mP}}$, $R^2_{\bar{\sA}}$, and $\mathrm{NMAE}_{\bar{\sA}}$.
With only $n_p=10$ restricted snapshot loading paths, the stochastic-fiber RVE already gives accurate out-of-range predictions, with a full-field stress error of $\varepsilon_{\mP}=1.92\%$ and a homogenized-stress error of $\mathrm{NMAE}_{\bar{\mP}}=0.09\%$. The hexagonal-fiber RVE is more sensitive in this low-snapshot regime, giving $\varepsilon_{\mP}=3.51\%$ and $\mathrm{NMAE}_{\bar{\mP}}=0.24\%$. This is consistent with the larger training error at the Q-DEIM points for the hexagonal-fiber case, which indicates that the restricted basis and limited number of snapshot paths make this setting more challenging.
Increasing the number of restricted snapshot loading paths to $n_p=20$ substantially improves both RVEs. The full-field stress error decreases to $\varepsilon_{\mP}=1.15\%$ for the stochastic-fiber RVE and to $\varepsilon_{\mP}=1.07\%$ for the hexagonal-fiber RVE, while the homogenized-stress error is reduced to $\mathrm{NMAE}_{\bar{\mP}}=0.05\%$ in both cases. The material tangent is also recovered accurately in this out-of-range setting, with $\mathrm{NMAE}_{\bar{\sA}}$ below $0.30\%$ for both RVEs. The forward-pass speed-up factors remain of order $10^3$ for the Q-DEIM loss evaluation, while the reduced homogenization speed-up ranges from $7\times 10^3$ to $1\times 10^4$.
\begin{figure}[ht]
    \centering
    \begin{subfigure}{0.3\textwidth}
        \centering
        \includegraphics[height=0.85\textwidth, trim={0 0 300 0}, clip]{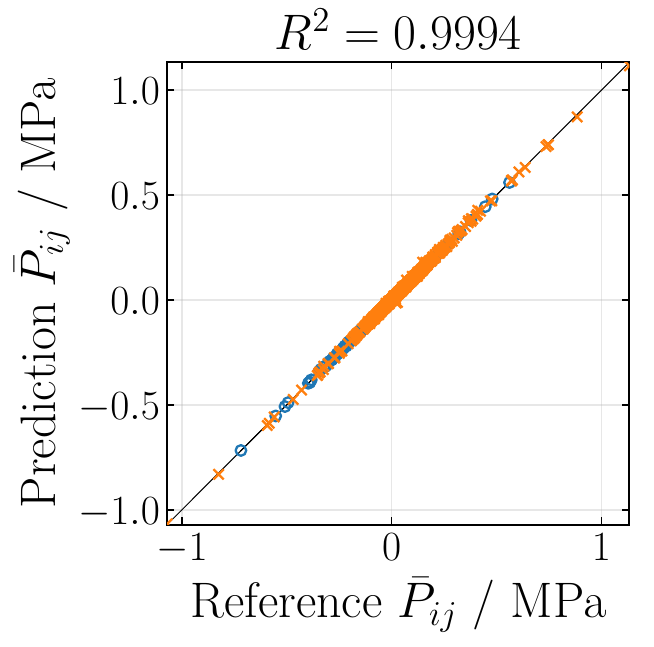}
        \caption{Stochastic-fiber RVE.}
    \end{subfigure}
    \begin{subfigure}{0.3\textwidth}
        \centering
        \includegraphics[height=0.85\textwidth, trim={0 0 300 0}, clip]{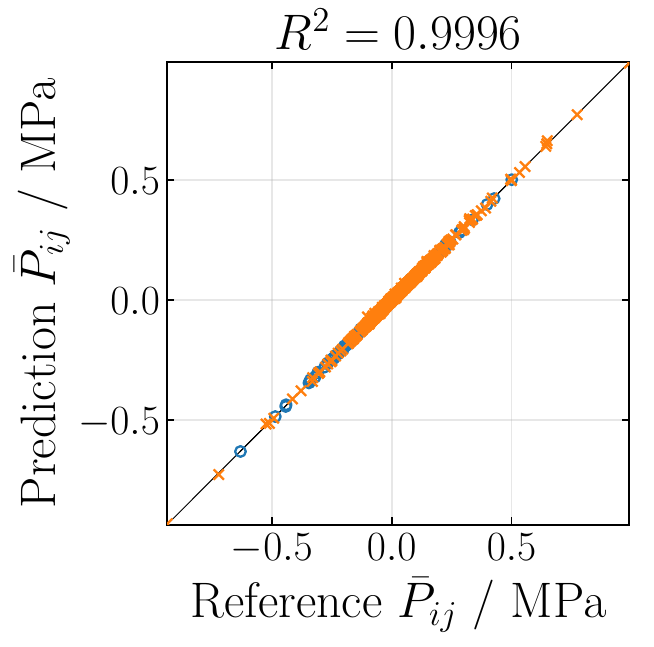}
        \caption{Hexagonal-fiber RVE.}
    \end{subfigure}
    \caption{Parity plots for the homogenized first Piola--Kirchhoff stress of the stochastic-fiber (left) and the hexagonal-fiber (right) RVEs for the out-of-range setting, showing the aggregated comparison over all first Piola--Kirchhoff stress components. Samples correspond to snapshot loading paths (blue) and unseen test loading paths (orange).}
    \label{fig:sto_parity}
\end{figure}

The corresponding parity plots for the homogenized first Piola--Kirchhoff stress in the out-of-range setting are shown in \cref{fig:sto_parity} for the $n_p=20$ models. The unseen test samples remain close to the diagonal for both RVEs, complementing the homogenized first Piola--Kirchhoff stress and tangent metrics in \cref{tab:errors_out_range}. These results show that EquiNO remains robust even when the test trajectories leave the strain range covered by the snapshot loading paths used for basis construction.

\section{Summary and conclusions}
\label{sec:conclusions}

In this work, we extended the Equilibrium Neural Operator (EquiNO) to three-dimensional finite-strain hyperelastic RVEs and combined it with Q-DEIM to obtain an efficient physics-informed surrogate for local fields and homogenized first Piola--Kirchhoff stresses. The formulation retains the central idea of EquiNO, namely to represent the displacement fluctuation and first Piola--Kirchhoff stress in reduced bases that satisfy periodicity and equilibrium by construction, and transfers it to the finite-strain setting. The numerical results for stochastic-fiber and hexagonal-fiber RVEs show that the resulting model delivers accurate in-range and out-of-range predictions of microscopic stress fields and homogenized first Piola--Kirchhoff stresses, while requiring only a limited number of offline snapshot loading paths for basis construction.

The main computational gain of the present framework is enabled by three ingredients. First, Q-DEIM reduces the cost of the physics-informed loss evaluation so that full-batch training with L-BFGS becomes practical for large sets of sampled loading states in three-dimensional RVEs. Second, the reduced operator representation confines the learned problem to modal coefficients rather than full-field degrees of freedom, which lowers the optimization cost while preserving the relevant mechanics by construction. Third, homogenized first Piola--Kirchhoff stresses are obtained directly from the reduced stress representation by averaging the modes offline, so macroscopic outputs can be evaluated without reconstructing the full-field stress. Together, these steps yield training-side speed-up factors of order $10^3$ and reduced-homogenization speed-up factors of order $10^3$ to $10^4$ relative to the corresponding full-field computations. In the present examples, physics-informed training on $233$ unsupervised loading paths takes approximately half the time required for one finite-element simulation of a single loading path with the periodic homogenization solver. This makes physics-informed operator learning substantially more practical for microstructure problems in which direct full-field training and evaluation would otherwise be limited by computational cost and memory usage.

A natural next step is the extension of the present framework to inelastic microstructures. There, the main challenge is the introduction of history dependence through internal variables and path-dependent constitutive updates. The results of the present work indicate that the combination of constrained reduced representations, hyper-reduced loss evaluation, and reduced homogenization is a promising basis for addressing this additional complexity. 


\section*{Data Availability}
The source code, datasets, trained models, and supplementary materials associated with this study will be available on GitHub.

\bibliographystyle{jabbrv_elsarticle_model1-num-names}
\small{
\bibliography{References}
}
\clearpage

\appendix

\section{Supplementary results}
\label{app:supplementary-results}

\begin{figure}[ht]
    \centering
    \begin{subfigure}{0.45\textwidth}
        \centering
        \includegraphics[width=\textwidth]{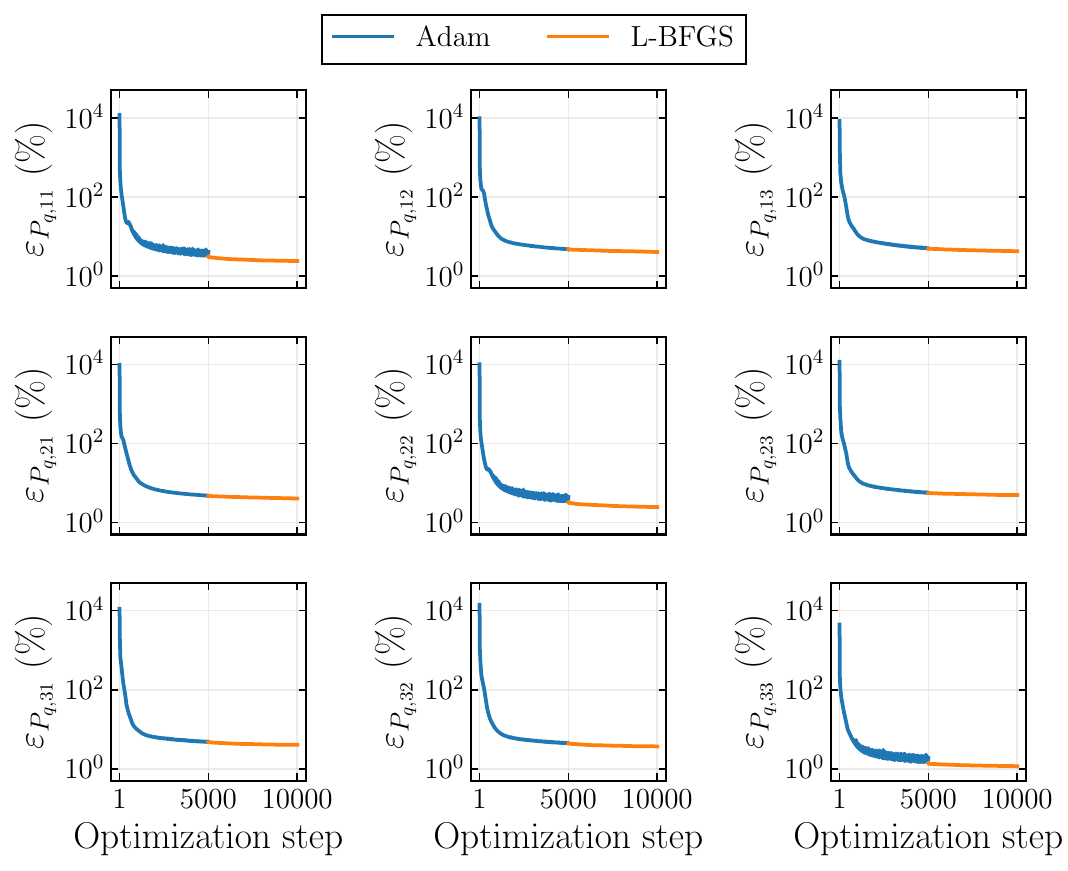}
        \caption{Stochastic-fiber RVE.}
    \end{subfigure}
    \hfil
    \begin{subfigure}{0.45\textwidth}
        \centering
        \includegraphics[width=\textwidth]{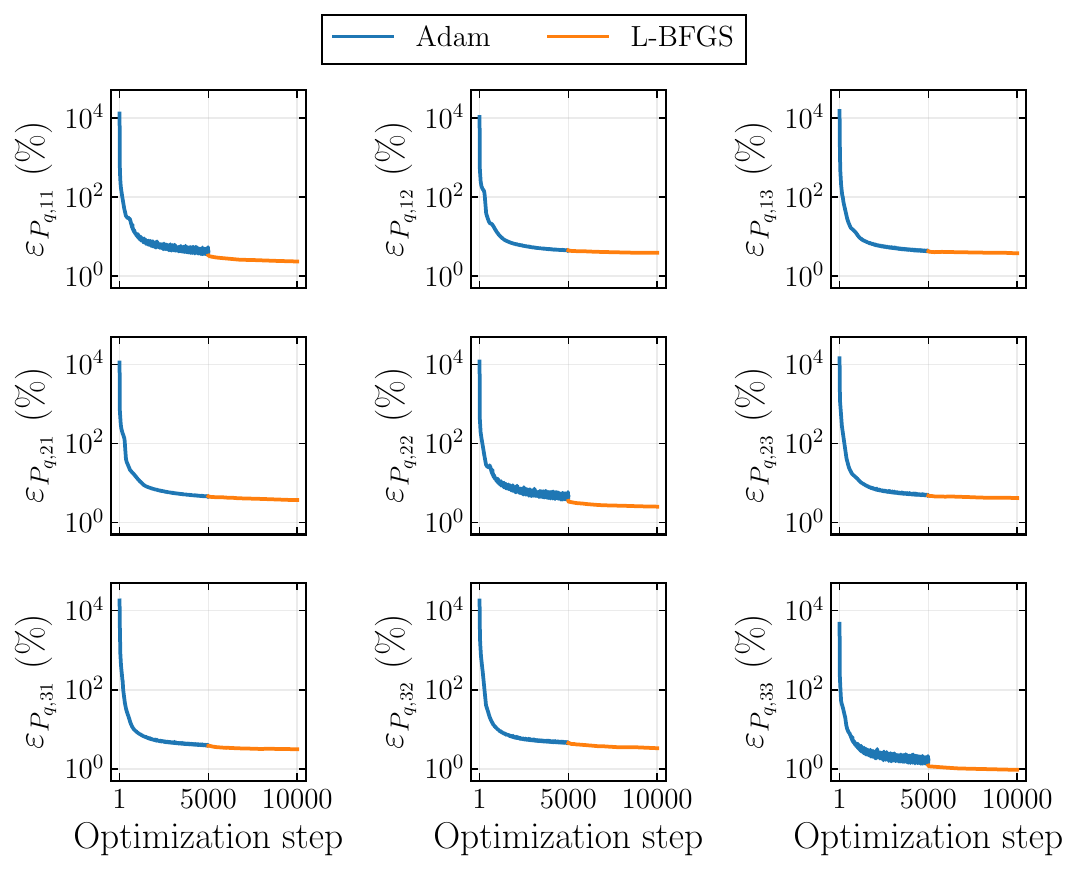}
        \caption{Hexagonal-fiber RVE.}
    \end{subfigure}
    \caption{Component-wise learning curves for the first Piola--Kirchhoff stress on the Q-DEIM points in the out-of-range setting corresponding to \cref{fig:learning_curve_hex}. The results are shown for the stochastic-fiber and hexagonal-fiber RVEs with $n_p=20$, $\beta=0.4$, and a $2\times64$ network.}
    \label{fig:learning_curve_components}
\end{figure}

This appendix collects additional diagnostics that complement the main in-range and out-of-range results. First, we report component-wise learning curves for the first Piola--Kirchhoff stress at the Q-DEIM points in \cref{fig:learning_curve_components}, which show how each stress component contributes to the reduced physics-informed loss during training. 

\begin{figure}[htb]
    \centering
    \begin{subfigure}{0.2\textwidth}
        \centering
        \includegraphics[height=0.85\textwidth]{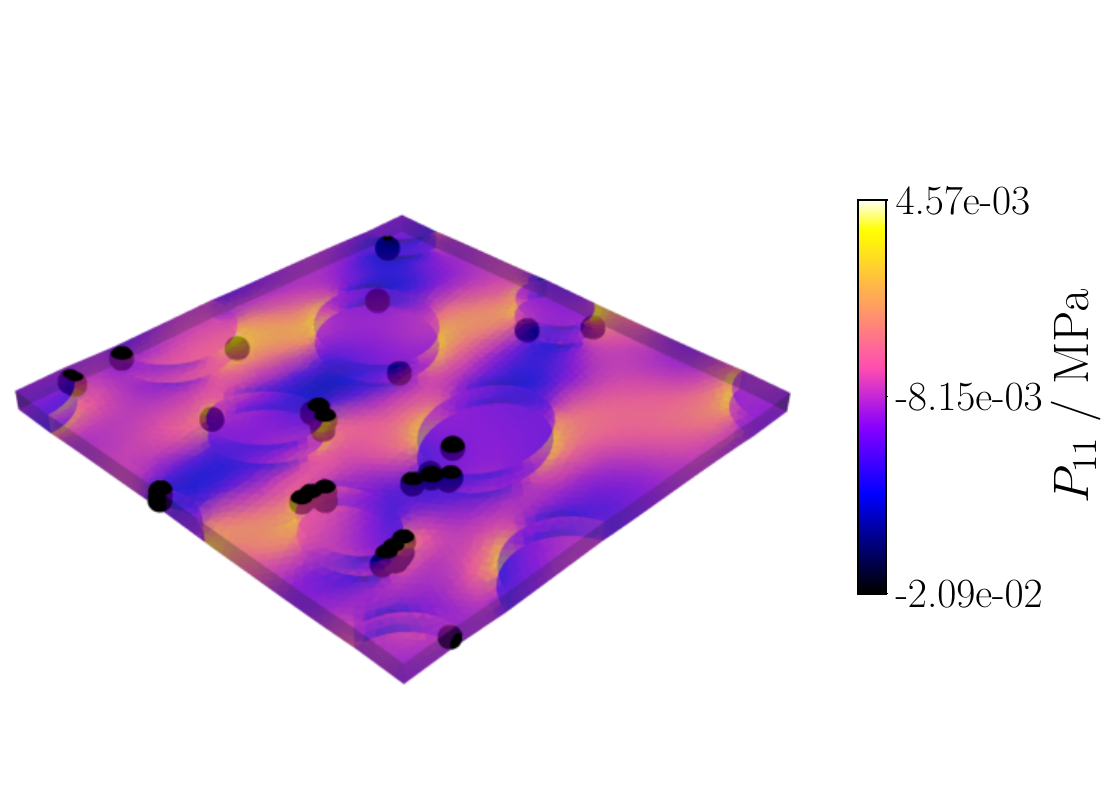}
    \end{subfigure}
    \hfil
    \begin{subfigure}{0.2\textwidth}
        \centering
        \includegraphics[height=0.85\textwidth]{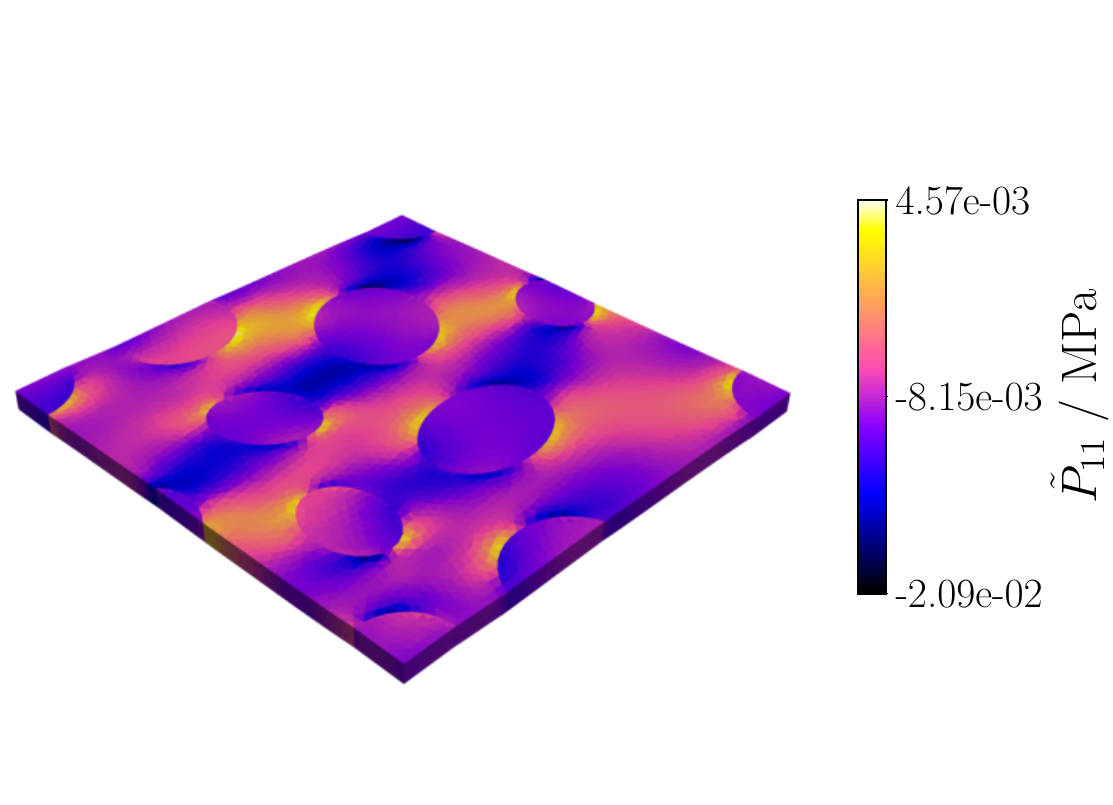}
    \end{subfigure}
    \hfil
    \begin{subfigure}{0.2\textwidth}
        \centering
        \includegraphics[height=0.85\textwidth]{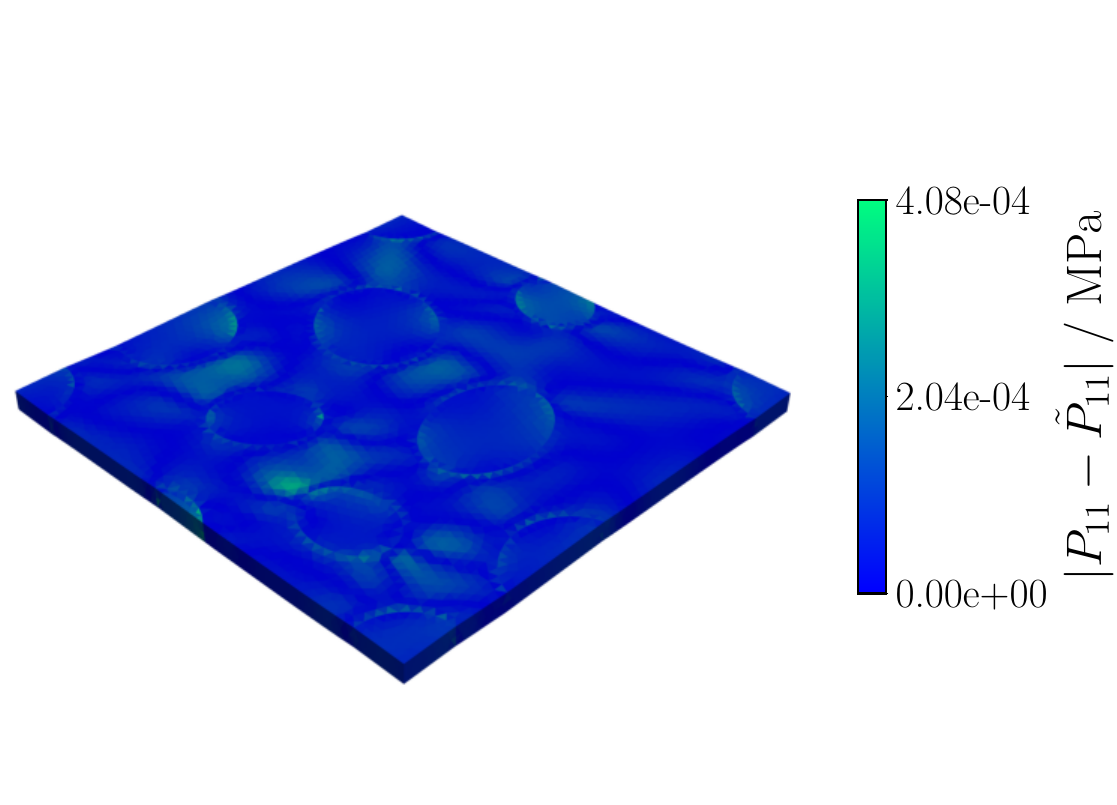}
    \end{subfigure}
    \hfil

    \begin{subfigure}{0.2\textwidth}
        \centering
        \includegraphics[height=0.85\textwidth]{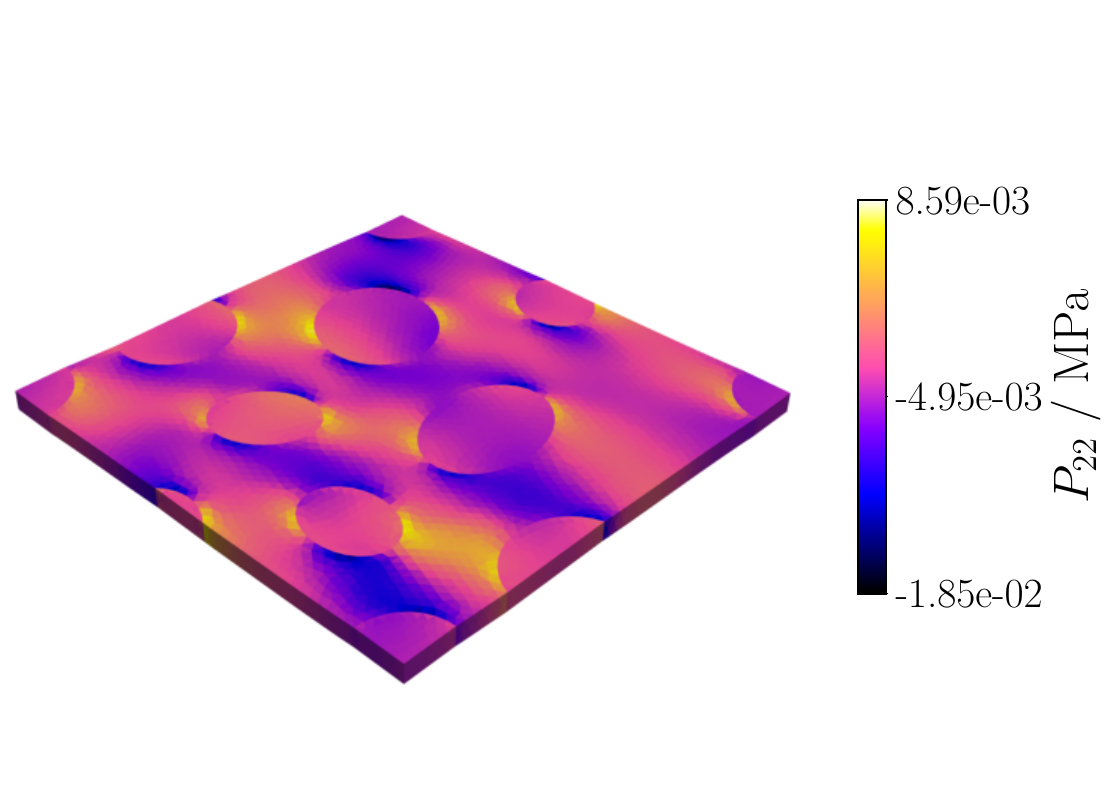}
    \end{subfigure}
    \hfil
    \begin{subfigure}{0.2\textwidth}
        \centering
        \includegraphics[height=0.85\textwidth]{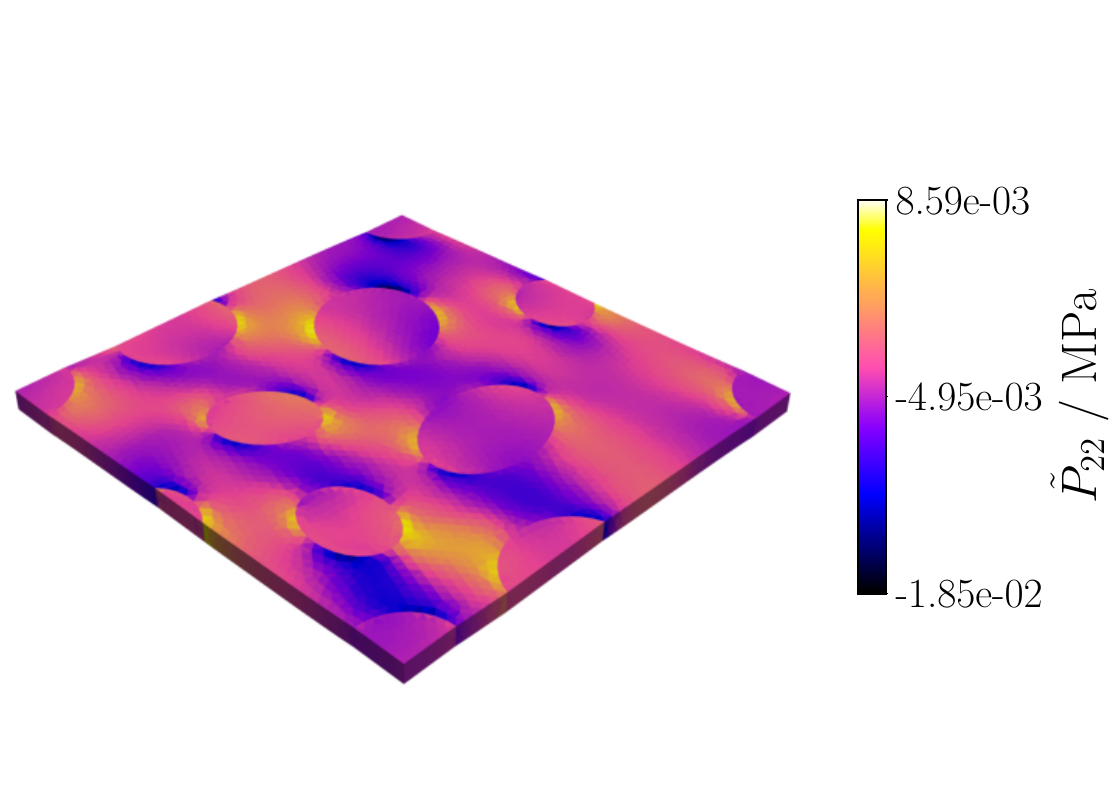}
    \end{subfigure}
    \hfil
    \begin{subfigure}{0.2\textwidth}
        \centering
        \includegraphics[height=0.85\textwidth]{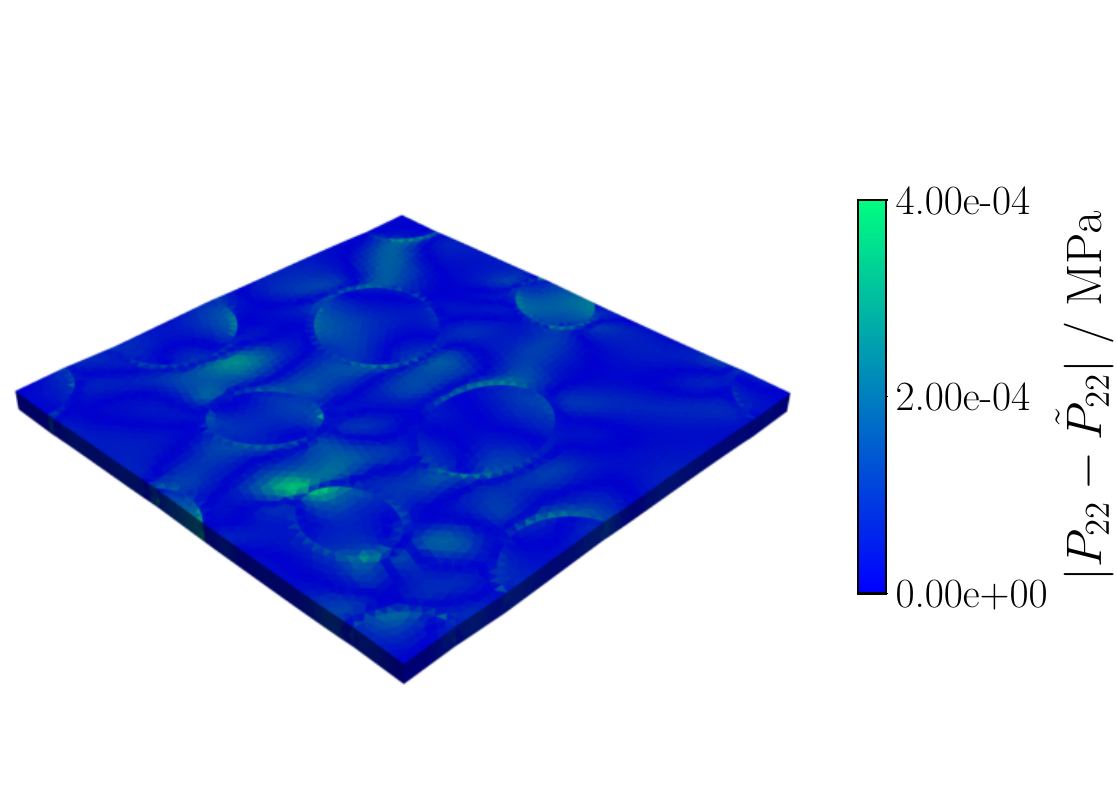}
    \end{subfigure}
    \hfil

    \begin{subfigure}{0.2\textwidth}
        \centering
        \includegraphics[height=0.85\textwidth]{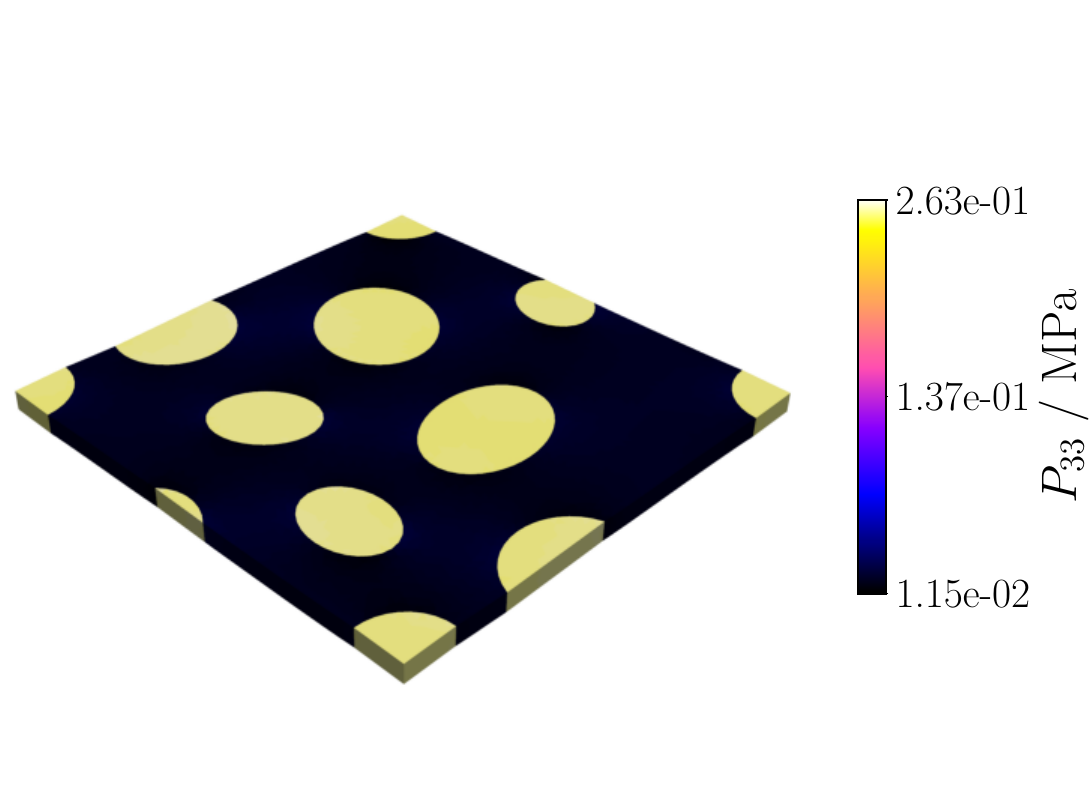}
    \end{subfigure}
    \hfil
    \begin{subfigure}{0.2\textwidth}
        \centering
        \includegraphics[height=0.85\textwidth]{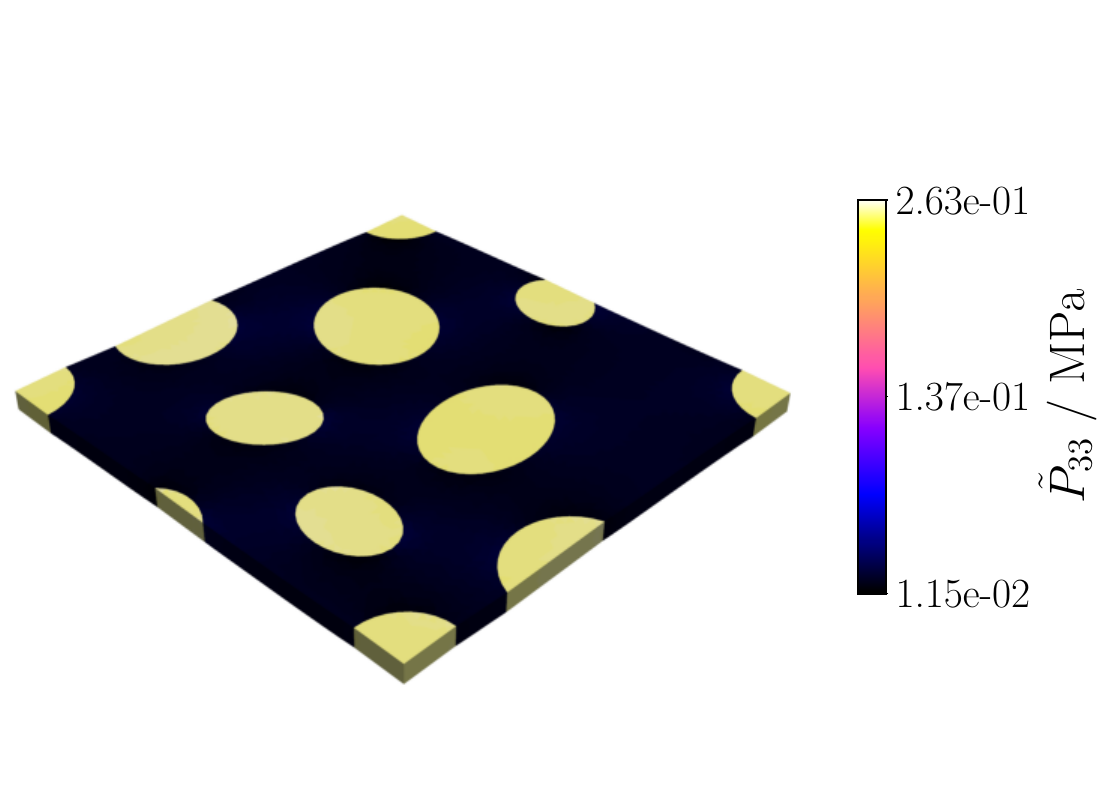}
    \end{subfigure}
    \hfil
    \begin{subfigure}{0.2\textwidth}
        \centering
        \includegraphics[height=0.85\textwidth]{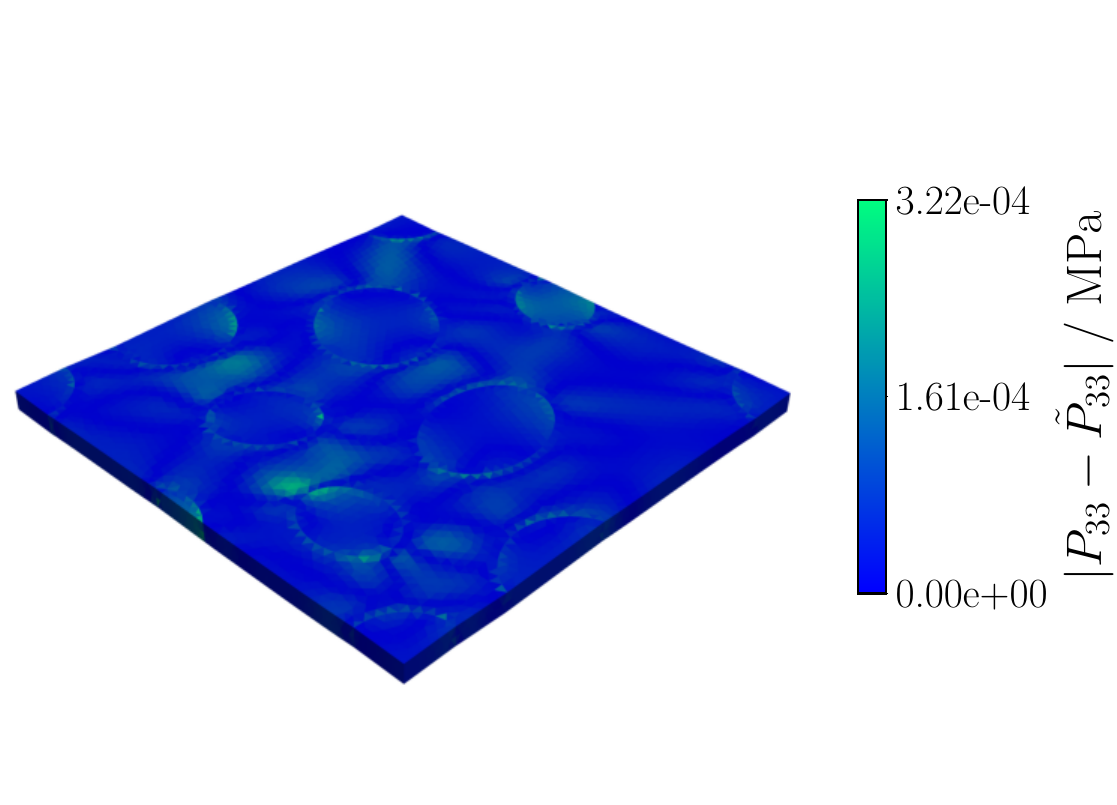}
    \end{subfigure}
    \hfil
    \caption{Comparison of reference, predicted, and absolute-error fields for the microscopic first Piola--Kirchhoff stress components of the stochastic-fiber RVE for the test sample with the minimum relative $L_2$ error among all in-range test samples. See \cref{fig:sto_vis} for the description of the visualization.}
    \label{fig:sto_vis_min}
\end{figure}

\begin{figure}[htb]
    \centering
    \begin{subfigure}{0.2\textwidth}
        \centering
        \includegraphics[height=0.85\textwidth]{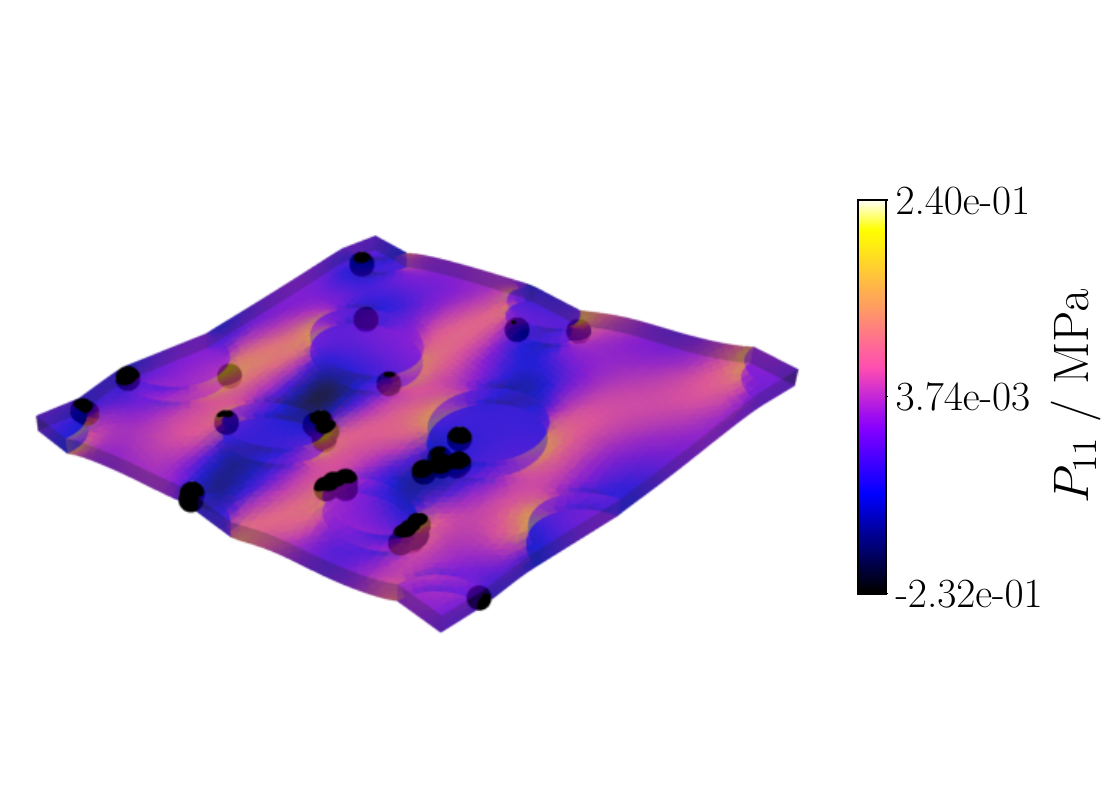}
    \end{subfigure}
    \hfil
    \begin{subfigure}{0.2\textwidth}
        \centering
        \includegraphics[height=0.85\textwidth]{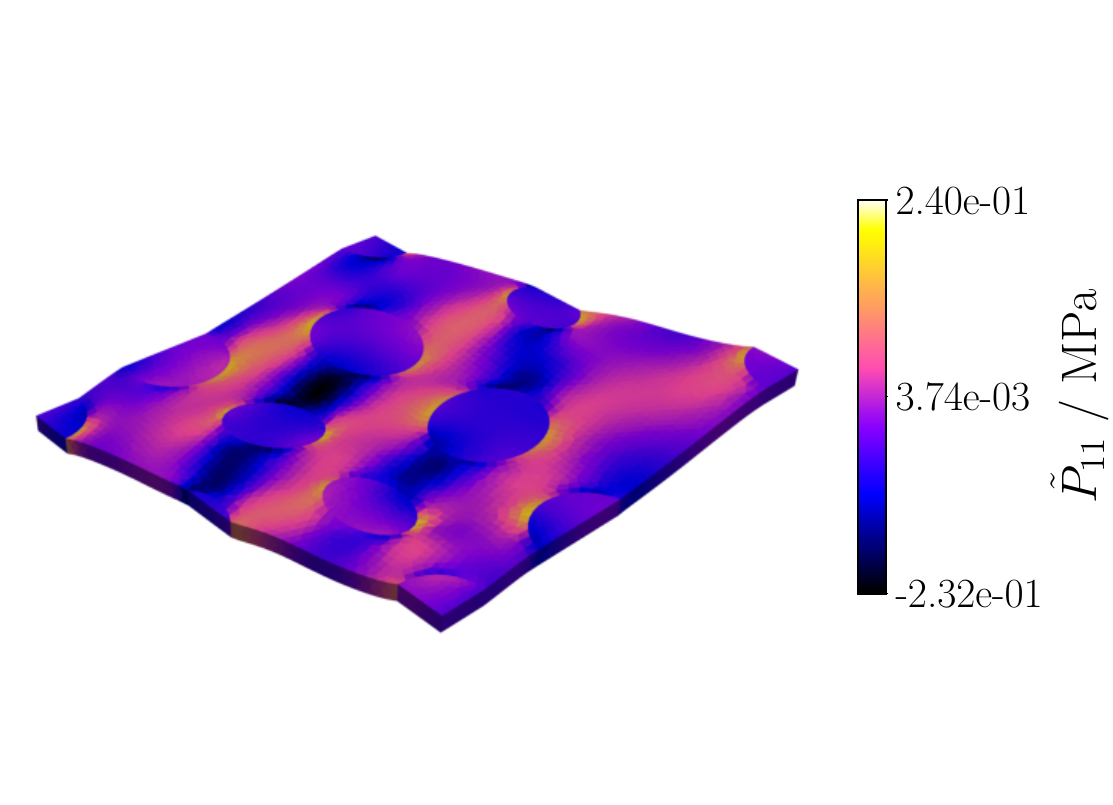}
    \end{subfigure}
    \hfil
    \begin{subfigure}{0.2\textwidth}
        \centering
        \includegraphics[height=0.85\textwidth]{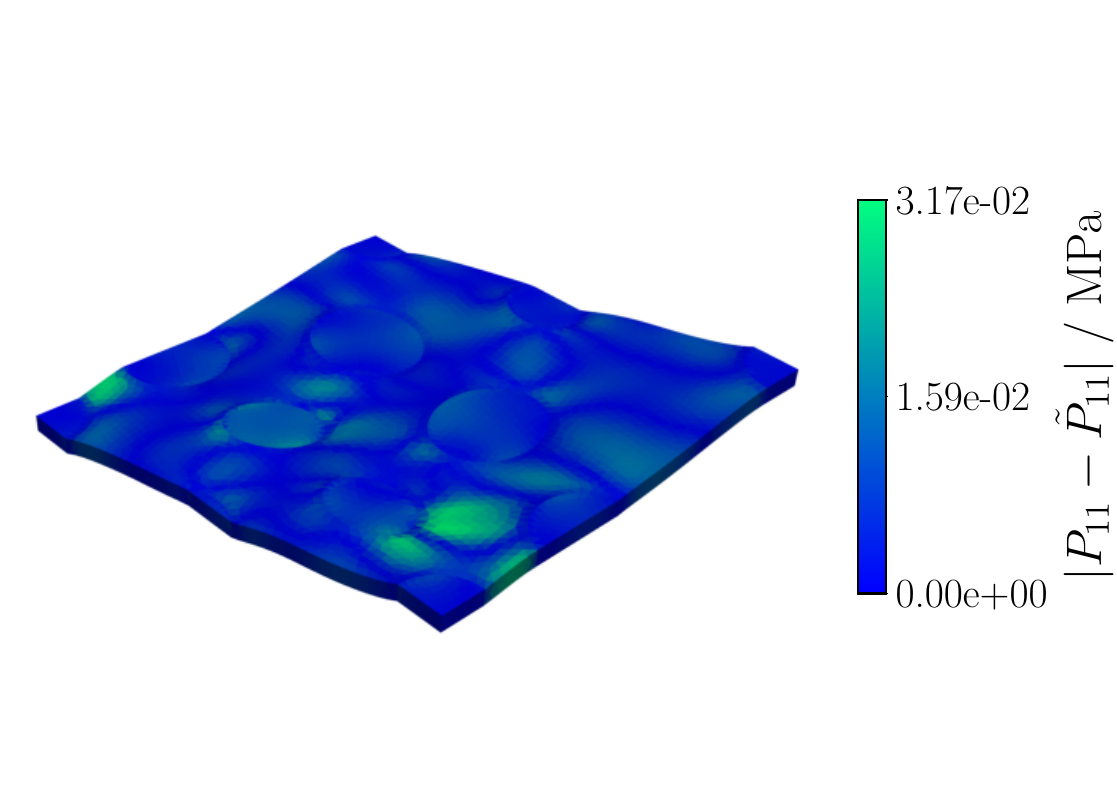}
    \end{subfigure}
    \hfil

    \begin{subfigure}{0.2\textwidth}
        \centering
        \includegraphics[height=0.85\textwidth]{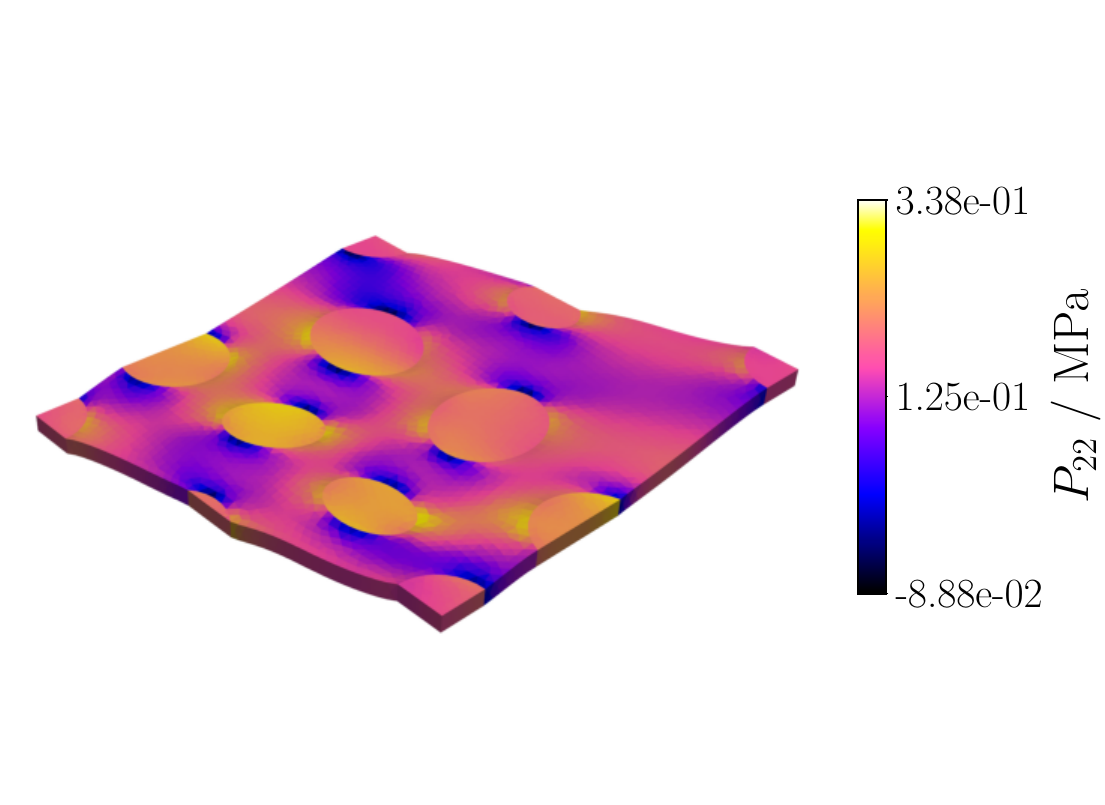}
    \end{subfigure}
    \hfil
    \begin{subfigure}{0.2\textwidth}
        \centering
        \includegraphics[height=0.85\textwidth]{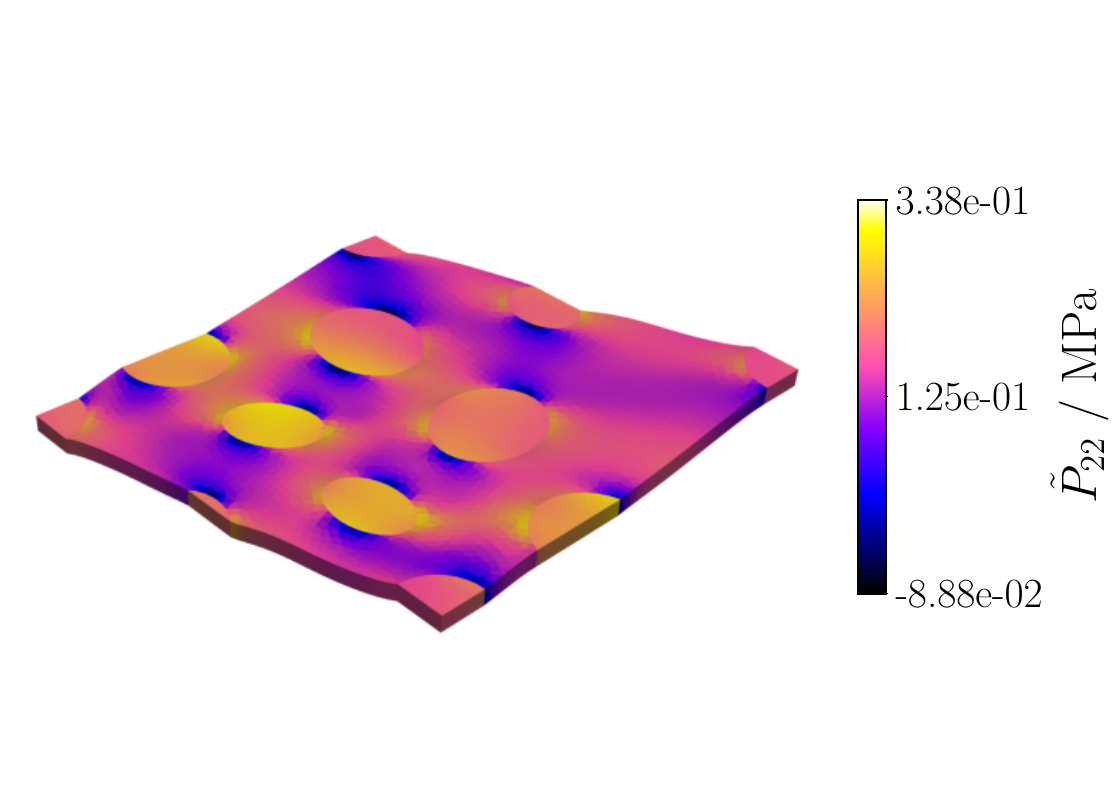}
    \end{subfigure}
    \hfil
    \begin{subfigure}{0.2\textwidth}
        \centering
        \includegraphics[height=0.85\textwidth]{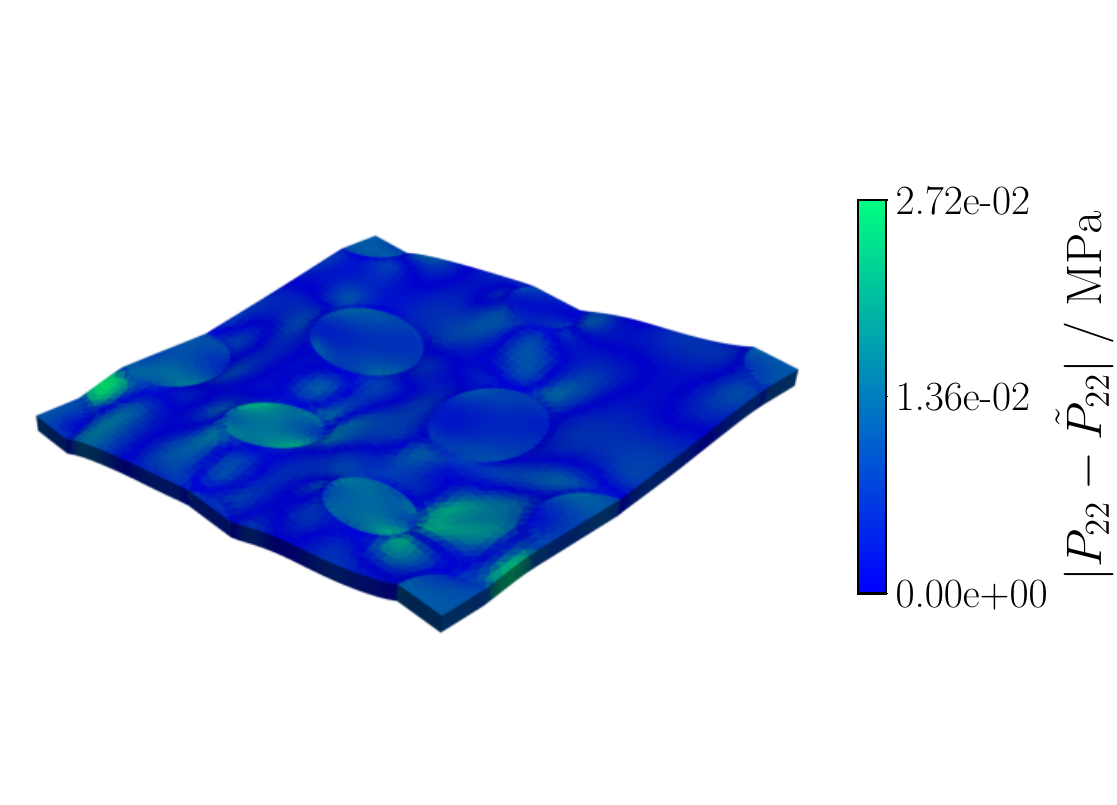}
    \end{subfigure}
    \hfil

    \begin{subfigure}{0.2\textwidth}
        \centering
        \includegraphics[height=0.85\textwidth]{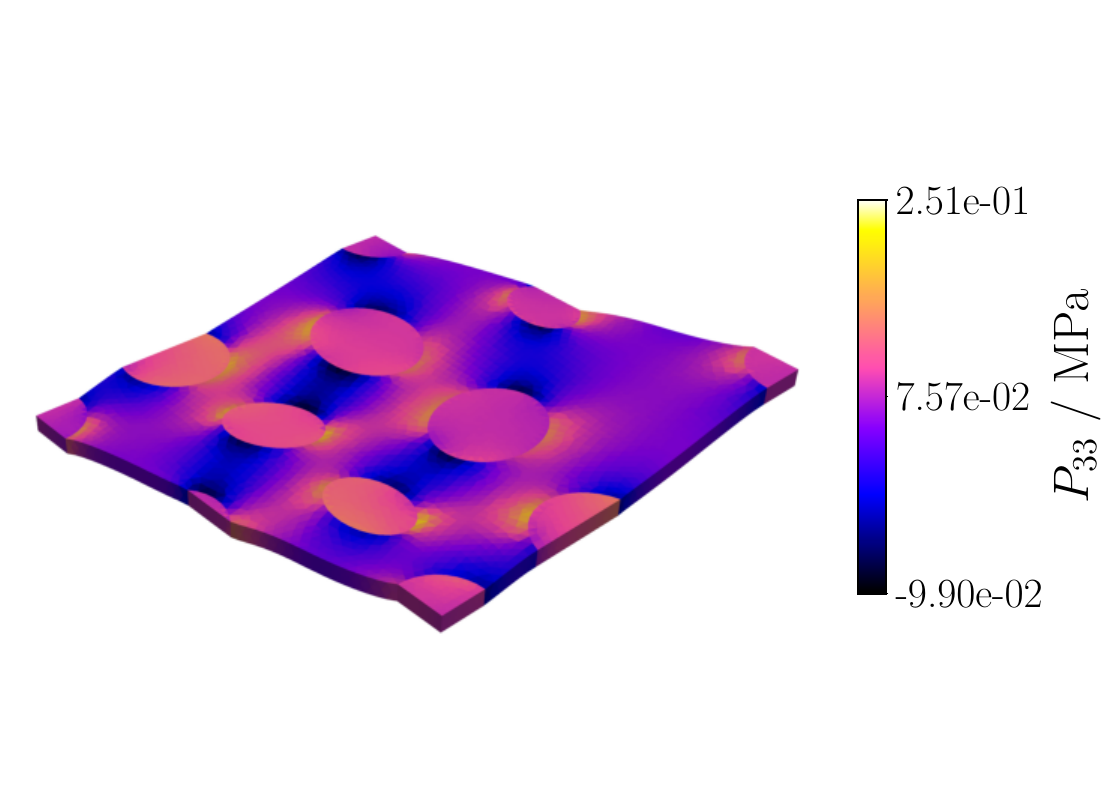}
    \end{subfigure}
    \hfil
    \begin{subfigure}{0.2\textwidth}
        \centering
        \includegraphics[height=0.85\textwidth]{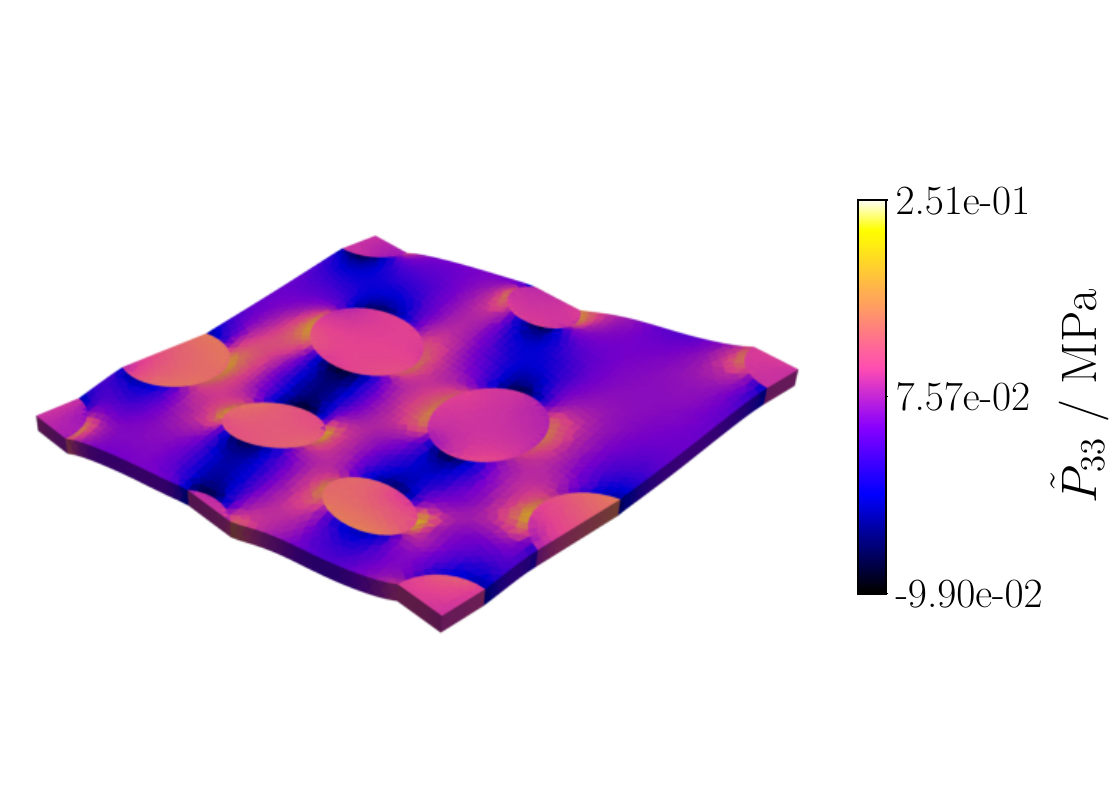}
    \end{subfigure}
    \hfil
    \begin{subfigure}{0.2\textwidth}
        \centering
        \includegraphics[height=0.85\textwidth]{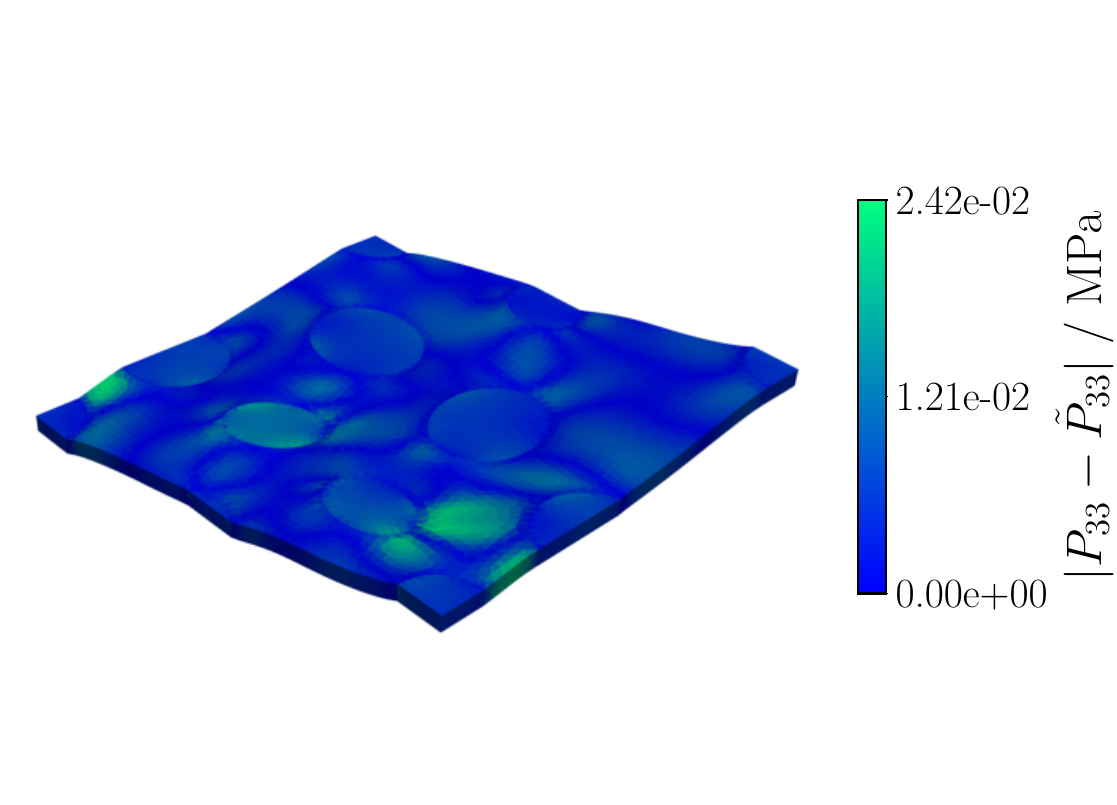}
    \end{subfigure}
    \hfil
    \caption{Comparison of reference, predicted, and absolute-error fields for the microscopic first Piola--Kirchhoff stress components of the stochastic-fiber RVE for the test sample with the maximum relative $L_2$ error among all in-range test samples. See \cref{fig:sto_vis} for the description of the visualization.}
    \label{fig:sto_vis_max}
\end{figure}

\begin{figure}[htb]
    \centering
    \begin{subfigure}{0.2\textwidth}
        \centering
        \includegraphics[height=0.85\textwidth]{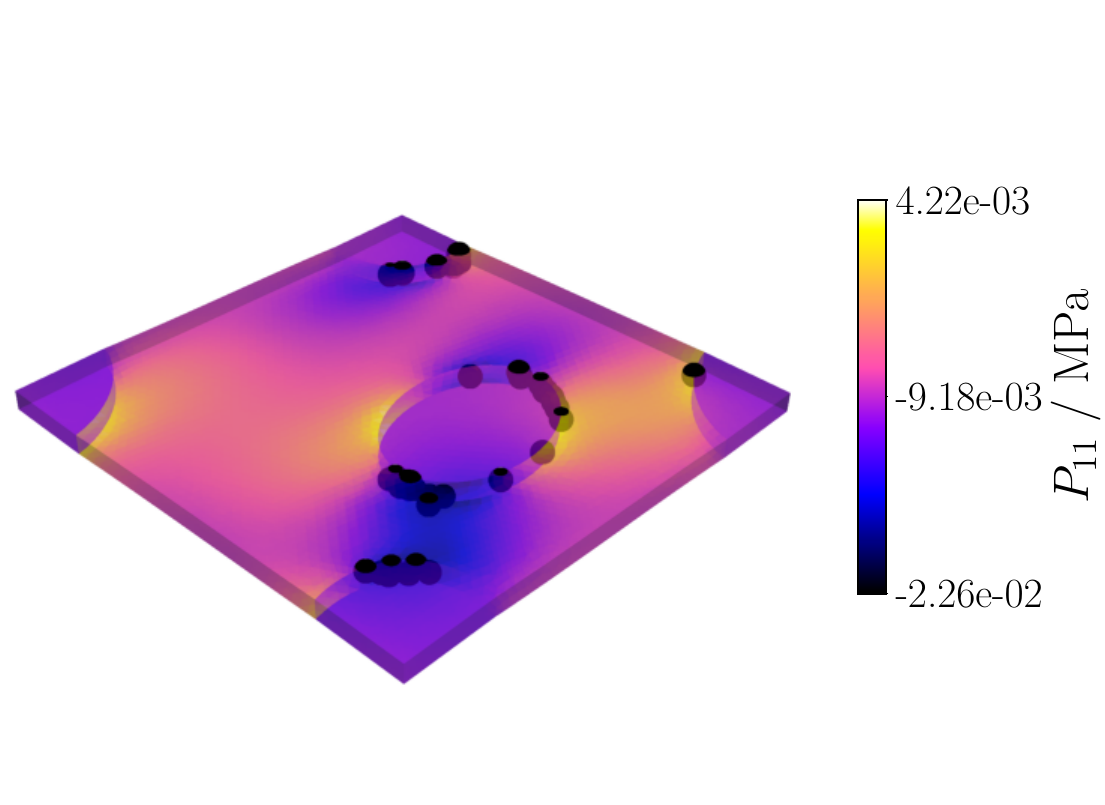}
    \end{subfigure}
    \hfil
    \begin{subfigure}{0.2\textwidth}
        \centering
        \includegraphics[height=0.85\textwidth]{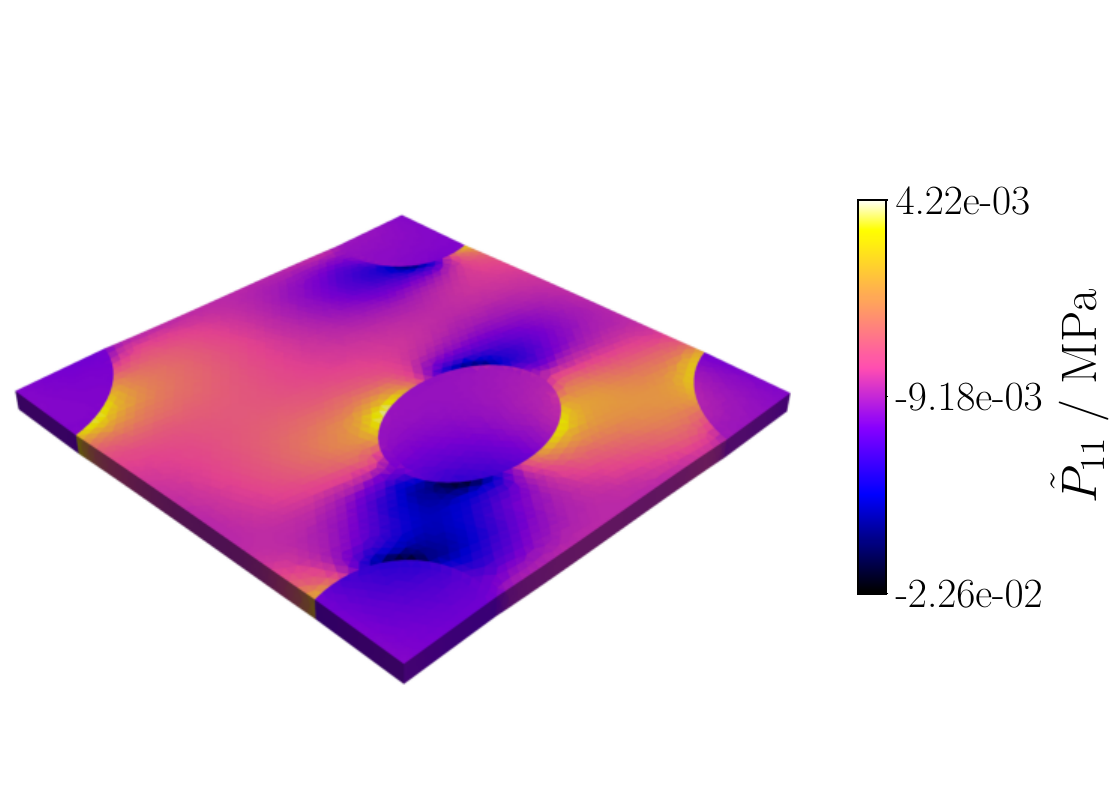}
    \end{subfigure}
    \hfil
    \begin{subfigure}{0.2\textwidth}
        \centering
        \includegraphics[height=0.85\textwidth]{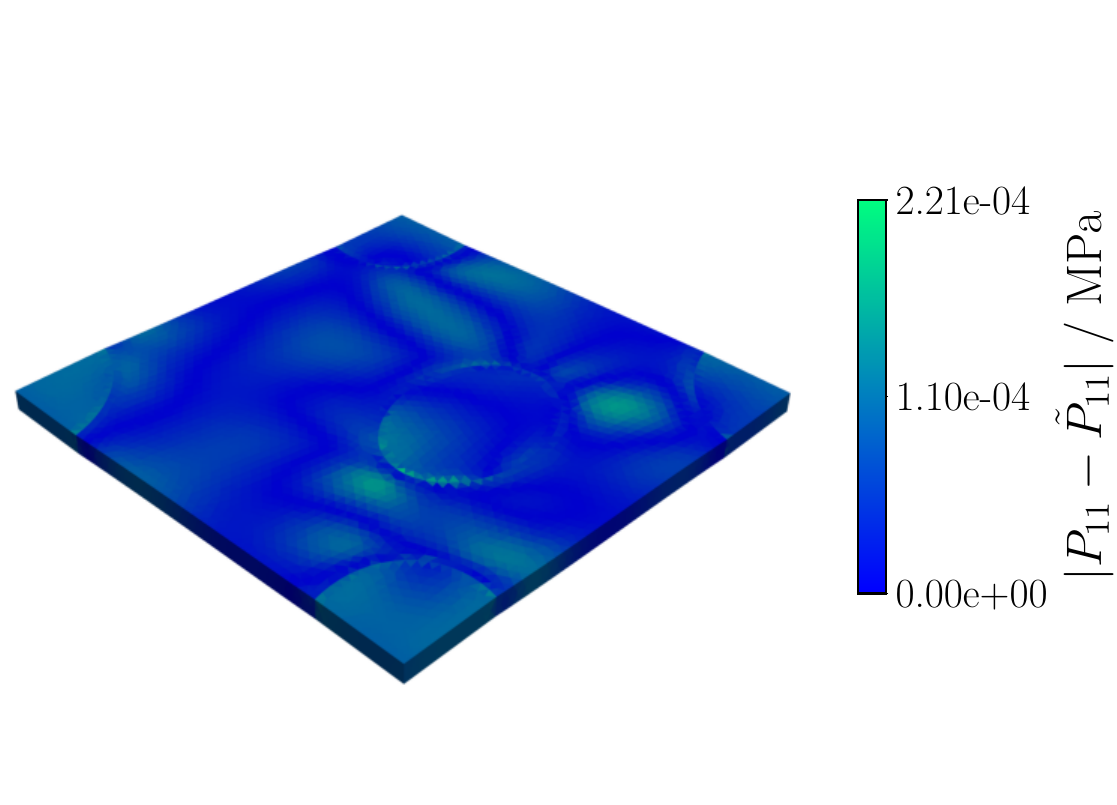}
    \end{subfigure}
    \hfil

    \begin{subfigure}{0.2\textwidth}
        \centering
        \includegraphics[height=0.85\textwidth]{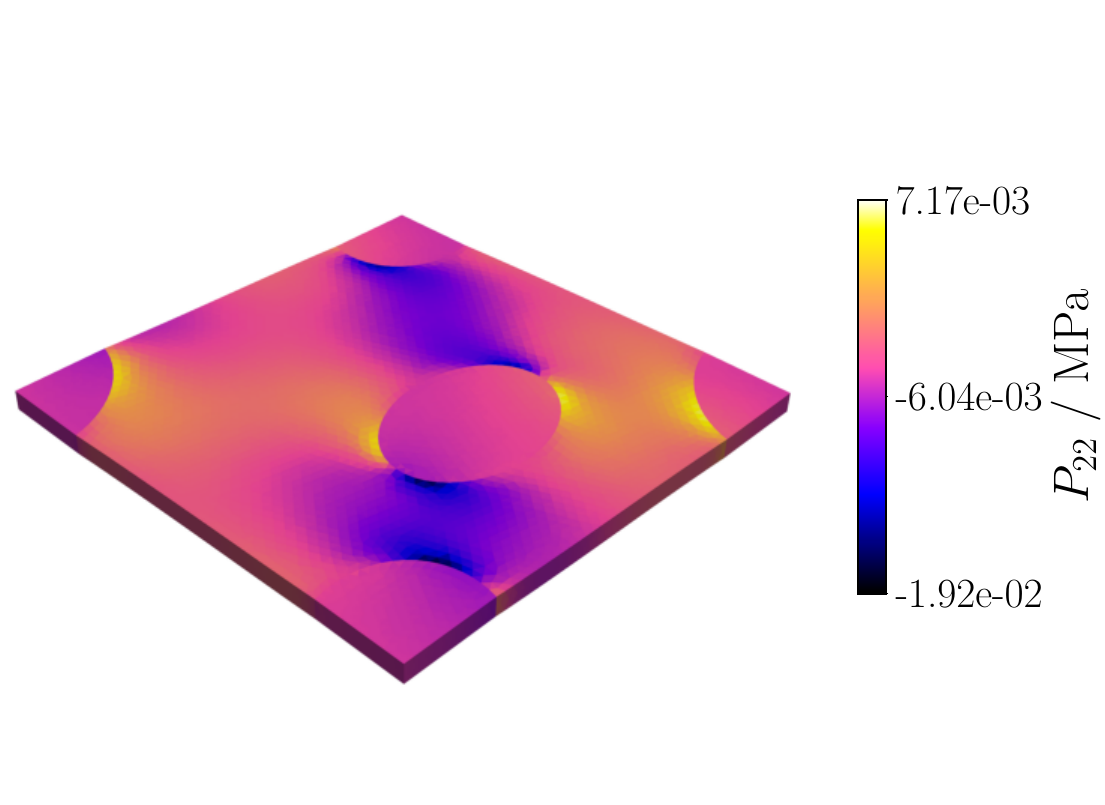}
    \end{subfigure}
    \hfil
    \begin{subfigure}{0.2\textwidth}
        \centering
        \includegraphics[height=0.85\textwidth]{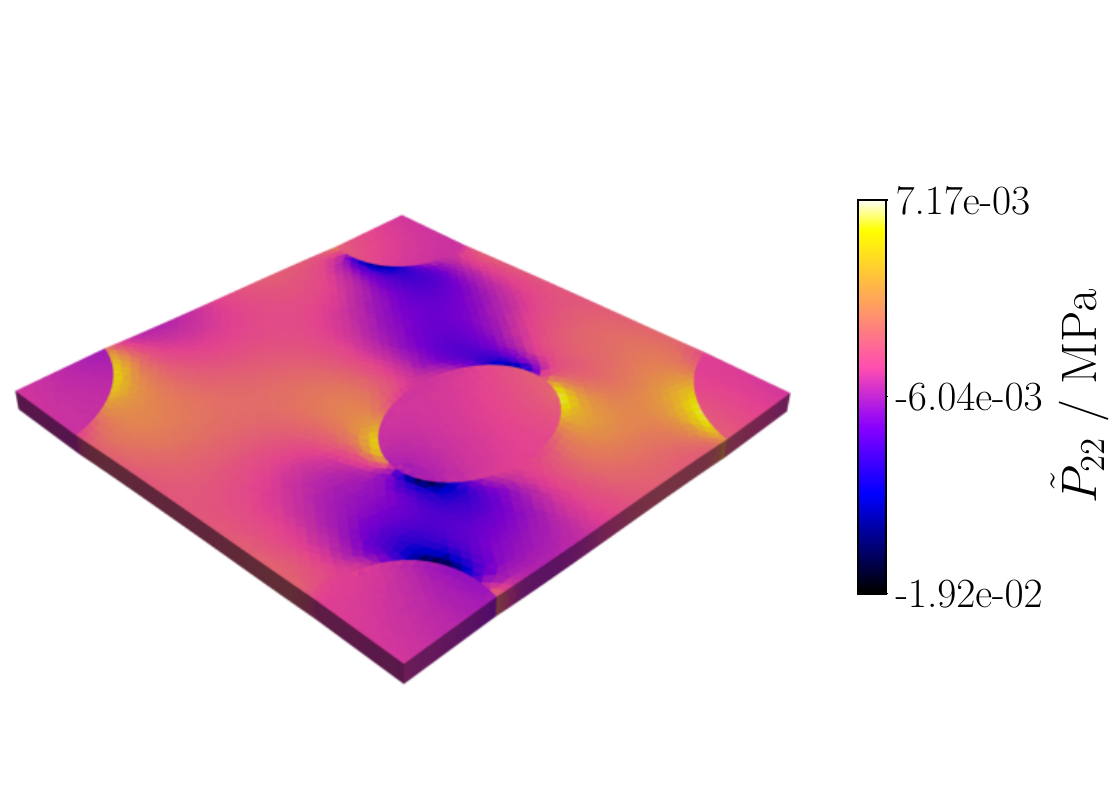}
    \end{subfigure}
    \hfil
    \begin{subfigure}{0.2\textwidth}
        \centering
        \includegraphics[height=0.85\textwidth]{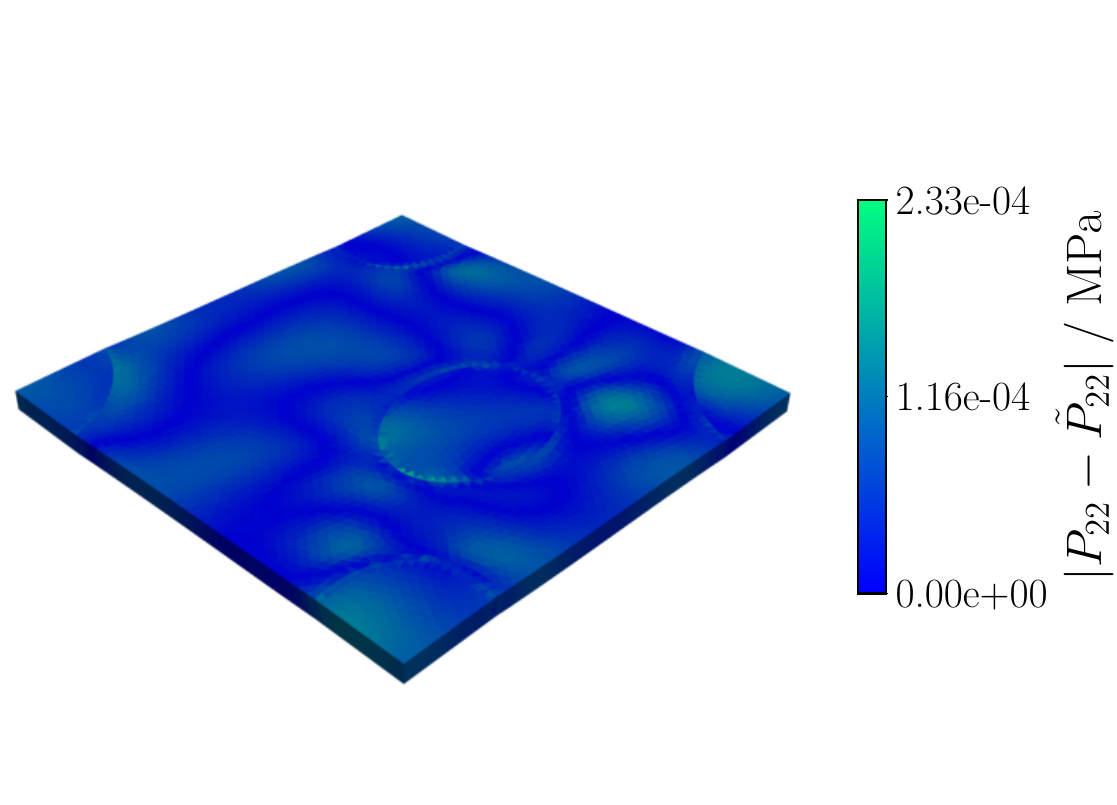}
    \end{subfigure}
    \hfil

    \begin{subfigure}{0.2\textwidth}
        \centering
        \includegraphics[height=0.85\textwidth]{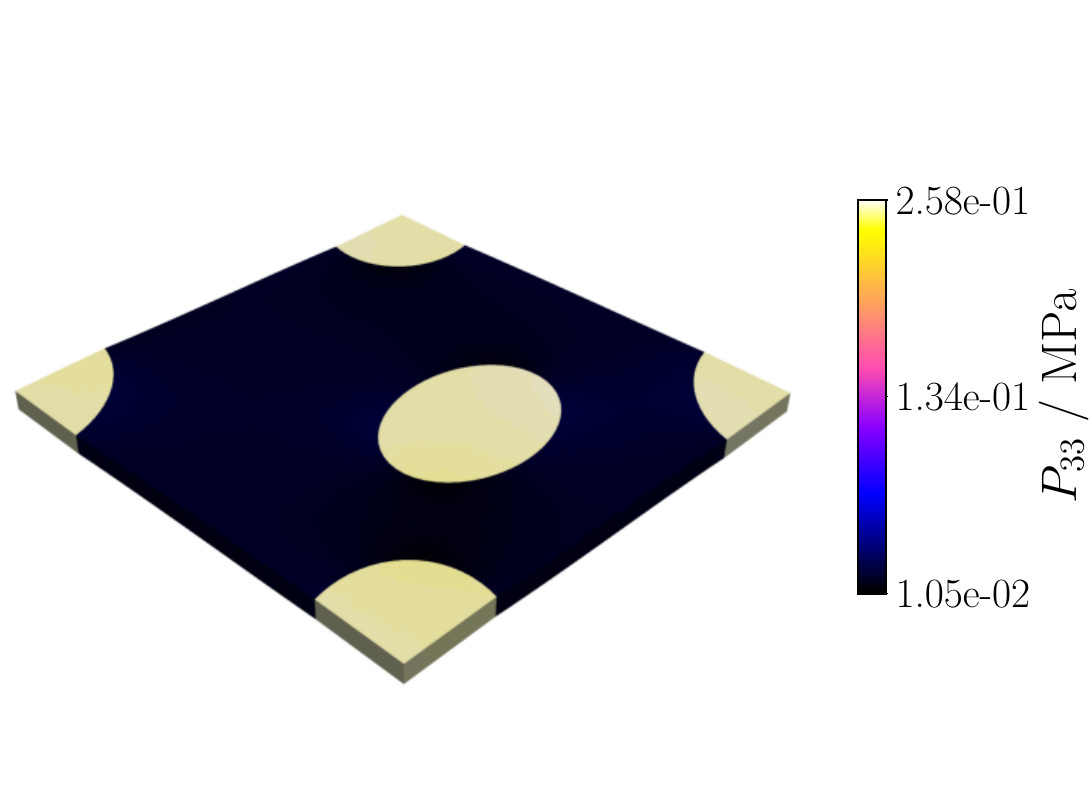}
    \end{subfigure}
    \hfil
    \begin{subfigure}{0.2\textwidth}
        \centering
        \includegraphics[height=0.85\textwidth]{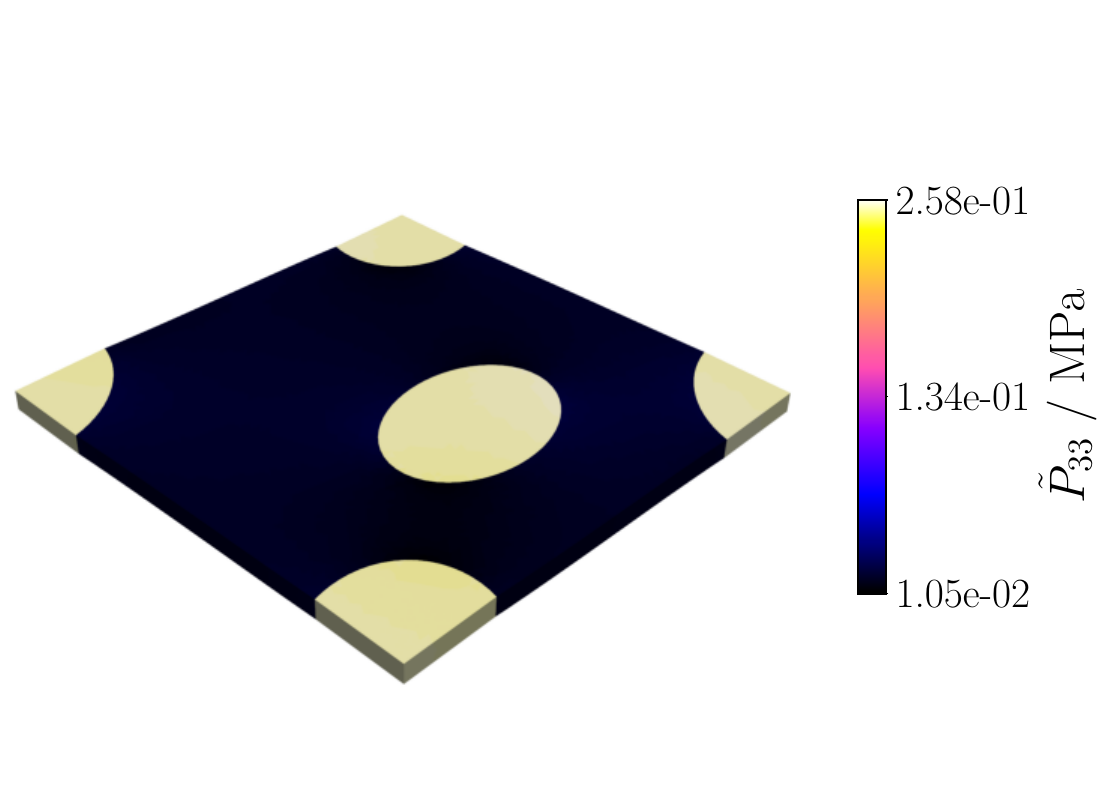}
    \end{subfigure}
    \hfil
    \begin{subfigure}{0.2\textwidth}
        \centering
        \includegraphics[height=0.85\textwidth]{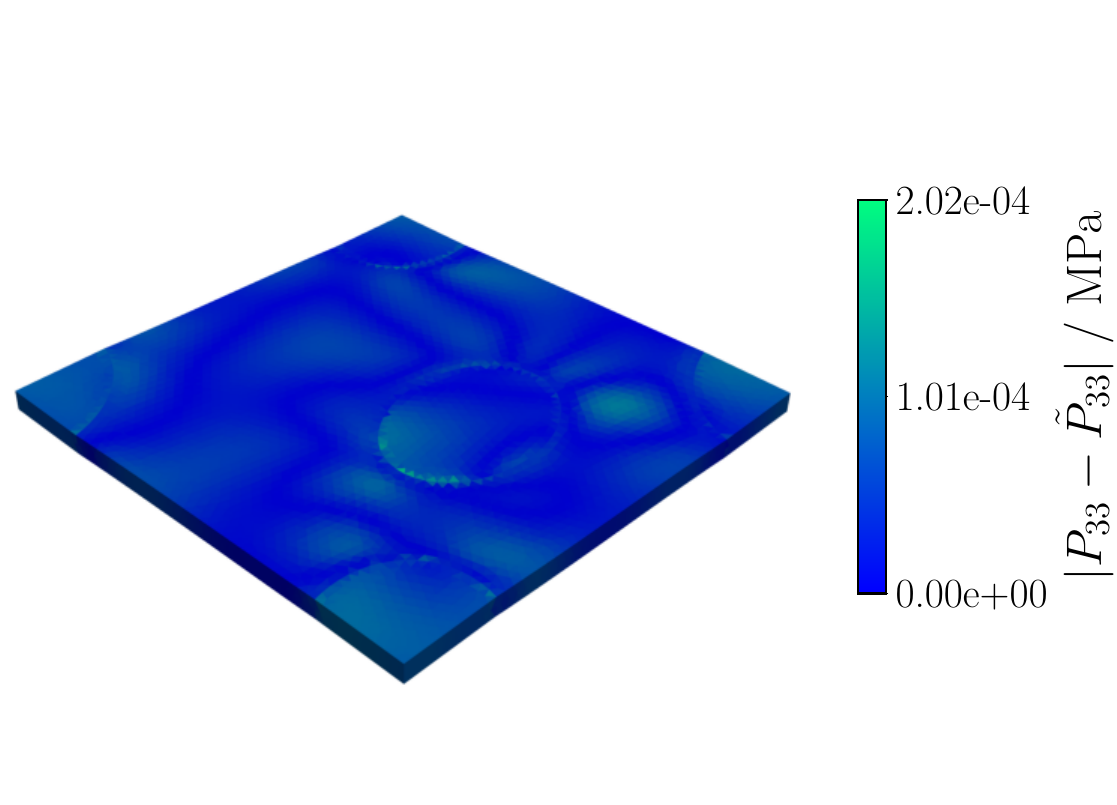}
    \end{subfigure}
    \hfil
    \caption{Comparison of reference, predicted, and absolute-error fields for the microscopic first Piola--Kirchhoff stress components of the hexagonal-fiber RVE for the test sample with the minimum relative $L_2$ error among all in-range test samples. See \cref{fig:hex_vis} for the description of the visualization.}
    \label{fig:hex_vis_min}
\end{figure}

\begin{figure}[htb]
    \centering
    \begin{subfigure}{0.2\textwidth}
        \centering
        \includegraphics[height=0.85\textwidth]{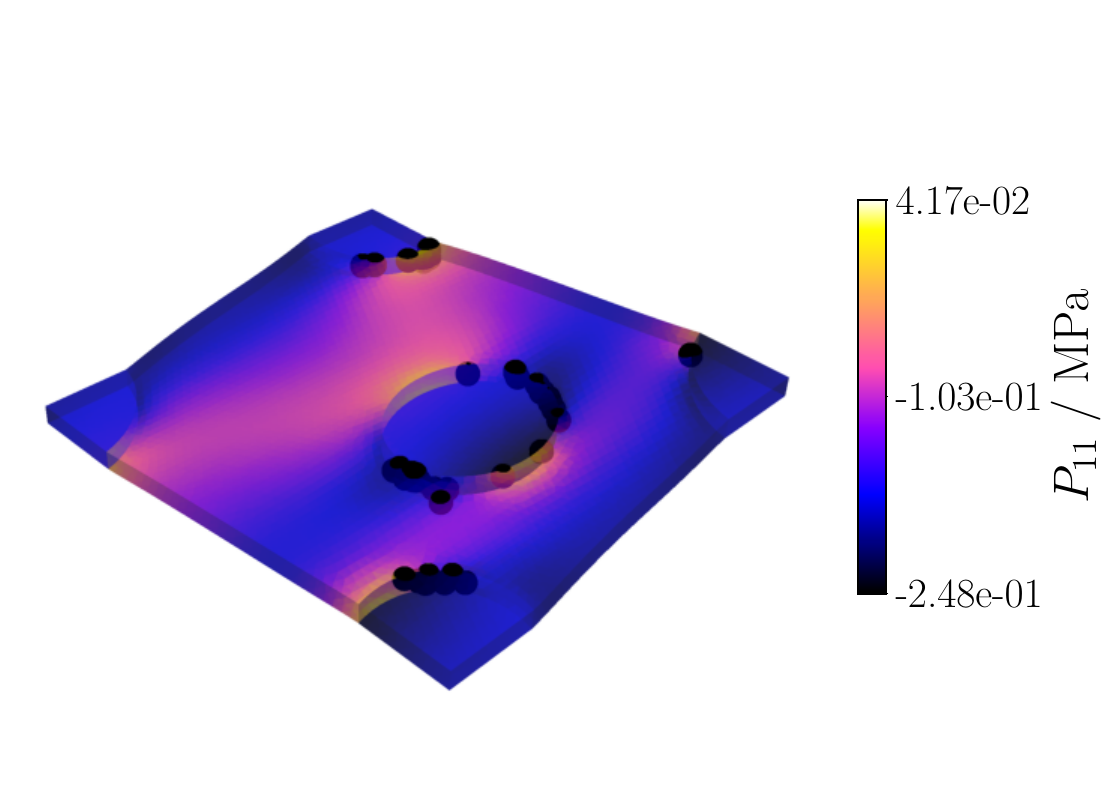}
    \end{subfigure}
    \hfil
    \begin{subfigure}{0.2\textwidth}
        \centering
        \includegraphics[height=0.85\textwidth]{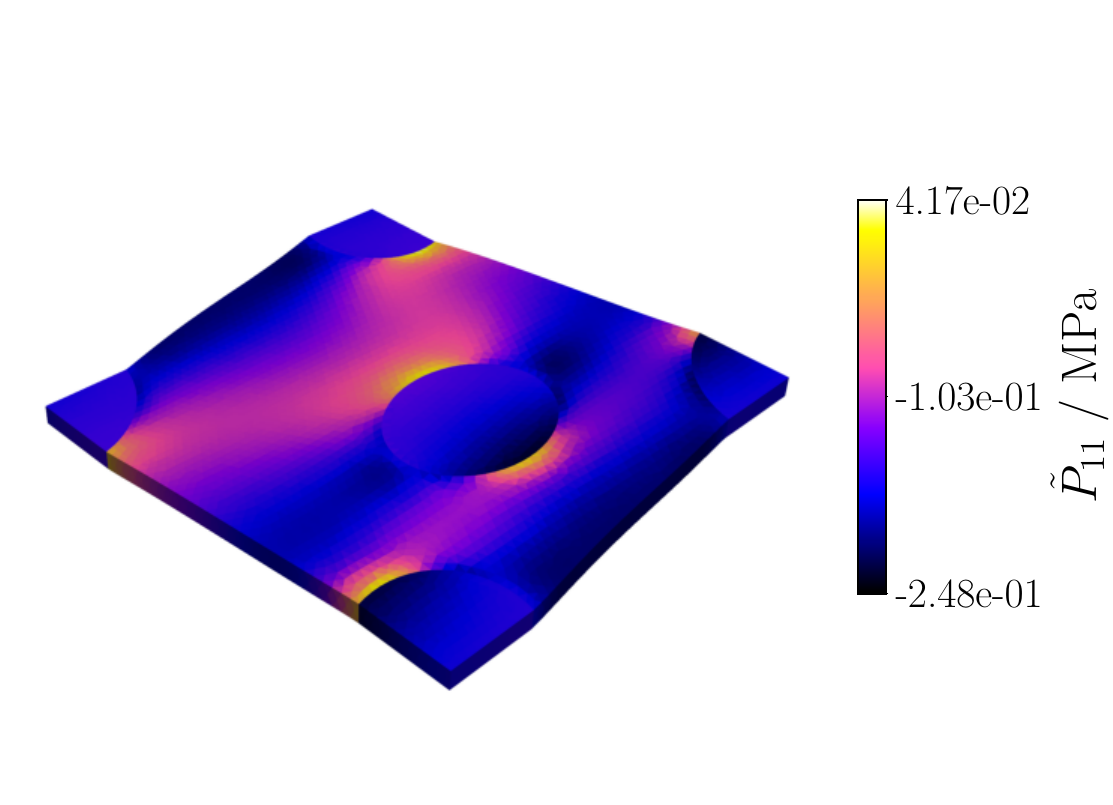}
    \end{subfigure}
    \hfil
    \begin{subfigure}{0.2\textwidth}
        \centering
        \includegraphics[height=0.85\textwidth]{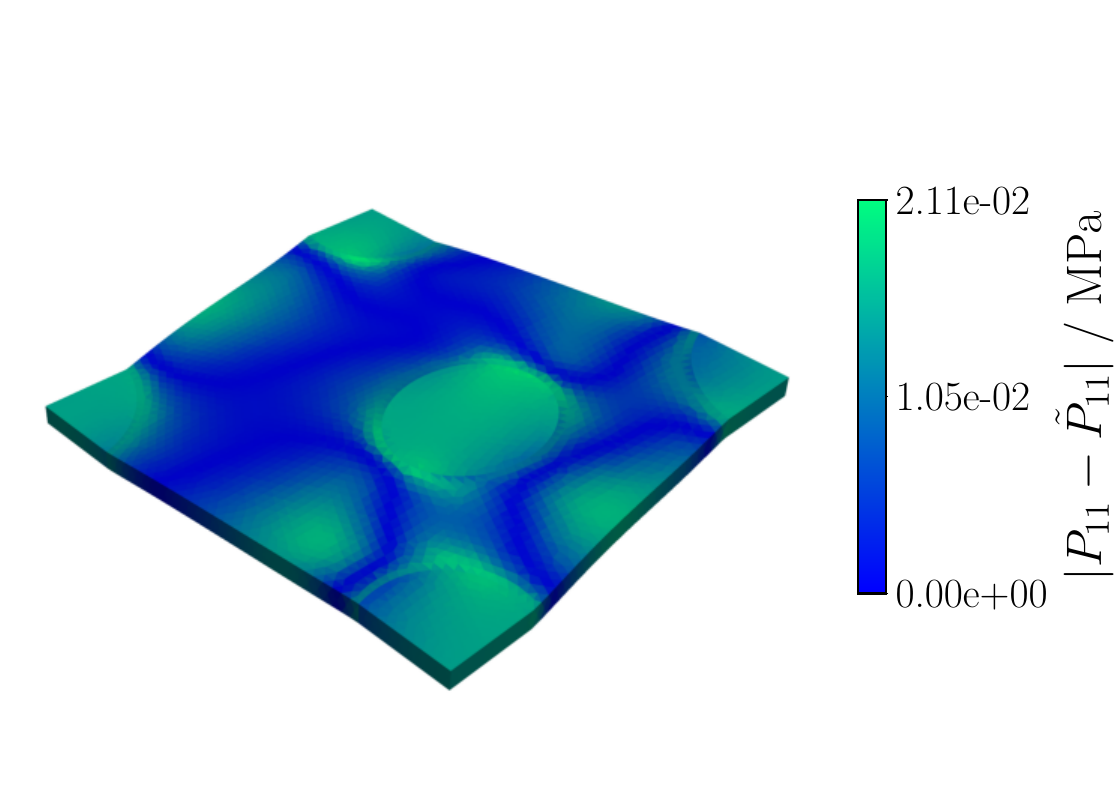}
    \end{subfigure}
    \hfil

    \begin{subfigure}{0.2\textwidth}
        \centering
        \includegraphics[height=0.85\textwidth]{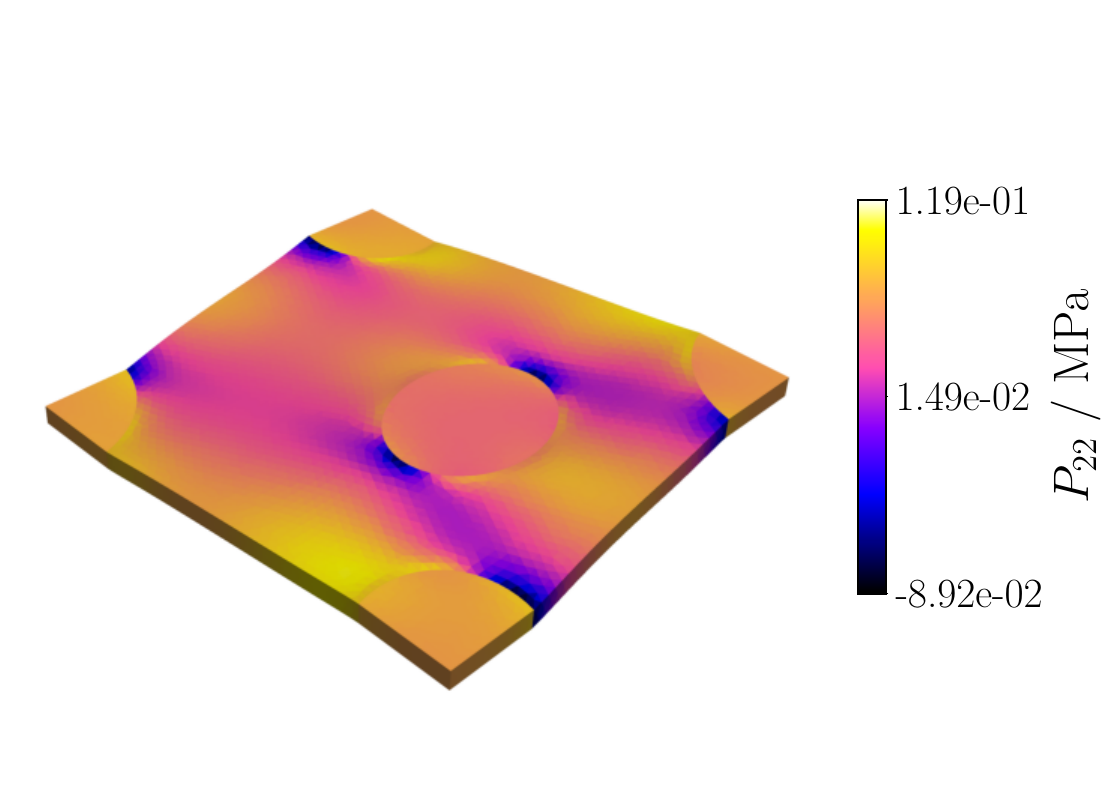}
    \end{subfigure}
    \hfil
    \begin{subfigure}{0.2\textwidth}
        \centering
        \includegraphics[height=0.85\textwidth]{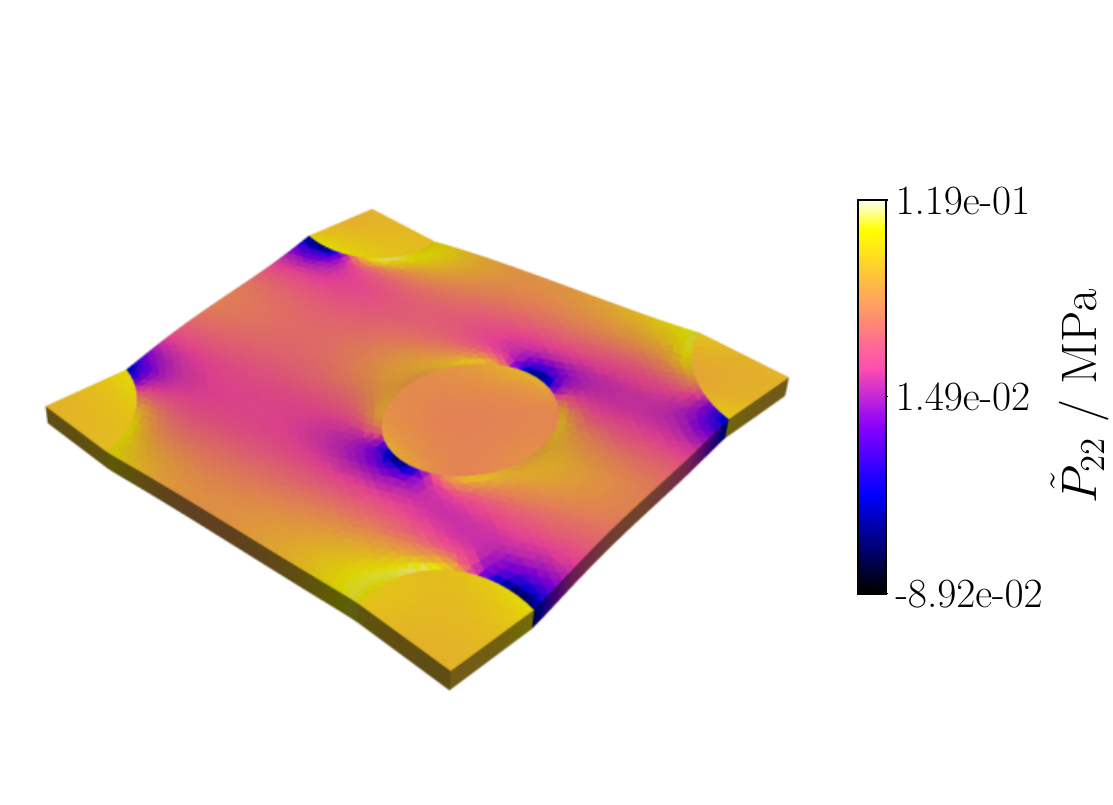}
    \end{subfigure}
    \hfil
    \begin{subfigure}{0.2\textwidth}
        \centering
        \includegraphics[height=0.85\textwidth]{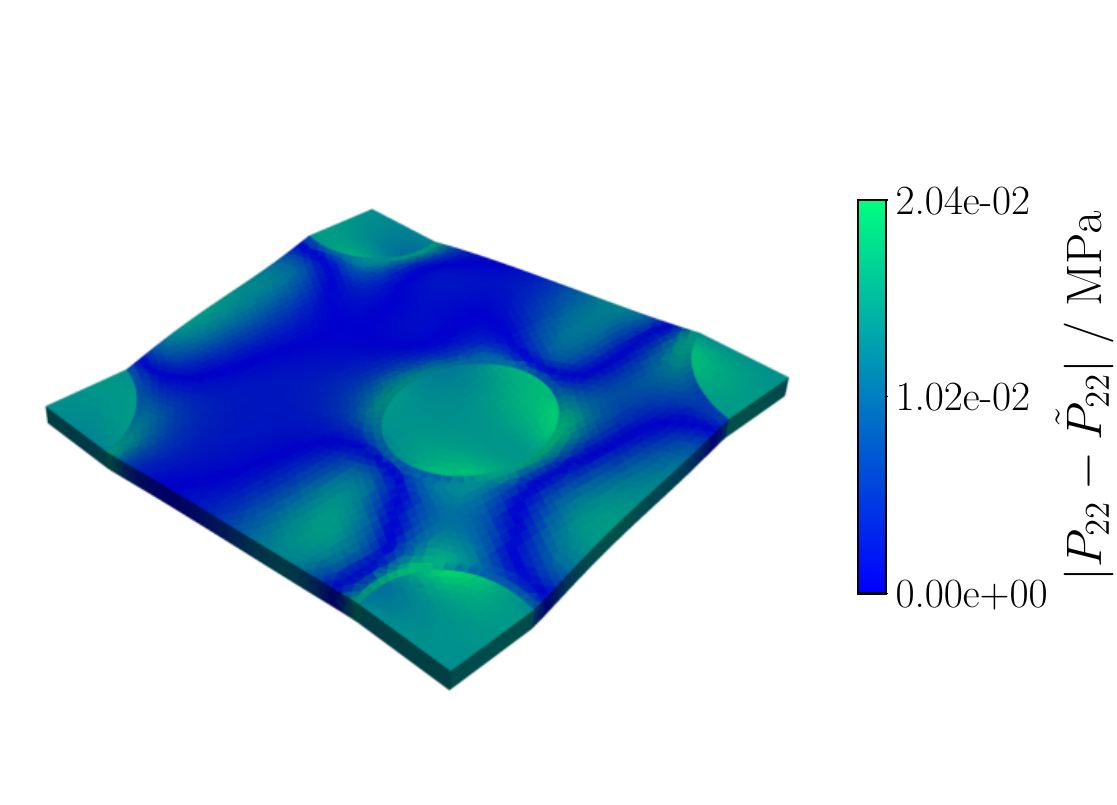}
    \end{subfigure}
    \hfil

    \begin{subfigure}{0.2\textwidth}
        \centering
        \includegraphics[height=0.85\textwidth]{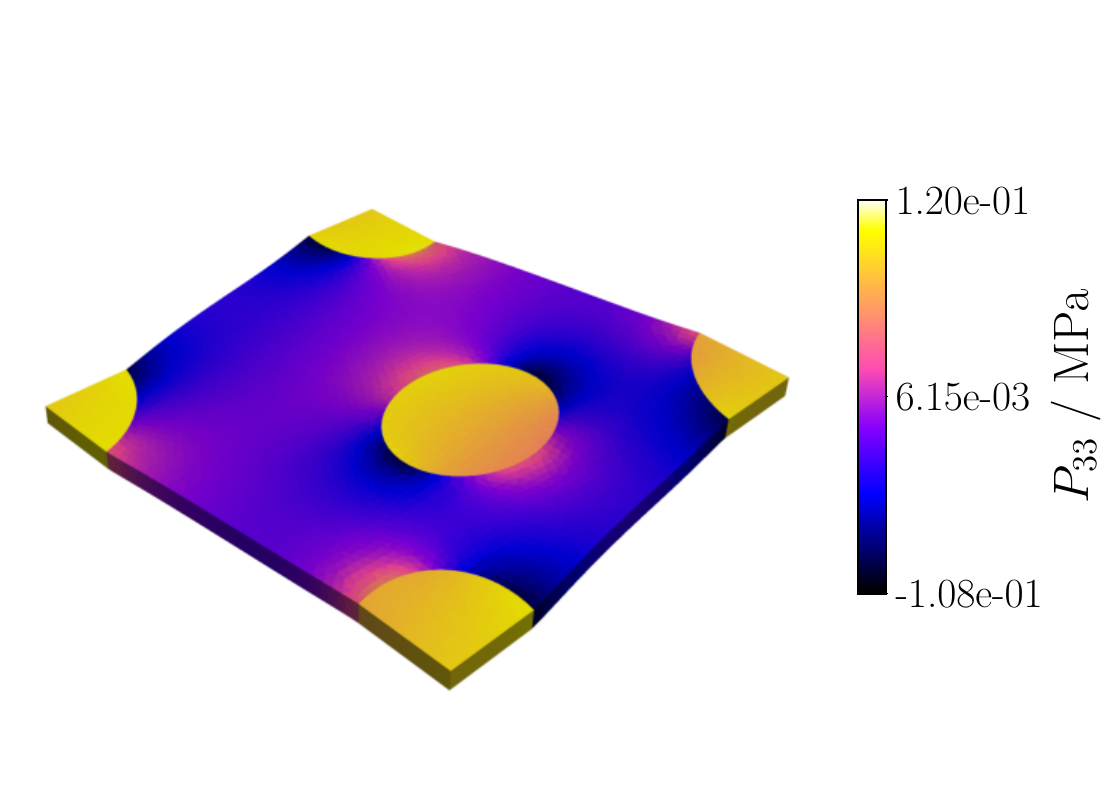}
    \end{subfigure}
    \hfil
    \begin{subfigure}{0.2\textwidth}
        \centering
        \includegraphics[height=0.85\textwidth]{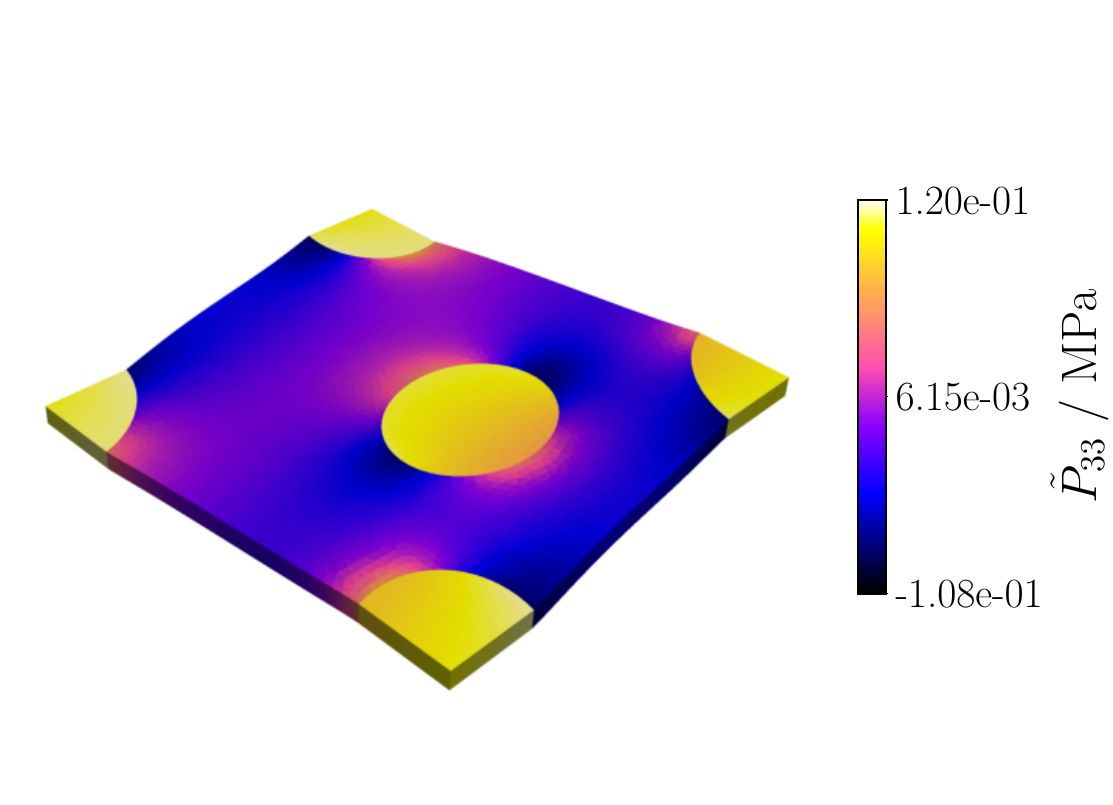}
    \end{subfigure}
    \hfil
    \begin{subfigure}{0.2\textwidth}
        \centering
        \includegraphics[height=0.85\textwidth]{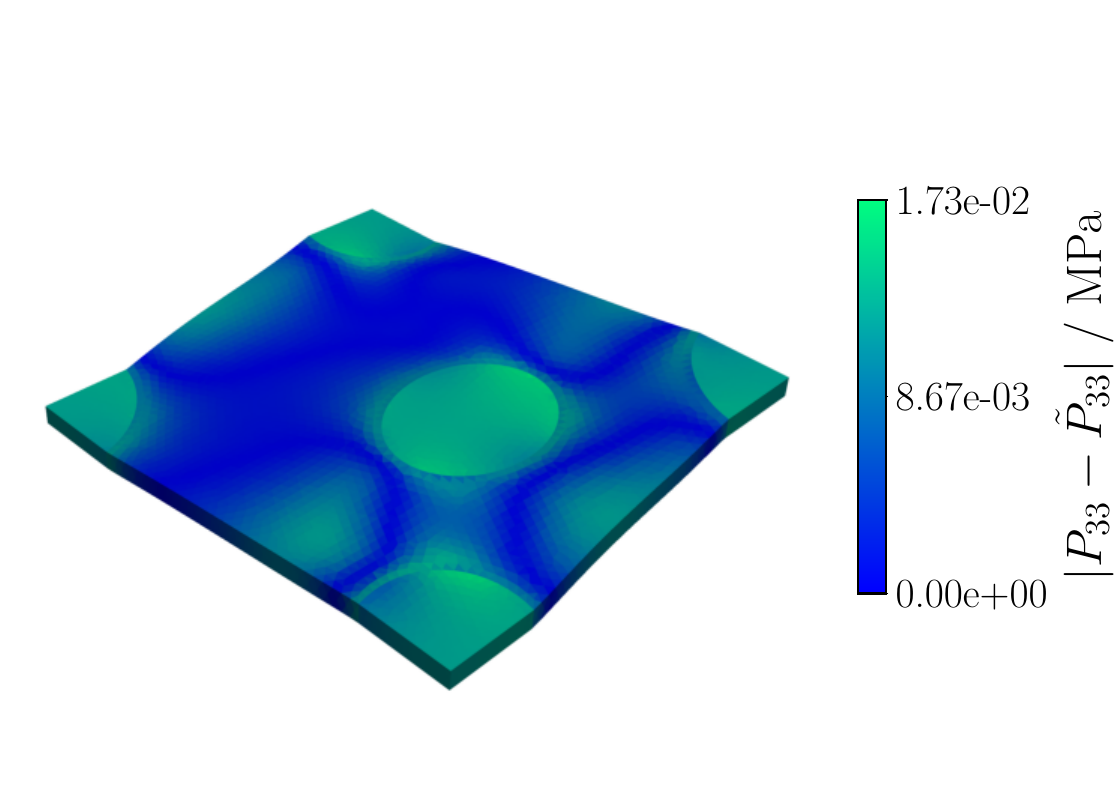}
    \end{subfigure}
    \hfil
    \caption{Comparison of reference, predicted, and absolute-error fields for the microscopic first Piola--Kirchhoff stress components of the hexagonal-fiber RVE for the test sample with the maximum relative $L_2$ error among all in-range test samples. See \cref{fig:hex_vis} for the description of the visualization.}
    \label{fig:hex_vis_max}
\end{figure}

We also include additional field visualizations for the stochastic-fiber and hexagonal-fiber RVEs, showing the in-range test samples with the minimum and maximum relative $L_2$ errors in \cref{fig:sto_vis_min,fig:sto_vis_max,fig:hex_vis_min,fig:hex_vis_max}. These cases complement the median-error examples shown in the main text and provide a broader view of the spatial stress-prediction accuracy.

\end{document}

%% file: math_commands.tex
\usepackage{amsmath,amsfonts,bm}

\def\eqref#1{equation~\ref{#1}}

\def\1{\bm{1}}

\def\vzero{{\bm{0}}}
\def\vone{{\bm{1}}}

\def\vtheta{{\bm{\theta}}}

\def\vvarphi{{\bm{\varphi}}}
\def\vxi{{\bm{\xi}}}
\def\vzeta{{\bm{\zeta}}}

\def\vb{{\bm{b}}}

\def\vu{{\bm{u}}}

\def\vx{{\bm{x}}}
\def\vy{{\bm{y}}}

\def\mC{{\bm{C}}}

\def\mE{{\bm{E}}}
\def\mF{{\bm{F}}}

\def\mI{{\bm{I}}}

\def\mP{{\bm{P}}}
\def\mQ{{\bm{Q}}}
\def\mR{{\bm{R}}}
\def\mS{{\bm{S}}}

\def\mX{{\bm{X}}}
\def\mY{{\bm{Y}}}

\def\mPhi{{\bm{\Phi}}}

\DeclareMathAlphabet{\mathsfit}{\encodingdefault}{\sfdefault}{m}{sl}
\SetMathAlphabet{\mathsfit}{bold}{\encodingdefault}{\sfdefault}{bx}{n}

\def\gD{{\mathcal{D}}}

\def\gG{{\mathcal{G}}}

\def\gL{{\mathcal{L}}}

\def\gN{{\mathcal{N}}}

\def\sA{{\mathbb{A}}}

